\documentclass{article}
\usepackage[utf8]{inputenc}

\usepackage{tikz}
\definecolor{colorSFPLAIC}{HTML}{cc0000}
\definecolor{colorSFPLBIC}{HTML}{f1c232}
\definecolor{colorPL}{HTML}{6aa84f}
\definecolor{colorPPL}{HTML}{3d85c6}

\usepackage[unicode,linktoc=all,hidelinks]{hyperref}
\usepackage[a4paper, margin=2cm]{geometry}
\usepackage{mathtools}

\usepackage{amsmath}
\usepackage{amssymb}
\usepackage{amsthm}
\usepackage{bm}
\usepackage{bbm}

\DeclareMathOperator*{\argmin}{arg\,min}

\DeclareSymbolFont{cmletters}{OT1}{cmr}{m}{n}
\DeclareMathSymbol{\Ups}{\mathalpha}{cmletters}{"7}

\usepackage{verbatim}

\usepackage{float} 
\usepackage{adjustbox}
\usepackage{threeparttable}
\usepackage{booktabs}

\usepackage{algorithm}
\usepackage{algorithmic}

\usepackage[english]{babel}
\usepackage{natbib}

\usepackage{graphicx}
\usepackage{multicol}

\usepackage[title]{appendix}

\newtheorem{assumption}{Assumption}

\newtheorem{lemma}{Lemma}
\newtheorem{theorem}{Theorem}
\newtheorem{corollary}{Corollary}

\newtheorem{proposition}{Proposition}

\title{Multi-Attribute Preferences: A Transfer Learning Approach}
\author{Sjoerd Hermes$^{1,2}$, Joost van Heerwaarden$^{1,2}$ and Pariya Behrouzi$^1$}
\date{%
    $^1$ Mathematical and Statistical Methods, Wageningen University\\%
    $^2$ Plant Production Systems, Wageningen University\\
}

\begin{document}

\maketitle
\begin{abstract}
\noindent We introduce a transfer-learning method based on the Bradley--Terry model for multi-attribute pairwise-comparison data. The aim is to estimate the log-worth parameters of one primary attribute while using information from related secondary attributes. The method first pools the primary data with data from informative secondary attributes. It then corrects this pooled estimate using the primary likelihood, with a ridge penalty controlling the size of the correction. When the informative set is unknown, we use held-out primary data to select secondary attributes. For a known informative set and under a pooled Bradley--Terry compatibility condition, we derive high-probability $\ell_\infty$ and $\ell_2$ error bounds. Under additional conditions, these bounds can have a smaller asymptotic order than the corresponding primary-only Bradley--Terry bounds. We also establish asymptotic normality for a one-step estimator. A simulation study evaluates the method under more general settings, and an application to consumer preferences for eba, a cassava-derived food product, illustrates its use and interpretation. An R package implementing the method is available at \url{https://CRAN.R-project.org/package=BTTL}.
\end{abstract}

\noindent%

\section{Introduction}
Preference data are widely used to rank a set of objects. Such data arise in economics (Palma, \citeyear{palma2017improving}), engineering (Franceschini et al., \citeyear{franceschini2022rankings}), information retrieval (Jeon \& Kim, \citeyear{jeon2013revisiting}), marketing (Yang et al., \citeyear{yang2002modeling}) and nutritional science (Hermes et al., \citeyear{hermes2024joint}). Preferences may be observed as pairwise comparisons, partial rankings or click-through data and can then be combined into an overall ranking. Many methods have been developed for this purpose. Classical examples include the Thurstone (Thurstone, \citeyear{thurstone1927law}), Babington Smith (\citeyear{babingtonsmith1950}), Bradley--Terry (Bradley \& Terry, \citeyear{bradley1952rank}), Mallows (\citeyear{mallows1957non}) and Plackett--Luce (Luce, \citeyear{luce1959}; Plackett, \citeyear{plackett1975analysis}) models. More recent methods include both statistical approaches (Deng et al., \citeyear{deng2014bayesian}; Xu et al., \citeyear{xu2018angle}; Li et al., \citeyear{li2022bayesian}) and machine-learning approaches (Dai et al., \citeyear{dai2021scalable}; Han et al., \citeyear{han2018robust}; Zhu et al., \citeyear{zhu2023principled}).

Most ranking methods focus on one attribute. Fewer methods consider preferences measured on several attributes, where an attribute is any property on which the objects can be compared. Existing multi-attribute methods usually model all attributes jointly (Dittrich et al., \citeyear{dittrich2006modelling}; Wang et al., \citeyear{wang2017variational}). This may be unnecessary when the main interest lies in only one attribute. For example, hotel guests may assess breakfast, hygiene, price and service in addition to overall satisfaction (Krivulin et al., \citeyear{krivulin2022using}; Bi et al., \citeyear{bi2022ranking}), and food studies may record taste, smell and appearance together with overall preference (Olaosebikan et al., \citeyear{olaosebikan2023drivers}). We call the main attribute of interest the primary attribute and the remaining attributes secondary attributes.

Machine learning distinguishes between learning several related tasks jointly and improving one task using information from other tasks. The first setting is usually called multi-task learning (Zhang \& Yang, \citeyear{zhang2018overview}); the second is called transfer learning (Weiss et al., \citeyear{weiss2016survey}; Zhuang et al., \citeyear{zhuang2020comprehensive}). Our aim fits the transfer-learning setting because we use secondary attributes to improve estimation for one primary attribute. In the general transfer-learning literature these would often be called the target and source tasks. We use primary and secondary attributes because these terms better match the preference setting considered here.

Transfer learning has been studied in machine learning for more than a decade (Dai et al., \citeyear{dai2007boosting}; Pan et al., \citeyear{pan2008transfer}; Taylor \& Stone, \citeyear{taylor2009transfer}; Konidaris et al., \citeyear{konidaris2012transfer}). More recent statistical work has focused on properties such as convergence rates and asymptotic normality. Examples include linear regression (Li et al., \citeyear{li2022transfer}; \citeyear{li2023targeting}; Liu, \citeyear{liu2023unified}; Zhao et al., \citeyear{zhao2023residual}; Fan et al., \citeyear{fan2024fast}), generalised linear models (Li et al., \citeyear{li2023estimation}; Sun \& Zhang, \citeyear{sun2023robust}; Tian \& Feng, \citeyear{tian2023transfer}), Cox models (Li et al., \citeyear{li2023accommodating}), functional linear regression (Lin et al., \citeyear{lin2022transfer}), quantile regression (Huang et al., \citeyear{huang2022estimation}; Jin et al., \citeyear{jin2024transfer}), Gaussian graphical models (He et al., \citeyear{he2022transfer}; Li et al., \citeyear{li2023transfer}) and Gaussian mixture models (Tian et al., \citeyear{tian2022unsupervised}).

We bring this transfer-learning idea to ranking models. The main method is based on the Bradley--Terry model, with a Plackett--Luce extension for partial rankings. Because only some secondary attributes may be useful, we adapt the procedure of Tian and Feng (\citeyear{tian2023transfer}) to screen the secondary attributes using held-out primary data. For a known informative set, we derive high-probability error bounds and give conditions under which the transfer bound is asymptotically smaller than the primary-only Bradley--Terry bound. A simulation study compares the method with primary-only estimation and with pooling all attributes. We also illustrate the method using consumer preference data for eba, a cassava-derived food product.

The rest of the paper is organised as follows. Section \ref{sec:methodology} introduces the Bradley--Terry model and the proposed transfer-learning procedures for known and unknown informative sets. Section \ref{sec:statistical-properties} gives high-probability error bounds for the oracle procedure and derives the asymptotic distribution of a one-step estimator based on a separate primary inference sample. Section \ref{sec:simulation} presents the simulation study, Section \ref{sec:application} gives the eba application, and Section \ref{sec:conclusion} concludes the paper.

\section{Methodology} \label{sec:methodology}
\noindent We use the following notation. The primary data are denoted by $\mathcal D_0$ and the data for secondary attribute $s$ by $\mathcal D_s$. Individual Bernoulli comparison outcomes are denoted by lower-case $y$, for example $y_{l,j}^{(i)}$. Matrices are written as bold capital letters (e.g.\ $\bm A$), vectors as bold lower-case letters (e.g.\ $\bm a$), and scalars as non-bold lower-case letters (e.g.\ $a$). A superscript $*$ denotes a population quantity and a hat denotes an estimate. Symbols without either are used as generic arguments or optimisation variables. For example, $\bm\alpha^*$ is the population log-worth vector for the primary attribute and $\hat{\bm\alpha}_{\lambda_\delta}$ is its ridge-corrected estimator. For a matrix $\bm A$, let $\|\bm A\|_2=\|\bm A\|$ denote the spectral operator norm and $\|\bm A\|_\infty:=\max_i\sum_j|A_{ij}|$ the induced matrix $\ell_\infty$ norm. We write $[\bm A]_j$ for its $j$th row and $[\bm A]_{jk}$ for its $(j,k)$ entry. For a vector $\bm a\in\mathbb R^M$, let $\|\bm a\|_2=\sqrt{\sum_{i=1}^M a_i^2}$ and $\|\bm a\|_\infty=\max_{1\le i\le M}|a_i|$. For a symmetric matrix $\bm A$, define $\lambda_{\min,\perp}(\bm A):=\min_{\bm v\perp\bm1_M,\,\|\bm v\|_2=1}\bm v^T\bm A\bm v$, the smallest Rayleigh quotient on the identifiable subspace $\bm1_M^\perp$. We write $\bm A\succeq\bm0$ when $\bm A$ is positive semidefinite. For a sequence of random variables $X_M$, $X_M\xrightarrow{d}X$ denotes convergence in distribution. For two objects $o_j$ and $o_l$, $o_j\succ o_l$ means that object $j$ is preferred to object $l$. Along an asymptotic sequence, $a_M=O(b_M)$ means that $a_M/b_M$ remains bounded and $a_M=o(b_M)$ means that $a_M/b_M\to0$. We write $a_M\lesssim b_M$ for $a_M=O(b_M)$, $a_M\gtrsim b_M$ for $b_M=O(a_M)$, and $a_M\asymp b_M$ when both hold. In probability statements, $O(M^{-c})$ denotes a quantity whose magnitude is bounded by a constant multiple of $M^{-c}$ for sufficiently large $M$.

\subsection{Bradley--Terry}
We use the Bradley--Terry (BT) model as the basis of the proposed method because its statistical properties are well studied (Simons \& Yao, \citeyear{simons1999asymptotics}; Chen et al., \citeyear{chen2019spectral}; Fan et al., \citeyear{fan2024uncertainty}; Liu et al., \citeyear{liu2023lagrangian}). The Bradley--Terry model (Bradley \& Terry, \citeyear{bradley1952rank}) describes pairwise preferences among $M\ge2$ objects. Each object has a latent log-worth, and differences in log-worth determine the pairwise preference probabilities. Formally,
\begin{equation}
\label{eq:preferencebt}
\mathbb{P}(o_j \succ o_l) = \frac{\exp(\theta_j)}{\exp(\theta_j) + \exp(\theta_l)} \quad \forall 1 \leq j \neq l \leq M,
\end{equation} 
where $\bm\theta\in\mathbb R^M$ is a generic vector of log-worth parameters. We impose $\bm1_M^T\bm\theta=0$ for identifiability. A pairwise comparison can be written as a Bernoulli outcome, with $\mathbb P(o_j\succ o_l)=\mathbb P(y_{l,j}=1)=\mathbb P(y_{j,l}=0)$.

For a pairwise-comparison data set $\mathcal D$, let $\mathcal G_{\mathcal D}=(\mathcal V,\mathcal E_{\mathcal D})$ be the corresponding undirected comparison graph. The vertices are $\mathcal V=\{1,\ldots,M\}$ and $(j,l)\in\mathcal E_{\mathcal D}$ when objects $j$ and $l$ are compared at least once.

Suppose first that every observed edge is compared $n$ times. Conditional on $\mathcal G_{\mathcal D}$, we use the following scaled negative log-likelihood:
\begin{gather}
L(\bm\theta) = -\sum_{(j,l) \in \mathcal E_{\mathcal D}, j > l}\left\{y_{l,j} \log\frac{\exp(\theta_j)}{\exp(\theta_j) + \exp(\theta_l)} + (1 - y_{l,j})\log\frac{\exp(\theta_l)}{\exp(\theta_j) + \exp(\theta_l)}\right\}\notag\\
= \sum_{(j,l) \in \mathcal E_{\mathcal D}, j > l}\left\{-y_{l,j} (\theta_j - \theta_l) + \log[1+\exp(\theta_j - \theta_l)]\right\} \label{eq:loglikbt},
\end{gather}
where $y_{l,j}=n^{-1}\sum_{i=1}^n y_{l,j}^{(i)}$ is the average outcome on edge $(j,l)$ and is a sufficient statistic (Fan et al., \citeyear{fan2024uncertainty}). The objective is smooth and convex and can therefore be minimised using standard convex optimisation methods.

\subsection{Transfer learning}
The previous section considers one attribute. We now allow the log-worth parameters to differ across attributes while keeping the primary attribute as the main object of interest. Let $\bm\alpha^*\in\mathbb R^M$ be the population Bradley--Terry log-worth vector for the primary attribute and let $\bm u^{(s)*}\in\mathbb R^M$, $s=1,\ldots,S$, be the population log-worth vectors for the secondary attributes. Because log-worths are identifiable only up to an additive constant, we use the common normalisation
\[
\bm1_M^T\bm\alpha^*=\bm1_M^T\bm u^{(s)*}=0.
\]
Following Li et al.\ (\citeyear{li2022transfer}) and Tian and Feng (\citeyear{tian2023transfer}), define the population difference between the primary attribute and secondary attribute $s$ as
\[
\bm\delta^{(s)*}:=\bm\alpha^*-\bm u^{(s)*},\qquad
\bm\alpha^*=\bm u^{(s)*}+\bm\delta^{(s)*}.
\]
This normalisation also gives $\bm1_M^T\bm\delta^{(s)*}=0$. We call secondary attribute $s$ informative when its population log-worth vector is sufficiently close to the primary population log-worth vector, and define
\begin{equation}
\mathcal S
=
\left\{1\le s\le S:
\|\bm\delta^{(s)*}\|_2^2\le h
\right\}.
\label{eq:informative_set}
\end{equation}
The threshold $h$ therefore controls how close a secondary attribute must be to the primary attribute to be called informative. Unlike much of the transfer-learning literature, we do not assume that the discrepancy vectors are sparse. We therefore use a squared $\ell_2$ penalty in the correction step introduced below.

We consider two settings: the informative set $\mathcal S$ is either known or unknown. We start with the known case, which we call the oracle setting. The practical method does not require the attributes to contain the same observed pairs; each data set may have its own comparison graph. We only require the same $M$ objects to be represented across attributes. Algorithm \ref{alg:oraclebtl} gives the oracle procedure.

The number of comparisons may differ across pairs and attributes. We therefore define the negative log-likelihood directly from the individual comparison outcomes. For a data set $\mathcal D$ with observed edge set $\mathcal E_{\mathcal D}$, let $n_{jl}(\mathcal D)$ be the number of observations on edge $(j,l)$ and let $|\mathcal D|=\sum_{(j,l)\in\mathcal E_{\mathcal D}}n_{jl}(\mathcal D)$. For a generic log-worth vector $\bm\theta\in\mathbb R^M$, define
\begin{equation}
L_{\mathcal D}(\bm\theta)
:=
\frac{|\mathcal E_{\mathcal D}|}{|\mathcal D|}
\sum_{(j,l)\in\mathcal E_{\mathcal D},\,j>l}
\sum_{i=1}^{n_{jl}(\mathcal D)}
\left\{-y_{l,j}^{(i)}(\theta_j-\theta_l)
+\log[1+\exp(\theta_j-\theta_l)]\right\}.
\label{eq:general_empirical_loss}
\end{equation}
Here $\bm\theta$ is a generic argument and is replaced by the relevant log-worth vector in each fit. The factor $|\mathcal E_{\mathcal D}|/|\mathcal D|$ only rescales the negative log-likelihood and therefore does not change an unregularised MLE. In a penalised fit, however, this scaling changes the likelihood relative to the penalty, so $\lambda_\delta$ must be rescaled if the same minimiser is to be retained. If every observed edge has exactly $n$ repetitions, Equation \eqref{eq:general_empirical_loss} reduces to Equation \eqref{eq:loglikbt}. We use $\sqcup$ to denote concatenation of data while retaining repeated observations. Let $\mathcal D_0$ be the primary data and $\mathcal D_{\mathrm{tr}}=\mathcal D_0\sqcup\bigsqcup_{s\in\mathcal S}\mathcal D_s$ the pooled data used in the transfer step.

\begin{algorithm}[H]
\caption{Oracle Trans-BT algorithm}\label{alg:oraclebtl}
    \hspace*{\algorithmicindent} \textbf{Input:} Primary data $\mathcal D_0$, secondary data $\{\mathcal D_s\}_{s \in \mathcal{S}}$ and penalty parameter $\lambda_\delta>0$\\
    \hspace*{\algorithmicindent} \textbf{Output:} $\hat{\bm{\alpha}}_{\lambda_\delta}$
  \begin{algorithmic}[1]
        \STATE \textbf{Transfer step:} Compute $\hat{\bm u}=\argmin_{\bm1_M^T\bm u=0}L_{\mathcal D_{\mathrm{tr}}}(\bm u)$.
        \STATE \textbf{Primary-attribute correction step:} Compute $\hat{\bm\delta}_{\lambda_\delta}=\argmin_{\bm1_M^T(\hat{\bm u}+\bm\delta)=0}\left\{L_{\mathcal D_0}(\hat{\bm u}+\bm\delta)+\frac{\lambda_\delta}{2}\|\bm\delta\|_2^2\right\}$.
                \STATE Set $\hat{\bm{\alpha}}_{\lambda_\delta} = \hat{\bm{u}} + \hat{\bm{\delta}}_{\lambda_\delta}$
  \end{algorithmic}
\end{algorithm}

\noindent The first step of Algorithm \ref{alg:oraclebtl} pools the primary data with all informative secondary data and estimates one log-worth vector $\hat{\bm u}$. Because the secondary population log-worth vectors need not equal $\bm\alpha^*$, this pooled estimate may differ systematically from the primary parameter. The second step therefore uses the primary data to correct $\hat{\bm u}$, while the squared $\ell_2$ penalty limits the size of that correction. For brevity, write $L_0:=L_{\mathcal D_0}$ and $L_{\mathrm{tr}}:=L_{\mathcal D_{\mathrm{tr}}}$. With $\bm\alpha=\hat{\bm u}+\bm\delta$, line 2 of Algorithm \ref{alg:oraclebtl} can be written as
\begin{equation}
\hat{\bm\alpha}_{\lambda_\delta}
=
\argmin_{\bm1_M^T\bm\alpha=0}
\left\{
L_0(\bm\alpha)
+
\frac{\lambda_\delta}{2}
\|\bm\alpha-\hat{\bm u}\|_2^2
\right\}.
\label{eq:alpha_ridge_form}
\end{equation}
Equation \eqref{eq:alpha_ridge_form} shows the role of $\lambda_\delta$. If $\lambda_\delta=0$, the pooled estimate drops out and, whenever a finite centred primary-only Bradley--Terry MLE exists, the result is the primary-only MLE. Larger values of $\lambda_\delta$ keep the final estimate closer to $\hat{\bm u}$, and $\hat{\bm\alpha}_{\lambda_\delta}\to\hat{\bm u}$ as $\lambda_\delta\to\infty$. Thus, $\lambda_\delta$ controls how strongly the final estimator is pulled toward the pooled estimate.

\subsection{Unknown set of informative secondary attributes}
The oracle procedure assumes that $\mathcal S$ is known. In practice, we need to decide which secondary attributes to use. We adapt the procedure of Tian and Feng (\citeyear{tian2023transfer}) and call the resulting method Discovery Trans-BT; see Algorithm \ref{alg:discbtl}. The primary data $\mathcal D_0$ are divided into three approximately equal folds $\{\mathcal D_0^{[i]}\}_{i=1}^3$. For each fold, we fit a Bradley--Terry model to the remaining primary data and, separately for each secondary attribute, a second model after adding that secondary data. We then compare the two fits on the held-out primary data. Other numbers of folds are possible, and related methods have also used a single training/test split (Sun and Zhang, \citeyear{sun2023robust}; Jin et al., \citeyear{jin2024transfer}).

The folds should respect the way the data were collected. If several pairwise comparisons come from the same ranking or respondent, they should be placed in the same fold. Algorithm \ref{alg:discbtl} is written for the case in which the primary and secondary samples are independent. If the same respondents provide both primary and secondary data, all observations from a held-out respondent should also be left out of the secondary data used in the corresponding training fit. Thus, in line 4, $\mathcal D_s$ should then contain only observations from respondents that are not held out in fold $q$.

For a set of held-out pairwise comparisons $\mathcal D$, write the negative log-likelihood evaluated on those data as
\[
L_{\mathrm{hold}}(\bm\theta;\mathcal D)
:=
\sum_{r\in\mathcal D}
\left\{-y_r(\theta_{j_r}-\theta_{l_r})+\log\bigl[1+\exp(\theta_{j_r}-\theta_{l_r})\bigr]\right\}.
\]
Here observation $r$ compares objects $j_r$ and $l_r$, with $y_r=1$ when object $j_r$ is preferred to object $l_r$ and $y_r=0$ otherwise. This is the same Bradley--Terry negative log-likelihood as above, evaluated on the held-out observations. We use its unscaled total rather than its mean. This keeps $\bar L_s-\bar L_0$, $\hat\sigma$ and the tolerance term $0.01|\mathcal D_0|/3$ on the same scale. The tolerance is one percent of the approximate fold size. Tian and Feng (\citeyear{tian2023transfer}) use mean held-out losses; with equal fold sizes, multiplying all terms in their rule by the fold size leaves the selection inequality unchanged.

Let $\mathcal D_0^{[q]}$ be fold $q$ and let $\mathcal D_0^{[-q]}:=\bigsqcup_{i\ne q}\mathcal D_0^{[i]}$ contain the other two folds. We average the held-out negative log-likelihoods over the three folds for the primary-only fit and for each secondary-attribute fit. We also compute the sample standard deviation $\hat\sigma$ of the three primary-only held-out negative log-likelihoods. A secondary attribute is included in $\hat{\mathcal S}$ when $\bar L_s-\bar L_0$ is no larger than $C_{\hat{\mathcal S}}\max\{\hat\sigma,0.01|\mathcal D_0|/3\}$. This rule is one-sided: any secondary attribute that lowers the held-out primary negative log-likelihood is included, while a small increase is tolerated. Larger values of $C_{\hat{\mathcal S}}$ allow more secondary attributes to enter the selected set. Tian and Feng (\citeyear{tian2023transfer}) use $C_0=2$ in their numerical experiments; we use $C_{\hat{\mathcal S}}=1$ in Section \ref{sec:simulation}. Other criteria, such as the normalised Kendall tau distance, could also be used (Kumar \& Vassilvitskii, \citeyear{kumar2010generalized}; Jin et al., \citeyear{jin2024transfer}).

\begin{algorithm}[H]
\caption{Discovery Trans-BT algorithm}\label{alg:discbtl}
    \hspace*{\algorithmicindent} \textbf{Input:} Primary data $\mathcal D_0$, secondary data $\{\mathcal D_s\}_{s = 1}^S$, a constant $C_{\hat{\mathcal{S}}}$ and penalty parameter $\lambda_\delta>0$\\
    \hspace*{\algorithmicindent} \textbf{Output:} $\hat{\bm{\alpha}}_{\lambda_\delta}$ and $\hat{\mathcal{S}}$
  \begin{algorithmic}[1]
  	\STATE \textbf{Discovery step:} Randomly divide $\mathcal D_0$ into three approximately equal folds $\{\mathcal D_0^{[i]}\}_{i = 1}^3$.
	      \FOR{$q = 1$ to $3$}
	      \STATE Compute $\hat{\bm{\alpha}}^{(0)[q]} = \argmin_{\bm1_M^T\bm{\alpha}=0} L_{\mathcal D_0^{[-q]}}(\bm{\alpha})$.
	      \STATE Compute $\hat{\bm u}_{\mathrm{pool}}^{(s)[q]} = \argmin_{\bm1_M^T\bm{u}=0} L_{\mathcal D_0^{[-q]}\sqcup\mathcal D_s}(\bm{u})$ for $s=1,\ldots,S$.
      	      \STATE Compute $L_{\mathrm{hold}}(\hat{\bm{\alpha}}^{(0)[q]};\mathcal D_0^{[q]})$.
      	      \STATE Compute $L_{\mathrm{hold}}(\hat{\bm u}_{\mathrm{pool}}^{(s)[q]};\mathcal D_0^{[q]})$ for $s = 1,\ldots,S$.
	\ENDFOR
            \STATE Compute $\bar L_0 = \frac{1}{3}\sum_{q=1}^3 L_{\mathrm{hold}}(\hat{\bm{\alpha}}^{(0)[q]};\mathcal D_0^{[q]})$
            \STATE Compute $\bar L_s = \frac{1}{3}\sum_{q=1}^3 L_{\mathrm{hold}}(\hat{\bm u}_{\mathrm{pool}}^{(s)[q]};\mathcal D_0^{[q]})$, $s=1,\ldots,S$
            \STATE Compute $\hat{\sigma} = \sqrt{\frac{1}{2}\sum_{q=1}^3\left\{L_{\mathrm{hold}}(\hat{\bm{\alpha}}^{(0)[q]};\mathcal D_0^{[q]})-\bar L_0\right\}^2}$
			\STATE Set $\hat{\mathcal{S}} = \left\{s = 1,\ldots,S: \bar L_s-\bar L_0 \leq C_{\hat{\mathcal{S}}}\max \{\hat{\sigma}, 0.01|\mathcal D_0|/3\}\right\}$
			\STATE Compute $\hat{\bm{\alpha}}_{\lambda_\delta}$ using Algorithm \ref{alg:oraclebtl} with primary data $\mathcal D_0$, secondary data $\{\mathcal D_s\}_{s \in \hat{\mathcal{S}}}$ and penalty parameter $\lambda_\delta$
  \end{algorithmic}
\end{algorithm}

\noindent The negative log-likelihoods in Algorithms \ref{alg:oraclebtl} and \ref{alg:discbtl} can be minimised using any method for smooth convex optimisation. We use gradient descent in the theoretical analysis (Boyd \& Vandenberghe, \citeyear{boyd2004convex}) and BFGS in the simulations and application (Nocedal \& Wright, \citeyear{nocedal1999numerical}). Numerically, we impose the zero-sum restriction by optimising over $M-1$ coordinates and setting the last coordinate equal to minus their sum.

\noindent The discussion above assumes that useful secondary attributes have preferences in roughly the same direction as the primary attribute. A secondary attribute may instead contain useful information after reversing its preference direction. Such an attribute may not be selected by Algorithm \ref{alg:discbtl}. For example, consider the following observations on five objects:
\begin{equation*}
\mathcal D_0
=
\{o_1 \succ o_2,\;o_2 \succ o_3,\;o_3 \succ o_4,\;o_4 \succ o_5\}
\quad\text{and}\quad
\mathcal D_1
=
\{o_2 \succ o_1,\;o_3 \succ o_2,\;o_4 \succ o_3,\;o_5 \succ o_4\}.
\end{equation*}
For example, centred latent log-worth vectors that favour these opposite orderings, with coordinates for objects one through five listed from top to bottom, are
\begin{equation*}
\bm{\alpha}^{*} = \begin{pmatrix}
0.5\\
0.25\\
0 \\
-0.25 \\
-0.5
\end{pmatrix}
\text{ and }
\bm{u}^{(1)*} = \begin{pmatrix}
-0.5\\
-0.25\\
0 \\
0.25 \\
0.5
\end{pmatrix}.
\end{equation*}
Although this example is extreme, reversing all preferences in $\mathcal D_1$ makes their direction agree with $\mathcal D_0$. An optional extension is therefore to reverse each unselected secondary attribute and apply the Discovery rule again. If the reversed version is selected, those reversed observations can be used in the transfer fit and the interpretation of that attribute must be reversed as well. This optional step is not part of Algorithm \ref{alg:discbtl} or of the theory.

\section{Statistical properties} \label{sec:statistical-properties}
This section gives high-probability error bounds for the oracle procedure in Algorithm \ref{alg:oraclebtl}. We show when these bounds can have a smaller asymptotic order than the corresponding primary-only Bradley--Terry bound. We also introduce a one-step estimator and derive its asymptotic normality. Proofs are given in the Supplementary material. We begin with the assumptions used in the theory.

\begin{assumption}\label{ass:1}
For the theoretical analysis, the primary attribute and all informative secondary attributes are observed on a common comparison graph $\mathcal G=(\mathcal V,\mathcal E)$ with $|\mathcal V|=M$, where $\mathcal G\sim\mathcal G_{M,p}$ with $0<p\le1$. Conditional on $\mathcal G$, a primary comparison on edge $(j,l)$ follows the Bradley--Terry model with population log-worth vector $\bm\alpha^*$, while a comparison for informative secondary attribute $s$ follows the Bradley--Terry model with population log-worth vector $\bm u^{(s)*}$. All comparison outcomes are independent across repetitions, edges and attributes, conditional on $\mathcal G$.
\end{assumption}

\begin{assumption}\label{ass:2}
For every $(j,l)\in\mathcal E$, the primary attribute contributes $n_0\ge1$ independent comparisons and each informative secondary attribute $s\in\mathcal S$ contributes $n_s\ge1$ independent comparisons. Define
\[
n_{\mathcal S}:=\sum_{s\in\mathcal S}n_s,
\qquad
N_{\mathrm{tr}}:=n_0+n_{\mathcal S}.
\]
Thus, if every informative secondary attribute has the same per-edge sample size as the primary attribute, $n_s=n_0$, then
\[
N_{\mathrm{tr}}=(1+|\mathcal S|)n_0.
\]
\end{assumption}

\noindent Assumption \ref{ass:1} uses one common Erd\H{o}s--R\'enyi comparison graph for the primary and informative secondary attributes. This is stronger than the practical method, where each attribute may have a different observed edge set. The common graph allows us to control node degrees and Laplacian eigenvalues and keeps the leave-one-out proof close to Chen et al.\ (\citeyear{chen2019spectral}) and Fan et al.\ (\citeyear{fan2024uncertainty}). Assumption \ref{ass:2} also assumes a constant number of repetitions on every observed edge within an attribute. Allowing edge-specific counts would require weighted concentration and curvature arguments. We do not add a separate condition for existence of the unregularised pooled MLE. Instead, the proof of Theorem \ref{thm:alphaconvergence} constructs an interior stationary point on the same high-probability event, which gives existence and uniqueness of the centred pooled MLE.

The pooled data come from different attribute-specific Bradley--Terry models. Thus, $L_{\mathrm{tr}}$ is the negative log-likelihood of a single Bradley--Terry model fitted to the pooled observations, rather than the correctly specified likelihood for the separate attribute-specific models. Its population target is also not simply the average of the secondary log-worth vectors. We define a pooled population vector $\bm u^*$ through the average pairwise probabilities in the pooled data. For a pair $j\ne l$, index the $N_{\mathrm{tr}}$ primary and informative-secondary repetitions by $i=1,\ldots,N_{\mathrm{tr}}$. Let $\pi_{jl}^{(i)}$ be the conditional probability, given $\mathcal G$, that object $j$ is preferred to object $l$ in repetition $i$, and define
\begin{equation}
\bar\pi_{jl}
:=
\frac{1}{N_{\mathrm{tr}}}
\sum_{i=1}^{N_{\mathrm{tr}}}\pi_{jl}^{(i)}.
\label{eq:pooled_edge_probability}
\end{equation}

\begin{assumption}[Pooled Bradley--Terry compatibility and aggregate transferability]\label{ass:pooledtarget}
There exists a centred pooled population vector $\bm u^*\in\mathbb R^M$, $\bm1_M^T\bm u^*=0$, such that for every pair $j\ne l$,
\begin{equation}
\bar\pi_{jl}
=
\frac{\exp(u_j^*)}{\exp(u_j^*)+\exp(u_l^*)}.
\label{eq:pooled_btl_compatibility}
\end{equation}
With
\[
\bm\delta^*:=\bm\alpha^*-\bm u^*,
\]
assume
\begin{equation}
\|\bm\delta^*\|_2^2\le h_{\mathrm{agg}}.
\label{eq:aggregate_transferability}
\end{equation}
Here $h_{\mathrm{agg}}\ge0$ is an aggregate discrepancy threshold, distinct from the threshold $h$ in Equation \eqref{eq:informative_set}.
\end{assumption}

\noindent Assumption \ref{ass:pooledtarget} has two parts. First, the average pairwise probabilities must themselves be representable by one Bradley--Terry log-worth vector $\bm u^*$. This does not follow from Equation \eqref{eq:informative_set}: averaging probabilities from different Bradley--Terry models does not generally produce another Bradley--Terry model. We impose this condition because the leave-one-out proof requires one fixed population target for the pooled likelihood. Second, $\bm\delta^*=\bm\alpha^*-\bm u^*$ measures how far that pooled target is from the primary parameter. More pooled observations can reduce estimation error around $\bm u^*$, but they cannot remove a fixed difference between $\bm u^*$ and $\bm\alpha^*$. 

We use three quantities to describe the dynamic ranges of the pooled and primary log-worth vectors and of the discrepancy vector:
\begin{equation}
\kappa_1
=
\exp\!\left\{\max_{j\ne l}(u_j^*-u_l^*)\right\},
\qquad
\kappa_2
=
\exp\!\left\{\max_{j\ne l}(\delta_j^*-\delta_l^*)\right\},
\qquad
\kappa_3
=
\exp\!\left\{\max_{j\ne l}(\alpha_j^*-\alpha_l^*)\right\}.
\label{eq:kappas_correct}
\end{equation}
Since both $\bm\alpha^*$ and $\bm u^*$ are centred, $\bm\delta^*$ is centred. Hence
\begin{equation}
d_{\mathcal S,\infty}
:=
\|\bm\delta^*\|_\infty
\le
\min\{\sqrt{h_{\mathrm{agg}}},\log\kappa_2\}.
\label{eq:delta_inf_h}
\end{equation}

\subsection{Rate of convergence}

For notational convenience, define
\begin{equation}
r_{u,\infty}
:=
\kappa_1^2
\sqrt{\frac{\log M}{MpN_{\mathrm{tr}}}}
\label{eq:ruinf}
\end{equation}
and
\begin{equation}
r_{\mathrm{tr},\infty}
:=
d_{\mathcal S,\infty}+r_{u,\infty}
\le
\min\{\sqrt{h_{\mathrm{agg}}},\log\kappa_2\}
+
\kappa_1^2
\sqrt{\frac{\log M}{MpN_{\mathrm{tr}}}}.
\label{eq:rtrinf}
\end{equation}
The quantity $r_{\mathrm{tr},\infty}$ is the sum of two terms: the population difference $\|\bm u^*-\bm\alpha^*\|_\infty$ and the sampling error from estimating $\bm u^*$ with the pooled data.

\begin{theorem}[Transfer $\ell_\infty$ error bound]\label{thm:alphaconvergence}
Suppose Assumptions \ref{ass:1}, \ref{ass:2} and \ref{ass:pooledtarget} hold, $Mp\ge C_\kappa\kappa_1^4\log M$ for a sufficiently large $C_\kappa$, and $N_{\mathrm{tr}}\le c_2M^{c_3}$ for some constants $c_2,c_3>0$. Then, on an event of probability at least $1-O(M^{-5})$, the centred unregularised pooled MLE $\hat{\bm u}$ in line 1 of Algorithm \ref{alg:oraclebtl} exists uniquely and
\begin{equation}
\|\hat{\bm u}-\bm u^*\|_\infty
\lesssim
r_{u,\infty}.
\label{eq:u_main_rate}
\end{equation}
On the same event, the following bound holds simultaneously for all $\lambda_\delta>0$:
\begin{equation}
\|\hat{\bm\alpha}_{\lambda_\delta}-\bm\alpha^*\|_\infty
\lesssim
r_{\mathrm{tr},\infty}
+
\frac{1}{\lambda_\delta}
\sqrt{\frac{Mp\log M}{n_0}}.
\label{eq:ridge_interpolation_bound}
\end{equation}
In particular, if
\begin{equation}
\lambda_\delta
\ge
c_{\lambda_\delta}
\frac{\sqrt{Mp\log M/n_0}}
{r_{\mathrm{tr},\infty}},
\label{eq:lambda_transfer_dominant}
\end{equation}
for a sufficiently large constant $c_{\lambda_\delta}$, then
\begin{equation}
\boxed{
\|\hat{\bm\alpha}_{\lambda_\delta}-\bm\alpha^*\|_\infty
\lesssim
d_{\mathcal S,\infty}
+
\kappa_1^2
\sqrt{\frac{\log M}{Mp(n_0+n_{\mathcal S})}}
}
\label{eq:main_linf_transfer}
\end{equation}
with probability at least $1-O(M^{-5})$.
\end{theorem}


\begin{proposition}[Primary-only Bradley--Terry rate]\label{prop:target_only_rate}
Suppose $\mathcal G\sim\mathcal G_{M,p}$ and, conditional on $\mathcal G$, each observed edge contributes $n_0\ge1$ independent primary Bradley--Terry comparisons with centred population log-worth vector $\bm\alpha^*$. If $Mp\ge C_\kappa\kappa_3^4\log M$ for a sufficiently large $C_\kappa$, and $n_0\le c_2M^{c_3}$ for some constants $c_2,c_3>0$, then, with probability at least $1-O(M^{-5})$, the centred unregularised Bradley--Terry MLE $\hat{\bm\alpha}^{(0)}$ fitted only to the primary observations exists uniquely and
\begin{equation}
\|\hat{\bm\alpha}^{(0)}-\bm\alpha^*\|_\infty
\lesssim
r_{\mathrm{BT},\infty}
:=
\kappa_3^2\sqrt{\frac{\log M}{Mpn_0}}.
\label{eq:target_only_rate}
\end{equation}
\end{proposition}

\begin{proof}[Proof of Proposition \ref{prop:target_only_rate}]
Apply Theorem \ref{thm:convergence_normal_step1} with no secondary attributes, so that $\mathcal S=\varnothing$, $n_{\mathcal S}=0$, $N_{\mathrm{tr}}=n_0$, $\bm u^*=\bm\alpha^*$, $h_{\mathrm{agg}}=0$ and $\kappa_1=\kappa_3$. The pooled Bradley--Terry compatibility and discrepancy conditions are then automatic, and the assumptions of Theorem \ref{thm:convergence_normal_step1} reduce exactly to those of the proposition. Its conclusion therefore gives existence and uniqueness of the centred unregularised primary MLE together with
\[
\|\hat{\bm\alpha}^{(0)}-\bm\alpha^*\|_\infty
\lesssim
\kappa_3^2\sqrt{\frac{\log M}{Mpn_0}},
\]
as claimed.
\end{proof}

\begin{corollary}[Comparison with primary-only Bradley--Terry]\label{crl:rate_improvement}
Suppose the conditions of Theorem \ref{thm:alphaconvergence} and Proposition \ref{prop:target_only_rate} hold, and suppose that $\lambda_\delta$ satisfies the transfer-dominant condition in Equation \eqref{eq:lambda_transfer_dominant}. Then the primary-only estimator satisfies \eqref{eq:target_only_rate}. Therefore the transfer-dominant upper bound in Equation \eqref{eq:main_linf_transfer} is asymptotically strictly smaller than the primary-only upper bound whenever
\begin{equation}
d_{\mathcal S,\infty}
+
\kappa_1^2
\sqrt{\frac{\log M}{Mp(n_0+n_{\mathcal S})}}
=
o\!\left(
\kappa_3^2
\sqrt{\frac{\log M}{Mpn_0}}
\right).
\label{eq:improvement_condition_general}
\end{equation}
If every informative secondary attribute has the same per-edge sample size as the primary attribute, $n_s=n_0$, then $n_{\mathcal S}=|\mathcal S|n_0$ and the stochastic transfer term becomes
\begin{equation}
\frac{\kappa_1^2}{\sqrt{1+|\mathcal S|}}
\sqrt{\frac{\log M}{Mpn_0}}.
\label{eq:equal_source_rate}
\end{equation}
A simple sufficient condition for a strict rate improvement is
\begin{equation}
d_{\mathcal S,\infty}
=
o(r_{\mathrm{BT},\infty}),
\qquad
\frac{\kappa_1^2}{\sqrt{1+|\mathcal S|}}
=
o(\kappa_3^2).
\label{eq:equal_source_improvement}
\end{equation}
\end{corollary}

\noindent Corollary \ref{crl:rate_improvement} shows when transfer can improve the error bound. If every informative secondary attribute has the same per-edge sample size as the primary attribute, the stochastic term decreases with the number of informative attributes through the factor $(1+|\mathcal S|)^{-1/2}$. However, the pooled population vector must also remain close to the primary parameter. The discrepancy term $d_{\mathcal S,\infty}$ does not decrease simply by adding more secondary observations. Equation \eqref{eq:improvement_condition_general} makes this precise. If the primary-only rate $r_{\mathrm{BT},\infty}$ tends to zero, then a strictly faster transfer rate requires $d_{\mathcal S,\infty} = o(r_{\mathrm{BT},\infty})$, so the pooled-to-primary discrepancy must also tend to zero. A fixed positive discrepancy can still help in finite samples, but it prevents a strictly smaller asymptotic order. This improvement comes from the ridge correction. The ridge penalty keeps the final estimate close to an accurately estimated pooled vector.

The same ridge argument also gives an $\ell_2$ result. Define
\[
r_{u,2}
:=
\kappa_1^2
\sqrt{\frac{\log M}{pN_{\mathrm{tr}}}},
\qquad
r_{\mathrm{tr},2}
:=
\sqrt{h_{\mathrm{agg}}}+r_{u,2}.
\]
\begin{corollary}[$\ell_2$ transfer bound]\label{crl:alphaconvergencel2}
Suppose the assumptions of Theorem \ref{thm:alphaconvergence} hold and the ridge parameter satisfies
\begin{equation}
\lambda_\delta
\gtrsim
\frac{M\sqrt{p\log M/n_0}}{r_{\mathrm{tr},2}}.
\label{eq:lambda_l2}
\end{equation}
Then, with probability at least $1-O(M^{-5})$,
\begin{equation}
\|\hat{\bm\alpha}_{\lambda_\delta}-\bm\alpha^*\|_2
\lesssim
\sqrt{h_{\mathrm{agg}}}
+
\kappa_1^2
\sqrt{\frac{\log M}{p(n_0+n_{\mathcal S})}}.
\label{eq:main_l2_transfer}
\end{equation}
If, in addition, the conditions of Proposition \ref{prop:target_only_rate} hold, its conservative primary-only $\ell_2$ order is $\kappa_3^2\sqrt{\log M/(pn_0)}$. Under these additional conditions, the transfer upper bound is asymptotically smaller whenever
\begin{equation}
\sqrt{h_{\mathrm{agg}}}+\kappa_1^2\sqrt{\frac{\log M}{p(n_0+n_{\mathcal S})}}
=
o\!\left(\kappa_3^2\sqrt{\frac{\log M}{pn_0}}\right).
\label{eq:l2_improvement_condition}
\end{equation}
\end{corollary}

\subsection{Asymptotic distribution}
Regularisation introduces bias, so asymptotic inference is commonly based on a debiased or one-step estimator. We use the same idea here to quantify uncertainty in the estimated primary log-worths.

After constructing $\hat{\bm\alpha}_{\lambda_\delta}$, we use a separate set of primary comparisons for inference. Under the common-graph assumption, these observations are collected on the same graph $\mathcal G=(\mathcal V,\mathcal E)$. Conditional on $\mathcal G$, they follow the primary Bradley--Terry model with population vector $\bm\alpha^*$, are independent across edges and repetitions, and are independent of the data used to construct $\hat{\bm\alpha}_{\lambda_\delta}$. Let $\mathcal D_{\mathrm{inf}}$ denote this inference sample, let $n_{\mathrm{inf}}\ge1$ be the number of independent inference repetitions per observed edge, and write $L_{\mathrm{inf}}:=L_{\mathcal D_{\mathrm{inf}}}$. If only one primary data set is available, repeated observations on each edge can instead be split into disjoint training and inference samples. Theorem \ref{thm:alphaconvergence} then uses the training repetitions and the results below use $n_{\mathrm{inf}}$.

Adding the same constant to every Bradley--Terry log-worth does not change the comparison probabilities. The Hessian therefore has the zero direction $\bm1_M$, so the inverse Fisher information is defined only on the identifiable subspace $\bm1_M^\perp$. For any centred $\bm\alpha$, define
\begin{equation}
\bm\Theta(\bm\alpha)
:=
\left[\nabla^2L_{\mathrm{inf}}(\bm\alpha)\right]^+,
\label{eq:theta_pseudoinverse}
\end{equation}
where $A^+$ denotes the Moore--Penrose pseudoinverse. Whenever the comparison graph is connected, and in particular on the graph event used in the inference theorem,
\[
\bm\Theta(\bm\alpha)
=
\left[
\begin{pmatrix}
\nabla^2L_{\mathrm{inf}}(\bm\alpha)&\bm1_M\\
\bm1_M^T&0
\end{pmatrix}^{-1}
\right]_{1:M,1:M}.
\]
We write
$\hat{\bm\Theta}_{11}=\bm\Theta(\hat{\bm\alpha}_{\lambda_\delta})$ and
$\bm\Theta_{11}^*=\bm\Theta(\bm\alpha^*)$, and define the one-step estimator
\begin{equation}
\hat{\bm\alpha}^{\mathrm{db}}
=
\hat{\bm\alpha}_{\lambda_\delta}
-
\hat{\bm\Theta}_{11}\nabla L_{\mathrm{inf}}(\hat{\bm\alpha}_{\lambda_\delta}).
\label{eq:debiased_corrected}
\end{equation}

\noindent For inference we use the full preliminary-estimator bound rather than only the transfer-dominant special case. Define
\begin{equation}
R_{\alpha,\lambda}
:=
r_{\mathrm{tr},\infty}
+
\frac{1}{\lambda_\delta}
\sqrt{\frac{Mp\log M}{n_0}},
\label{eq:Ralpha_lambda}
\end{equation}
and
\begin{equation}
D_{\Theta,\lambda}
:=
\frac{\kappa_3^2}{Mp}R_{\alpha,\lambda}.
\label{eq:Dtheta_lambda}
\end{equation}
By Theorem \ref{thm:alphaconvergence}, $\|\hat{\bm\alpha}_{\lambda_\delta}-\bm\alpha^*\|_\infty\lesssim R_{\alpha,\lambda}$ with probability at least $1-O(M^{-5})$. If $\lambda_\delta$ also satisfies the transfer-dominant condition in Equation \eqref{eq:lambda_transfer_dominant}, then $R_{\alpha,\lambda}\lesssim r_{\mathrm{tr},\infty}$, so the inference conditions below reduce to conditions on the transfer rate itself.

\begin{assumption}[Inference remainder conditions]\label{ass:rmax}
In addition to the conditions of Theorem \ref{thm:alphaconvergence} and the inference-sample construction described above, suppose that $\lambda_\delta>0$ is a deterministic tuning sequence (allowed to depend on $M$) and that
\begin{equation}
\frac{\kappa_3}{\sqrt{Mpn_{\mathrm{inf}}}}=o(1),
\qquad
\kappa_3^2R_{\alpha,\lambda}\sqrt{M\log M}=o(1),
\qquad
M\kappa_3R_{\alpha,\lambda}^2\sqrt{pn_{\mathrm{inf}}}=o(1).
\label{eq:inference_conditions_simple}
\end{equation}
\end{assumption}
\noindent The three conditions in Equation \eqref{eq:inference_conditions_simple} control the Gaussian approximation, estimation of the inverse information matrix, and the second-order Taylor remainder, respectively. The second condition implies $R_{\alpha,\lambda}=o(1)$. Because $R_{\alpha,\lambda}$ contains $d_{\mathcal S,\infty}$, the pooled-to-primary difference must vanish sufficiently quickly for the inference result. These conditions can be satisfied. For example, they all hold if $p$ is bounded away from zero, $\kappa_1$ and $\kappa_3$ are bounded, $d_{\mathcal S,\infty}=0$, $n_0\asymp N_{\mathrm{tr}}\asymp n_{\mathrm{inf}}\asymp M^a$ for some fixed $a>0$, and $\lambda_\delta\asymp M$. The result is stated for deterministic theoretical choices of $\lambda_\delta$; a data-selected value would require an additional argument.

\begin{theorem}[Asymptotic normality]\label{thm:asympnormal}
Suppose Assumption \ref{ass:rmax} holds. Then, for every fixed coordinate $j$,
\begin{equation}
\frac{\sqrt{n_{\mathrm{inf}}}(\hat{\alpha}_j^{\mathrm{db}}-\alpha_j^*)}
{\sqrt{[\bm\Theta_{11}^*]_{jj}}}
\xrightarrow{d}N(0,1).
\label{eq:asympnormal_corrected}
\end{equation}
The quantity $[\bm\Theta_{11}^*]_{jj}/n_{\mathrm{inf}}$ is the conditional variance of the leading likelihood-score term given the comparison graph. The theorem accounts for randomness in both the graph and the comparison outcomes. On the vanishing-probability event where the graph is not connected or the displayed quantities are not defined, the standardised statistic may be assigned any fixed value without changing the limiting distribution.
\end{theorem}

\begin{corollary}[Consistent studentisation]\label{crl:asympnormal2}
Under the assumptions of Theorem \ref{thm:asympnormal},
\begin{equation}
\|\hat{\bm\Theta}_{11}-\bm\Theta_{11}^*\|_2
\lesssim D_{\Theta,\lambda}
\label{eq:theta_rate_corrected}
\end{equation}
with probability at least $1-O(M^{-5})$. Moreover, using the same convention on the vanishing-probability exceptional event as in Theorem \ref{thm:asympnormal}, for every fixed coordinate $j$,
\begin{equation}
\frac{\sqrt{n_{\mathrm{inf}}}(\hat\alpha_j^{\mathrm{db}}-\alpha_j^*)}
{\sqrt{[\hat{\bm\Theta}_{11}]_{jj}}}
\xrightarrow{d}N(0,1).
\label{eq:studentized_asympnormal}
\end{equation}
\end{corollary}

\section{Simulation study} \label{sec:simulation}
We use simulations to study how well the proposed method recovers the primary population log-worth vector. We compare Oracle BT and Discovery BT with a Bradley--Terry model fitted only to the primary data (BT) and a model that pools all attributes without screening (PBT). For $M\in\{10,20\}$, we first draw $\widetilde{\bm\alpha}\sim U(-2,2)^M$ and centre it to obtain
\[
\bm\alpha^*
:=
\widetilde{\bm\alpha}-\frac{\bm1_M^T\widetilde{\bm\alpha}}{M}\bm1_M.
\]
This gives the same zero-sum identifiability constraint used in the fitted models. Let $\mathcal S_{\mathrm{gen}}\subseteq\{1,\ldots,S\}$, with $S\in\{5,10,20\}$, be the set of secondary attributes designated as informative in the simulation, and let $\mathcal S_{\mathrm{gen}}^c$ contain the remaining attributes. For $s\in\mathcal S_{\mathrm{gen}}$, draw $\widetilde{\bm\delta}^{(s)}$ from $U(-a_h,a_h)^M$ and resample until $\|\widetilde{\bm\delta}^{(s)}\|_2^2\le h$. For $s\in\mathcal S_{\mathrm{gen}}^c$, draw $\widetilde{\bm\delta}^{(s)}$ from the wider distribution $U(-2,2)^M$ without truncation. We then centre each discrepancy vector:
\[
\bm\delta^{(s)*}
:=
\widetilde{\bm\delta}^{(s)}-\frac{\bm1_M^T\widetilde{\bm\delta}^{(s)}}{M}\bm1_M.
\]
Centring is an orthogonal projection, so $\|\bm\delta^{(s)*}\|_2\le\|\widetilde{\bm\delta}^{(s)}\|_2$. Therefore every $s\in\mathcal S_{\mathrm{gen}}$ belongs to the population informative set $\mathcal S$ in Equation \eqref{eq:informative_set}. The secondary population log-worth vectors are
\[
\bm u^{(s)*}
=
\bm\alpha^*-\bm\delta^{(s)*},
\]
which ensures $\bm1_M^T\bm u^{(s)*}=0$ for every secondary attribute. The bound $a_h$ is defined as follows
\begin{equation*}
a_h = \begin{cases}
  0.1, & \text{if } h = 0.1;\\
  0.45, & \text{if } h = 1;\\
  0.85, & \text{if } h = 3,
\end{cases}
\end{equation*}
where $h\in\{0.1,1,3\}$. Even $h=1$ can allow noticeable differences between the primary and secondary log-worth vectors. For each attribute, we sample an unordered pair uniformly from the $\binom{M}{2}$ possible pairs and draw the outcome from the Bradley--Terry model, using $\bm\alpha^*$ for the primary attribute and $\bm u^{(s)*}$ for secondary attribute $s$. We sample $N\in\{500,1000,1500\}$ independent comparisons for every attribute $s=0,\ldots,S$.

Each attribute has the same sample size, but Oracle BT uses more observations as $|\mathcal S_{\mathrm{gen}}|$ increases. Ignoring differences in the realised edge sets, the pooled fit uses about $(1+|\mathcal S_{\mathrm{gen}}|)N$ observations, compared with $N$ for the primary-only fit. Here $N$ is the total number of sampled comparisons per attribute, not a common number of repetitions on every observed edge as assumed in the theory. The simulation setting is therefore more general than that of the theory shown in Section \ref{sec:statistical-properties}.

The condition $\|\bm\delta^{(s)*}\|_2^2\le h$ for $s\in\mathcal S_{\mathrm{gen}}$ does not guarantee the pooled Bradley--Terry compatibility condition in Equation \eqref{eq:pooled_btl_compatibility}. The simulations therefore also test the method when this theoretical condition is not imposed exactly. We generate 50 data sets for each parameter combination.

\begin{figure}[H]
\centering
\begin{minipage}{0.33\linewidth}
\centering $h = 0.1$\\ \textcolor{white}{A}\\ \includegraphics[width=\textwidth]{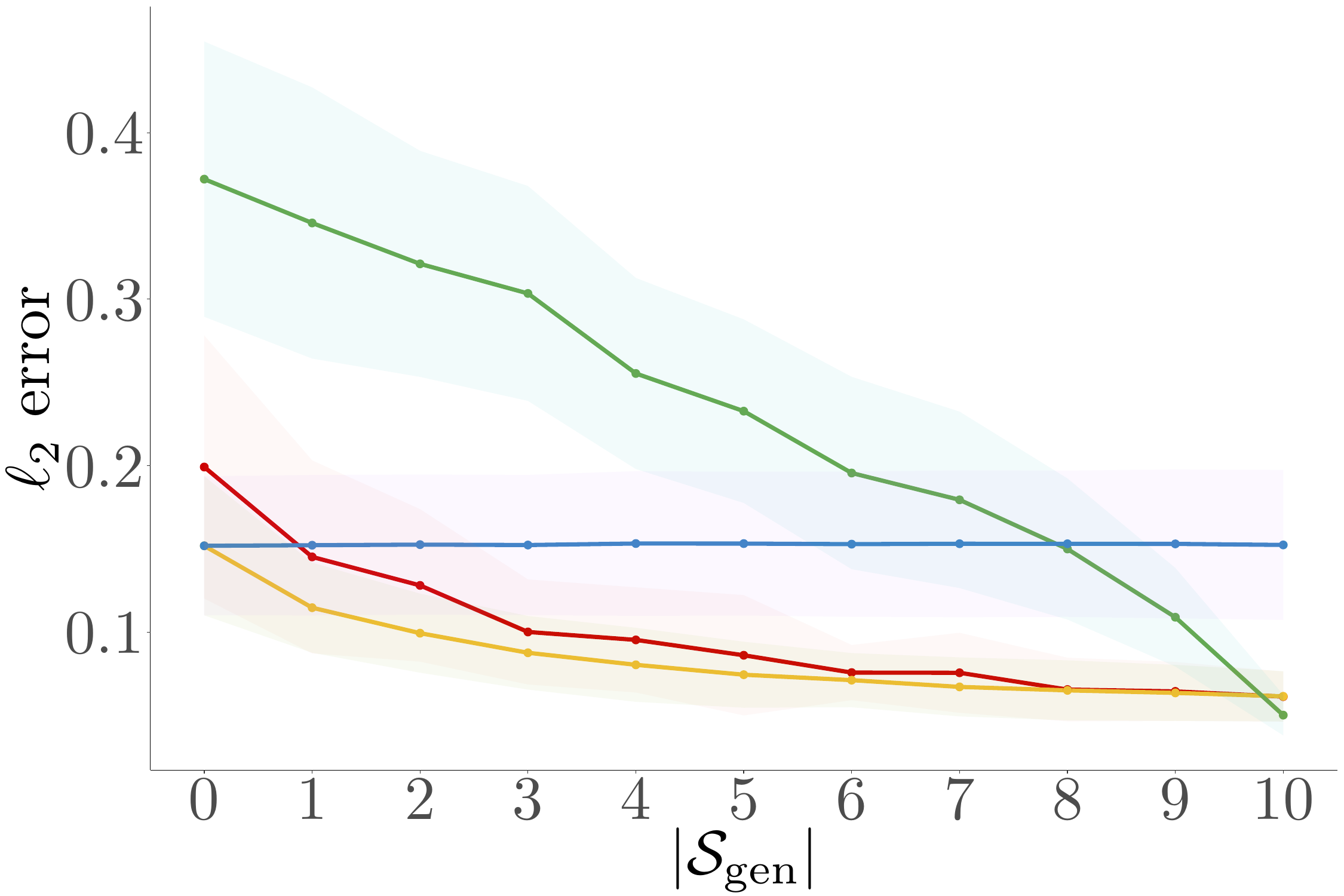}
\end{minipage}%
\hspace{1px}%
\begin{minipage}{0.33\linewidth}
  \centering $h = 1$\\$M = 10$\\  \includegraphics[width=\textwidth]{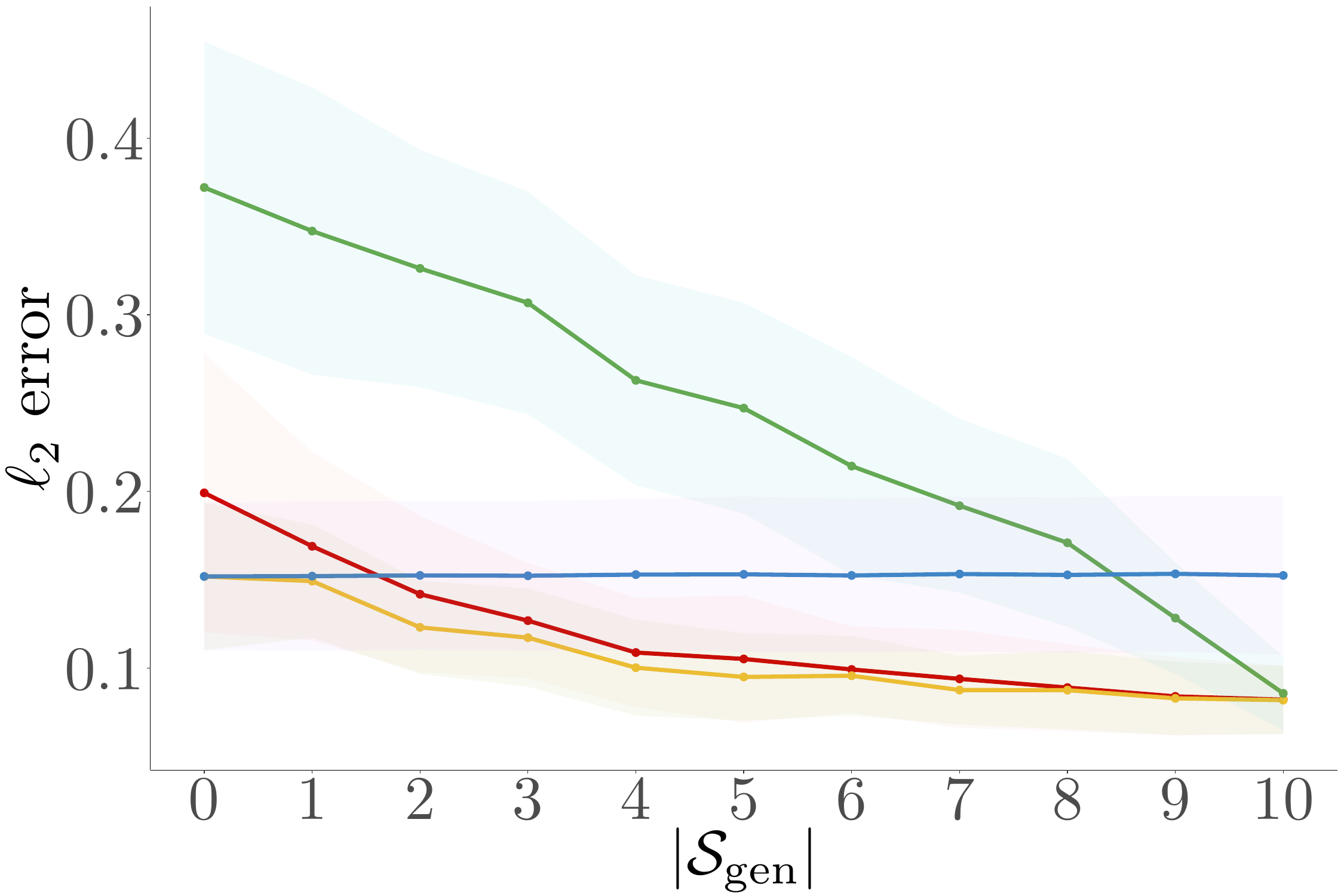}
\end{minipage}%
\hspace{1px}%
\begin{minipage}{0.33\linewidth}
  \centering $h = 3$\\\textcolor{white}{A}\\  \includegraphics[width=\textwidth]{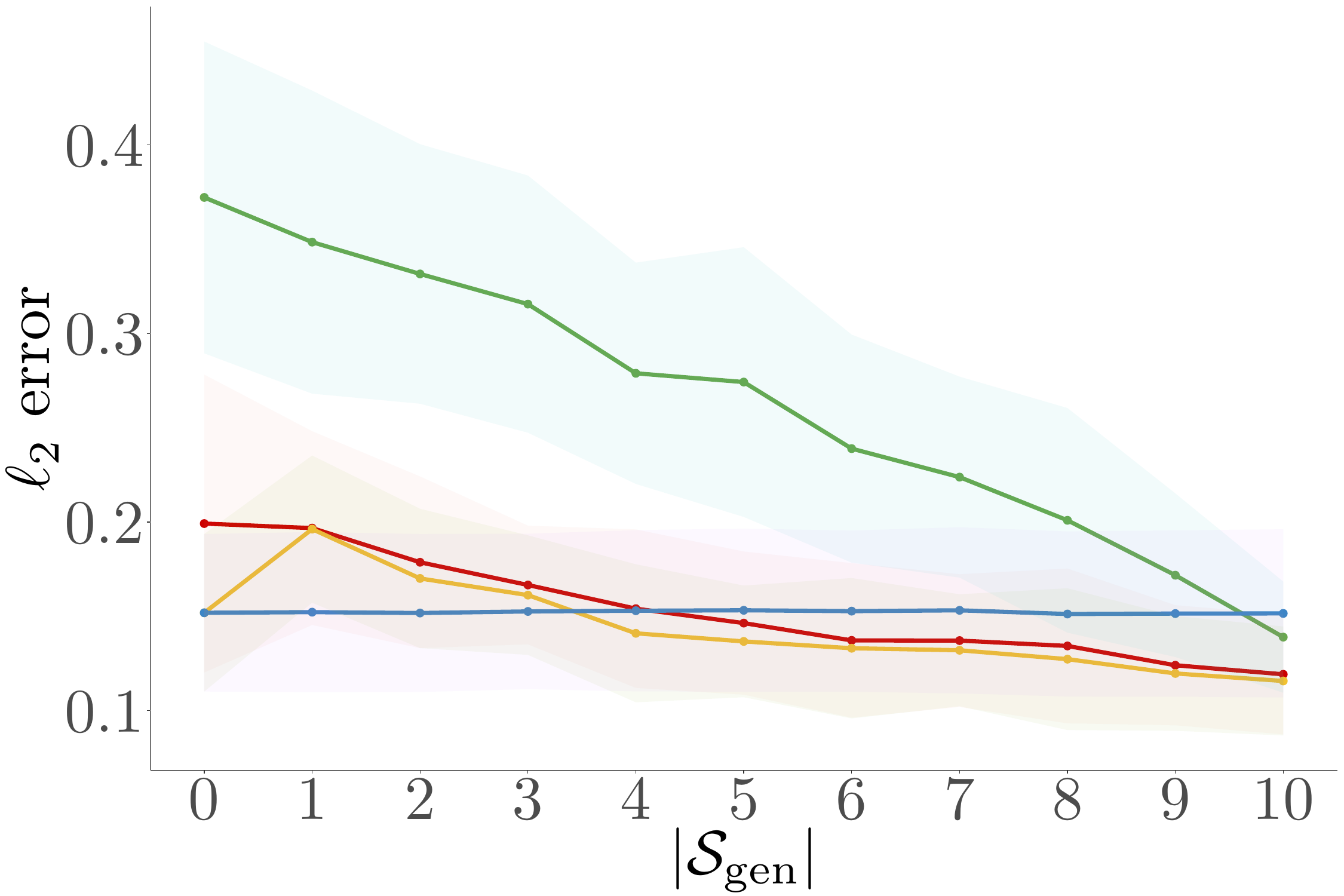}
\end{minipage}\\
\text{$M = 20$}\\
\includegraphics[width=0.33\textwidth]{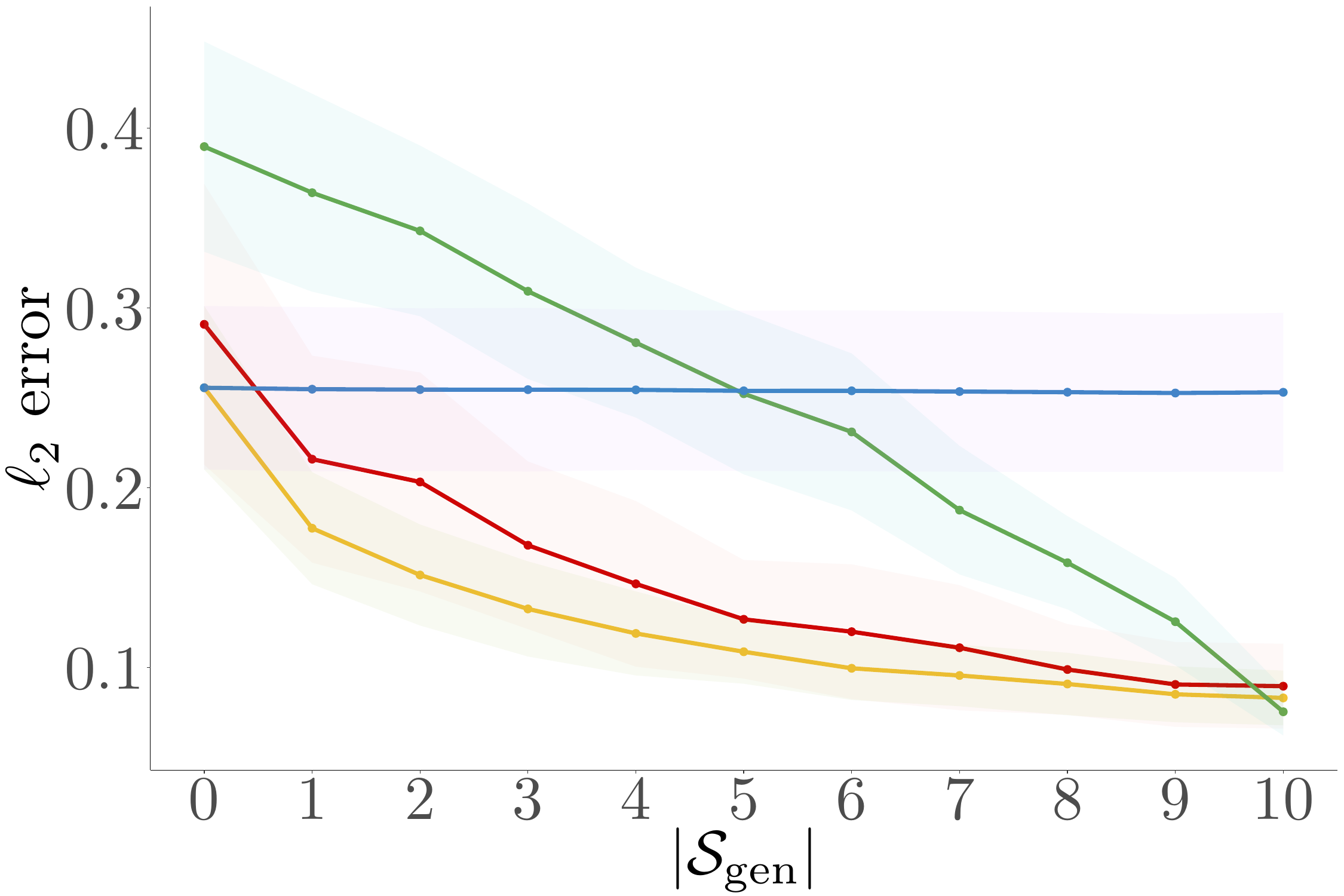}\hfill
\includegraphics[width=0.33\textwidth]{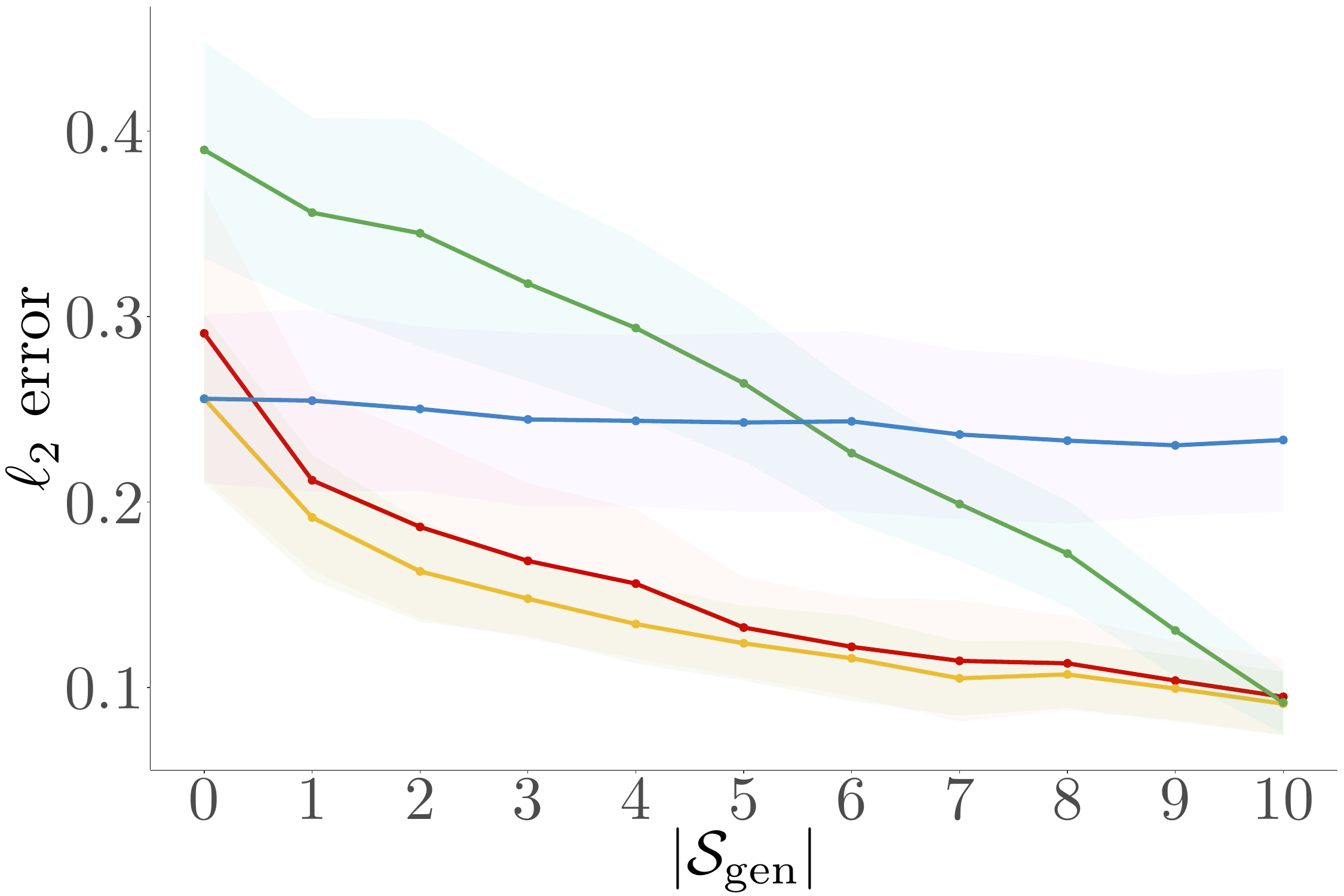}\hfill
\includegraphics[width=0.33\textwidth]{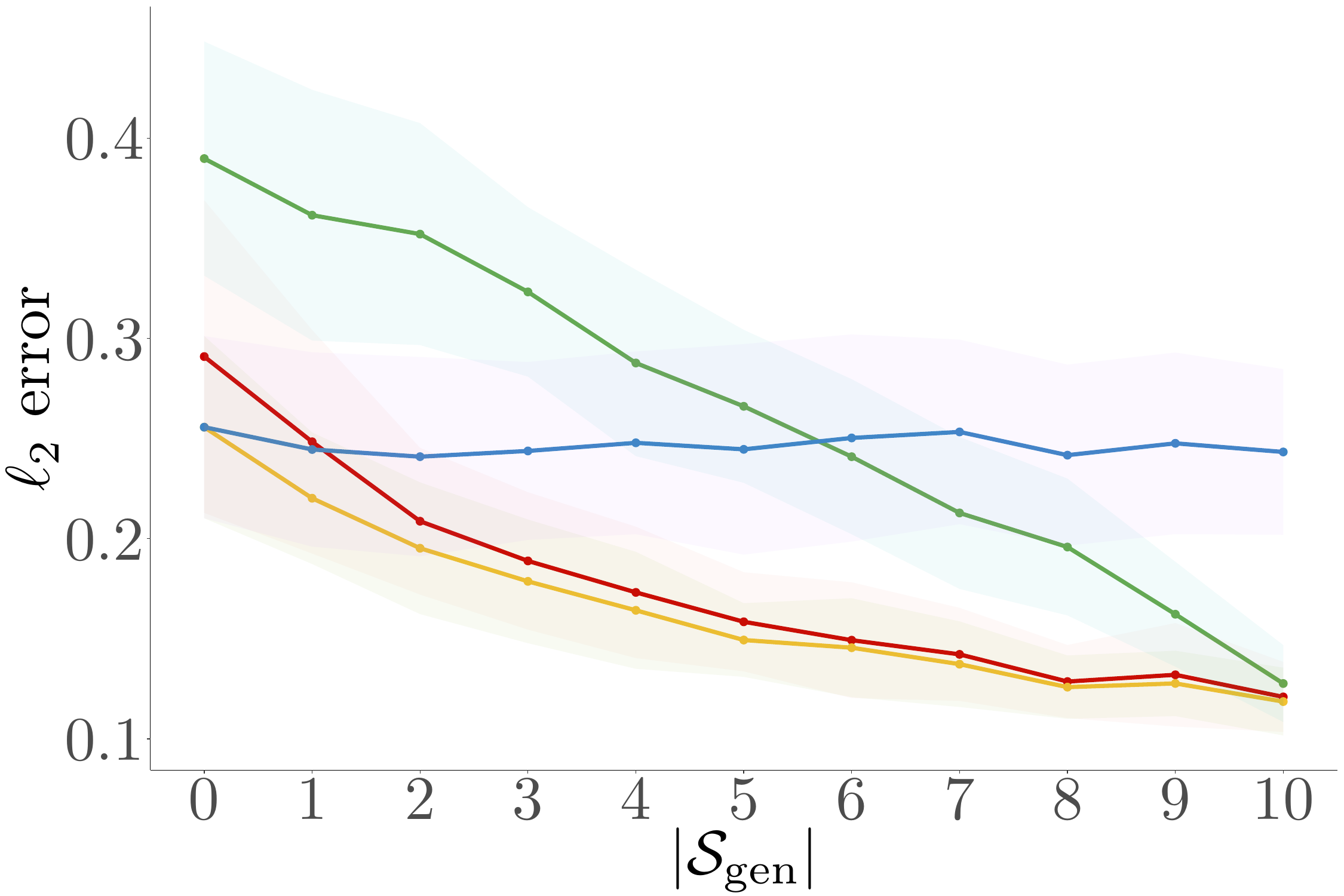}\hfill
\caption[]{Simulation results for the $\ell_2$ errors. From left to right, results for the following values of $h$ are shown: 0.1, 1 and 3. The coloured lines represent the mean $\ell_2$ error across 50 simulated data sets with $S=10$ and $N = 1000$ for the following methods: Discovery BT \begin{tikzpicture}\draw [fill = colorSFPLAIC] (0,0) rectangle (0.5,0.25);\end{tikzpicture}, Oracle BT \begin{tikzpicture}\draw [fill = colorSFPLBIC] (0,0) rectangle (0.5,0.25);\end{tikzpicture}, Pooled BT \begin{tikzpicture}\draw [fill = colorPL] (0,0) rectangle (0.5,0.25);\end{tikzpicture} and Primary attribute data only BT \begin{tikzpicture}\draw [fill = colorPPL] (0,0) rectangle (0.5,0.25);\end{tikzpicture}. The shaded areas represent $\pm1$ estimated standard deviation of the $\ell_2$ error across the 50 simulated data sets. In the panels, $|\mathcal S_{\mathrm{gen}}|$, the number of attributes designated as informative by the simulation mechanism.}
\label{fig:n250p11}
\end{figure}

For Oracle BT and Discovery BT, $\lambda_\delta$ is selected by cross-validation on the primary data. Performance is measured by the $\ell_2$ error used in Corollary \ref{crl:alphaconvergencel2}. Figure \ref{fig:n250p11} shows the results as $|\mathcal S_{\mathrm{gen}}|$ changes, and Table \ref{tab:simres1} reports errors averaged over the considered values of $|\mathcal S_{\mathrm{gen}}|$. Simulations for partial rankings using the Plackett--Luce model are given in Supplementary Material \ref{sec:pl-extension}.

\begin{table}[H]
\resizebox{\textwidth}{!}{%
\centering
  \begin{threeparttable}
  \caption{Simulation results for the Bradley--Terry methods. The $\ell_2$ errors are averaged over 50 simulated data sets for each parameter combination and over the considered values of $|\mathcal S_{\mathrm{gen}}|$, and are rounded to two decimals. Monte Carlo standard errors are shown in parentheses. Bold values indicate the smallest error among the four methods for that parameter combination; equal displayed values need not be exact ties before rounding. Oracle BT is given $\mathcal S_{\mathrm{gen}}$, Discovery BT uses Algorithm \ref{alg:discbtl}, BT uses only the primary data, and PBT pools all attributes.}
  \label{tab:simres1}
     \begin{tabular}{c | cc | cc }
        \toprule
        \midrule
         & \multicolumn{2}{c|}{$M = 10$} & \multicolumn{2}{c}{$M = 20$}\\ \midrule
         \textbf{$S, N, h$} & \textbf{Oracle BT/Discovery BT} & \textbf{BT/PBT} & \textbf{Oracle BT/Discovery BT} & \textbf{BT/PBT}\\ \midrule
$5, 500, 0.1$ & \textbf{0.14} (0.02)/0.17 (0.03) & 0.22 (0.00)/0.26 (0.05) & \textbf{0.22} (0.03)/0.29 (0.04) & 0.37 (0.00)/0.30 (0.05)\\ 
$5, 1000, 0.1$ & \textbf{0.10} (0.01)/0.11 (0.02) & 0.15 (0.00)/0.25 (0.05) & \textbf{0.15} (0.02)/0.17 (0.03) & 0.25 (0.00)/0.28 (0.05)\\ 
$5, 1500, 0.1$ & \textbf{0.08} (0.01)/0.09 (0.01) & 0.12 (0.00)/0.25 (0.05) & \textbf{0.13} (0.02)/0.13 (0.02) & 0.20 (0.00)/0.27 (0.05)\\ 
$5, 500, 1$ & \textbf{0.16} (0.02)/0.19 (0.02) & 0.22 (0.00)/0.27 (0.04) & \textbf{0.23} (0.03)/0.29 (0.04) & 0.37 (0.00)/0.31 (0.04)\\ 
$5, 1000, 1$ & \textbf{0.12} (0.01)/0.13 (0.01) & 0.15 (0.00)/0.26 (0.04) & \textbf{0.17} (0.02)/0.19 (0.02) & 0.25 (0.00)/0.29 (0.04)\\ 
$5, 1500, 1$ & \textbf{0.10} (0.01)/0.11 (0.01) & 0.12 (0.00)/0.26 (0.04) & \textbf{0.14} (0.01)/0.15 (0.02) & 0.20 (0.00)/0.28 (0.05)\\ 
$5, 500, 3$ & \textbf{0.20} (0.01)/0.22 (0.02) & 0.22 (0.00)/0.30 (0.03) & \textbf{0.25} (0.03)/0.30 (0.03) & 0.36 (0.00)/0.32 (0.04)\\ 
$5, 1000, 3$ & 0.16 (0.01)/0.17 (0.01) & \textbf{0.15} (0.00)/0.29 (0.03) & \textbf{0.20} (0.02)/0.21 (0.02) & 0.25 (0.00)/0.30 (0.04)\\ 
$5, 1500, 3$ & 0.14 (0.01)/0.14 (0.01) & \textbf{0.12} (0.00)/0.29 (0.03) & \textbf{0.17} (0.01)/0.17 (0.01) & 0.20 (0.00)/0.30 (0.04)\\ 
$10, 500, 0.1$ & \textbf{0.11} (0.01)/0.15 (0.02) & 0.24 (0.00)/0.24 (0.03) & \textbf{0.17} (0.02)/0.25 (0.02) & 0.37 (0.00)/0.26 (0.03)\\ 
$10, 1000, 0.1$ & \textbf{0.09} (0.01)/0.10 (0.01) & 0.15 (0.00)/0.23 (0.03) & \textbf{0.13} (0.02)/0.15 (0.02) & 0.25 (0.00)/0.25 (0.03)\\ 
$10, 1500, 0.1$ & \textbf{0.07} (0.01)/0.08 (0.01) & 0.12 (0.00)/0.23 (0.03) & \textbf{0.10} (0.01)/0.11 (0.01) & 0.23 (0.00)/0.24 (0.03)\\ 
$10, 500, 1$ & \textbf{0.14} (0.01)/0.16 (0.02) & 0.24 (0.00)/0.24 (0.03) & \textbf{0.19} (0.02)/0.25 (0.02) & 0.36 (0.00)/0.26 (0.03)\\ 
$10, 1000, 1$ & \textbf{0.11} (0.01)/0.12 (0.01) & 0.15 (0.00)/0.24 (0.03) & \textbf{0.14} (0.01)/0.15 (0.02) & 0.24 (0.00)/0.25 (0.03)\\ 
$10, 1500, 1$ & \textbf{0.09} (0.01)/0.10 (0.01) & 0.12 (0.00)/0.24 (0.03) & \textbf{0.12} (0.01)/0.12 (0.01) & 0.19 (0.00)/0.25 (0.03)\\ 
$10, 500, 3$ & \textbf{0.18} (0.01)/0.20 (0.01) & 0.23 (0.00)/0.27 (0.02) & \textbf{0.21} (0.02)/0.25 (0.02) & 0.36 (0.00)/0.28 (0.02)\\ 
$10, 1000, 3$ & \textbf{0.14} (0.00)/0.15 (0.00) & 0.15 (0.00)/0.26 (0.02) & \textbf{0.17} (0.01)/0.18 (0.02) & 0.25 (0.00)/0.27 (0.03)\\ 
$10, 1500, 3$ & 0.13 (0.01)/0.13 (0.01) & \textbf{0.12} (0.00)/0.26 (0.02) & \textbf{0.15} (0.01)/0.15 (0.01) & 0.20 (0.00)/0.26 (0.03)\\ 
$20, 500, 0.1$ & \textbf{0.09} (0.01)/0.13 (0.01) & 0.24 (0.00)/0.23 (0.02) & \textbf{0.14} (0.01)/0.23 (0.01) & 0.36 (0.00)/0.24 (0.02)\\ 
$20, 1000, 0.1$ & \textbf{0.07} (0.01)/0.09 (0.01) & 0.15 (0.00)/0.22 (0.02) & \textbf{0.10} (0.01)/0.13 (0.01) & 0.25 (0.00)/0.23 (0.02)\\ 
$20, 1500, 0.1$ & \textbf{0.06} (0.00)/0.07 (0.01) & 0.12 (0.00)/0.22 (0.02) & \textbf{0.09} (0.01)/0.09 (0.01) & 0.20 (0.00)/0.23 (0.02)\\ 
$20, 500, 1$ & \textbf{0.11} (0.01)/0.14 (0.01) & 0.24 (0.00)/0.23 (0.02) & \textbf{0.15} (0.01)/0.22 (0.01) & 0.35 (0.00)/0.24 (0.02)\\ 
$20, 1000, 1$ & \textbf{0.09} (0.01)/0.10 (0.01) & 0.15 (0.00)/0.23 (0.02) & \textbf{0.11} (0.01)/0.13 (0.01) & 0.24 (0.00)/0.23 (0.02)\\ 
$20, 1500, 1$ & \textbf{0.08} (0.00)/0.08 (0.01) & 0.12 (0.00)/0.23 (0.02) & \textbf{0.10} (0.01)/0.10 (0.01) & 0.19 (0.00)/0.23 (0.02)\\ 
$20, 500, 3$ & \textbf{0.15} (0.01)/0.17 (0.01) & 0.24 (0.00)/0.25 (0.01) & \textbf{0.17} (0.01)/0.25 (0.01) & 0.37 (0.00)/0.25 (0.02)\\ 
$20, 1000, 3$ & \textbf{0.13} (0.01)/0.14 (0.01) & 0.15 (0.00)/0.25 (0.01) & \textbf{0.14} (0.01)/0.15 (0.01) & 0.25 (0.00)/0.25 (0.02)\\ 
$20, 1500, 3$ & \textbf{0.11} (0.00)/0.12 (0.00) & 0.12 (0.00)/0.25 (0.01) & \textbf{0.12} (0.01)/0.13 (0.01) & 0.20 (0.00)/0.25 (0.02)\\ 
       \midrule
        \bottomrule
     \end{tabular}
  \end{threeparttable}}
\end{table}

\noindent Figure \ref{fig:n250p11} and Table \ref{tab:simres1} compare the proposed methods with primary-only BT and full pooling. The transfer methods perform best when the designated informative attributes are close to the primary attribute, corresponding to smaller $h$. The figure also shows the two endpoints of the simulation design. When $|\mathcal S_{\mathrm{gen}}|=0$, no secondary attribute is drawn from the concentrated informative distribution and primary-only BT is competitive, although a wider draw may occasionally satisfy Equation \eqref{eq:informative_set}. When $|\mathcal S_{\mathrm{gen}}|=S$, all secondary attributes satisfy the $h$-condition and full pooling is less harmful. For intermediate values, Oracle BT and the Discovery BT generally improve on both alternatives in the displayed results.

The main exception in Figure \ref{fig:n250p11} is $M=10$ and $h=3$, where primary-only BT remains competitive when only a few secondary attributes are designated informative. With only ten coordinates, the condition $\|\bm\delta^{(s)*}\|_2^2\le h$ can still allow fairly large differences between the primary and secondary log-worth vectors. More informative attributes are then needed before the extra observations compensate for this difference. This agrees qualitatively with Theorem \ref{thm:alphaconvergence}, whose bound contains both a pooled-to-primary discrepancy term and a sampling term that decreases as $n_{\mathcal S}$ increases when the other quantities are held fixed.

Table \ref{tab:simres1}, which averages over $|\mathcal S_{\mathrm{gen}}|$, shows the same general pattern. Transfer improves as more informative secondary attributes are available, while primary-only BT cannot use these extra observations. Oracle BT and the Discovery BT are usually similar, but the difference is larger in some sparse settings, such as $M=20$ and $N=500$. There, the primary and secondary training fits may involve different realised edge sets, which makes $\bar L_s-\bar L_0$ more variable and makes the screening step less stable.


\section{Eba consumer preferences} \label{sec:application}
Cassava (\textit{Manihot esculenta}) is an important source of carbohydrates in tropical regions, especially in Africa (Cock, \citeyear{cock1982cassava}). Because fresh cassava has a short shelf life, it is also processed into other foods. One example is eba, a West African staple made from cassava flour (Awoyale et al., \citeyear{awoyale2021review}). Identifying cassava varieties that consumers prefer for eba can therefore be useful for breeding and product development.

\begin{figure}[H]
\centering
\includegraphics[width=0.6\textwidth]{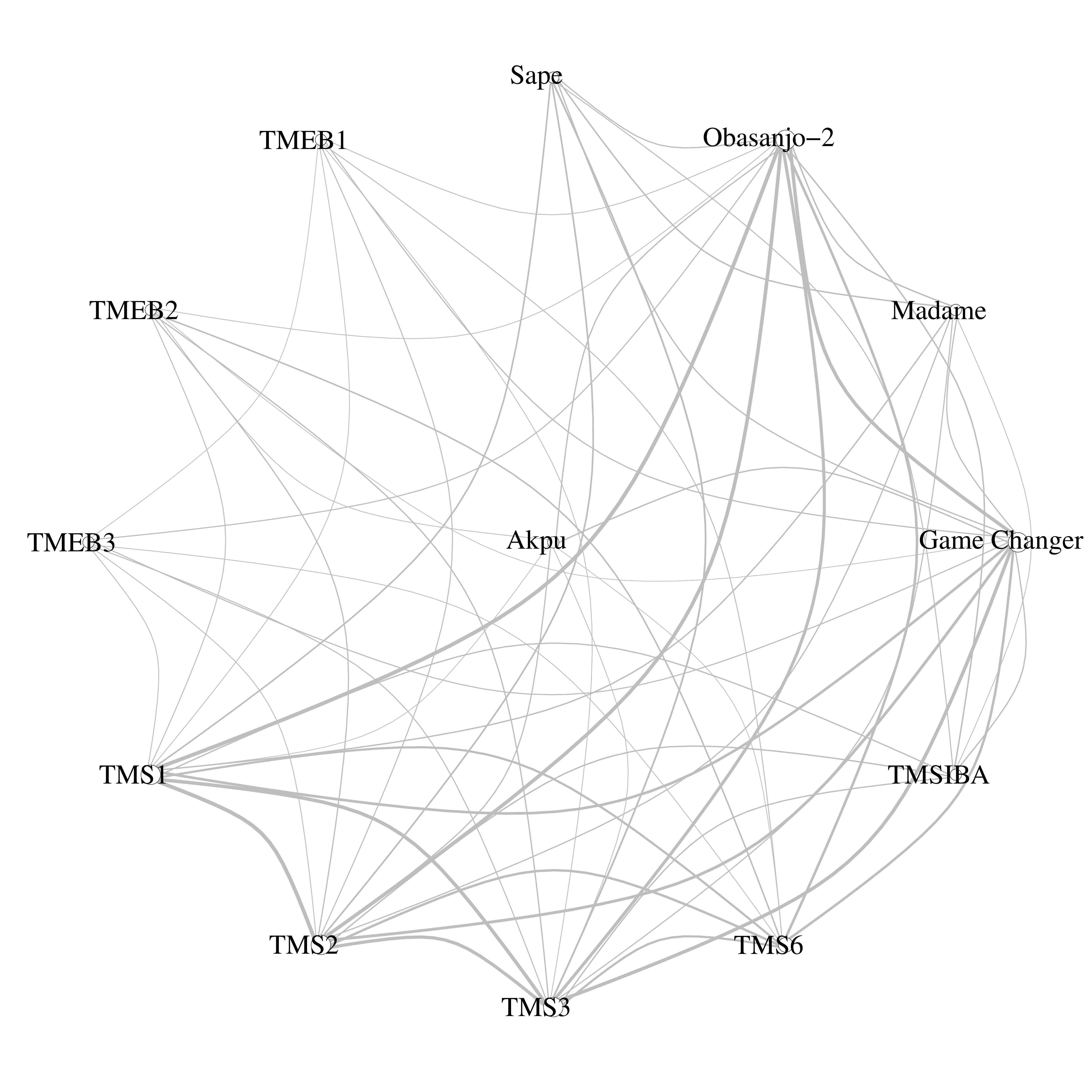}
\caption{Comparison graph for the primary attribute. An edge indicates that two cassava varieties were compared, and edge width represents the number of observed comparisons for that pair.}
\label{fig:connectgraph}
\end{figure}

We analyse the consumer preference data of Olaosebikan et al.\ (\citeyear{olaosebikan2023drivers}). The original partial rankings are converted to pairwise comparisons using full breaking. The data contain 13 cassava varieties and eight attributes: overall preference, colour, odour, firmness, stretchability, taste, smoothness and mouldability. Each attribute contains 1200 pairwise comparisons. We use overall preference as the primary attribute and the other seven attributes as secondary attributes. The consumers evaluated eba prepared from different cassava varieties in Cameroon and Nigeria.  Figure \ref{fig:connectgraph} shows the comparison graph for the primary attribute.

\noindent The graph in Figure \ref{fig:connectgraph} is not complete, so some cassava varieties are never compared directly. Nevertheless, it is relatively dense and contains many comparisons. We fit Discovery BT using the primary and secondary attributes and compare it with a Bradley--Terry model fitted only to the primary data. The value of $\lambda_\delta$ is selected by cross-validation. All secondary attributes satisfied the Discovery rule. Table \ref{tab:rankingssapp1} reports the estimates and rankings.

\begin{table}[H]
\centering
  \begin{threeparttable}
 \caption{Estimated log-worths and rankings using the proposed method for the cassava data. Log-worths are displayed to two decimal places, so equal displayed log-worths need not represent exact ties in the fitted values.}
  \label{tab:rankingssapp1}
     \begin{tabular}{|l| c | c | c | c |}
        \toprule
                \midrule
         & \multicolumn{2}{c|}{Discovery BT} & \multicolumn{2}{c|}{Primary attribute only BT}\\ \midrule
\textbf{Variety} & \textbf{Estimated log-worth} & \textbf{Estimated ranking} & \textbf{Estimated log-worth} & \textbf{Estimated ranking}\\ \midrule
Akpu & -0.67 & 13 & -0.88 & 13\\		
Game Changer & 0.56 & 1 & 0.26 & 5\\				
Madame & -0.03  & 8 & 0.37  & 3\\			
Obasanjo-2 & 0.45 & 3 & 0.01 & 8\\			
Sape  & 0.16 & 7 & 0.93 & 1\\			
TMEB1 & -0.49 &  10 & 0.27 &  4\\		
TMEB2 & -0.64 & 12 & -0.64 & 12\\	
TMEB3 & -0.49 & 11 & 0.21 & 6\\	
TMS1 &  0.30  & 5 &  -0.36  & 9\\			
TMS2  & 0.21 & 6 & -0.50 & 11\\			
TMS3 & 0.48 & 2 & 0.05  & 7\\			
TMS6 & 0.34 & 4 & 0.71 & 2\\			
TMSIBA & -0.20 & 9 & -0.42 & 10\\				
\midrule
\bottomrule
     \end{tabular}
 \end{threeparttable}
\end{table}

\noindent The two methods produce some similar estimates, especially among the lower-ranked varieties, but there are also clear differences. Discovery BT ranks Game Changer, TMS3 and Obasanjo-2 as the three most preferred varieties, whereas none of these is among the top four under primary-only BT. Popular local and regional varieties such as Akpu, TMEB1, TMEB2 and TMEB3 receive relatively low estimated overall preference, which may warrant further investigation (Olaosebikan et al., \citeyear{olaosebikan2023drivers}). Because all secondary attributes enter the current selected set, Discovery BT uses substantially more data than the primary-only fit, so differences between the rankings are not unexpected.

\section{Conclusion} \label{sec:conclusion}
We introduce a transfer-learning method for multi-attribute preference data when the main interest lies in one primary attribute. The method pools information from informative secondary attributes and then uses the primary data to correct the pooled estimate of the primary Bradley--Terry log-worths. For a known informative set and under a pooled Bradley--Terry compatibility condition, we derive high-probability error bounds and give conditions under which the transfer bound has a smaller asymptotic order than the corresponding primary-only bound. We also derive asymptotic normality for a one-step estimator based on a separate primary inference sample. The simulations show that the proposed method can improve estimation across a range of settings, and the eba example illustrates how the method can be used in practice. 

The same transfer construction can also be used with a Plackett--Luce likelihood for rankings of more than two objects. We study this extension empirically, but do not prove a corresponding transfer-learning theorem. Extending the sparse-graph theory to partial rankings is therefore one direction for future work. Another extension is to include characteristics of the individuals or objects. Such information could be used to recommend objects to new individuals or to predict preferences for new objects. Bilinear models can address both tasks (Schäfer \& Hüllermeier, \citeyear{schafer2018dyad}), although they can be computationally demanding because many user--object interactions must be estimated. Other theoretical extensions include relaxing exact pooled Bradley--Terry compatibility through a pseudo-true pooled parameter and establishing guarantees for the data-driven Discovery step.

\bibliographystyle{Chicago}
\bibliography{library}

\appendix 
\section*{Supplementary material}

\section{Extension to partial ranking data} \label{sec:pl-extension}

Although the main paper focuses on pairwise comparisons, the same transfer framework can be applied to partial rankings by replacing the Bradley--Terry likelihood with a Plackett--Luce likelihood. Suppose that $n_{\mathrm{PL}}$ rankings are observed. Ranking $i$ is based on a presented set $A_i\subseteq\{1,\ldots,M\}$ containing $m_i$ objects, where $2\le m_i\le M$, and gives a complete ordering of those objects:
\[
\bm\sigma_i=(\sigma_{i1},\ldots,\sigma_{im_i}),
\qquad \sigma_{ij}\in A_i,
\]
where $\sigma_{ij}$ is the object ranked in position $j$. Conditional on the presented set $A_i$, the Plackett--Luce probability of the observed ordering is
\begin{equation*}
\mathbb{P}(\bm\sigma_i\mid A_i,\bm\alpha)
=
\prod_{j=1}^{m_i}
\frac{\exp(\alpha_{\sigma_{ij}})}
{\sum_{l=j}^{m_i}\exp(\alpha_{\sigma_{il}})}.
\end{equation*}
Here $\alpha_k\in\mathbb R$ is the log-worth of object $k$ and $\exp(\alpha_k)>0$ is its Plackett--Luce worth. The expression above assumes that all objects in $A_i$ are ranked. If only the top $m$ objects from a larger choice set were observed, the unranked objects would remain in the denominators. The corresponding negative log-likelihood is
\begin{equation}
\label{eq:likpl}
L(\bm{\alpha}) = \sum_{i = 1}^{n_{\mathrm{PL}}}\sum_{j = 1}^{m_i}\left\{\log\left[\sum_{l = j}^{m_i} \exp(\alpha_{\sigma_{il}})\right] - \alpha_{\sigma_{ij}}\right\}.
\end{equation}
Replacing the Bradley--Terry negative log-likelihood in Algorithms \ref{alg:oraclebtl} and \ref{alg:discbtl} by Equation \eqref{eq:likpl} gives the corresponding Plackett--Luce transfer procedure. In Discovery PL, the screening criterion uses the total Plackett--Luce negative log-likelihood on the held-out primary rankings. As before, multiplying the negative log-likelihood by a positive constant does not change an unregularised fit, but $\lambda_\delta$ must be rescaled in a penalised fit if the same minimiser is to be retained.

We simulate ranking data in the same way as in Section \ref{sec:simulation}, except that each observation is now a complete ranking of a presented subset of $m=3$ objects rather than a pairwise comparison. Here $N$ denotes the number of sampled rankings per attribute. Table \ref{tab:simrespl} shows the results.

\begin{table}[H]
\resizebox{\textwidth}{!}{%
\centering
  \begin{threeparttable}
  \caption{Simulation results for the Plackett--Luce methods. The $\ell_2$ errors are averaged over 50 simulated data sets for each parameter combination and over the considered values of $|\mathcal S_{\mathrm{gen}}|$, and are rounded to two decimals. Monte Carlo standard errors are shown in parentheses. Bold values indicate the smallest error among the four methods for that parameter combination; equal displayed values need not be exact ties before rounding. Oracle PL is given $\mathcal S_{\mathrm{gen}}$, Discovery PL uses Algorithm \ref{alg:discbtl} with the Plackett--Luce likelihood, PL uses only the primary data, and PPL pools all attributes.}
  \label{tab:simrespl}
     \begin{tabular}{c | cc | cc }
        \toprule
        \midrule
         & \multicolumn{2}{c|}{$M = 10$} & \multicolumn{2}{c}{$M = 20$}\\ \midrule
         \textbf{$S, N, h$} & \textbf{Oracle PL/Discovery PL} & \textbf{PL/PPL} & \textbf{Oracle PL/Discovery PL} & \textbf{PL/PPL}\\ \midrule
$5, 500, 0.1$ & \textbf{0.09} (0.01)/0.10 (0.02) & 0.14 (0.00)/0.25 (0.05) & \textbf{0.14} (0.02)/0.15 (0.02) & 0.22 (0.00)/0.27 (0.05)\\ 
$5, 1000, 0.1$ & \textbf{0.07} (0.01)/0.08 (0.01) & 0.10 (0.00)/0.24 (0.05) & \textbf{0.10} (0.01)/0.10 (0.01) & 0.15 (0.00)/0.27 (0.05)\\ 
$5, 1500, 0.1$ & \textbf{0.06} (0.00)/0.06 (0.00) & 0.08 (0.00)/0.24 (0.05) & \textbf{0.09} (0.01)/0.09 (0.01) & 0.12 (0.00)/0.26 (0.06)\\ 
$5, 500, 1$ & \textbf{0.11} (0.01)/0.12 (0.01) & 0.14 (0.00)/0.26 (0.04) & \textbf{0.15} (0.00)/0.16 (0.00) & 0.22 (0.00)/0.29 (0.05)\\ 
$5, 1000, 1$ & \textbf{0.09} (0.00)/0.09 (0.01) & 0.10 (0.00)/0.26 (0.05) & \textbf{0.11} (0.01)/0.12 (0.01) & 0.15 (0.00)/0.28 (0.05)\\ 
$5, 1500, 1$ & \textbf{0.07} (0.00)/0.08 (0.00) & 0.08 (0.00)/0.26 (0.05) & \textbf{0.10} (0.01)/0.10 (0.01) & 0.12 (0.00)/0.28 (0.05)\\ 
$5, 500, 3$ & 0.15 (0.01)/0.16 (0.01) & \textbf{0.14} (0.00)/0.29 (0.03) & \textbf{0.18} (0.01)/0.19 (0.02) & 0.22 (0.00)/0.30 (0.04)\\ 
$5, 1000, 3$ & 0.11 (0.01)/0.12 (0.00) & \textbf{0.10} (0.00)/0.28 (0.03) & \textbf{0.14} (0.01)/0.15 (0.01) & 0.15 (0.00)/0.29 (0.04)\\ 
$5, 1500, 3$ & 0.10 (0.01)/0.09 (0.00) & \textbf{0.08} (0.00)/0.28 (0.03) & \textbf{0.12} (0.00)/0.12 (0.00) & 0.12 (0.00)/0.29 (0.04)\\ 
$10, 500, 0.1$ & \textbf{0.08} (0.01)/0.09 (0.01) & 0.14 (0.00)/0.23 (0.03) & \textbf{0.11} (0.01)/0.13 (0.01) & 0.21 (0.00)/0.25 (0.03)\\ 
$10, 1000, 0.1$ & \textbf{0.07} (0.00)/0.07 (0.01) & 0.10 (0.00)/0.23 (0.03) & \textbf{0.09} (0.01)/0.09 (0.01) & 0.15 (0.00)/0.24 (0.03)\\ 
$10, 1500, 0.1$ & \textbf{0.06} (0.00)/0.06 (0.00) & 0.08 (0.00)/0.23 (0.03) & \textbf{0.08} (0.01)/0.08 (0.01) & 0.12 (0.00)/0.24 (0.03)\\ 
$10, 500, 1$ & \textbf{0.10} (0.01)/0.11 (0.01) & 0.14 (0.00)/0.24 (0.03) & \textbf{0.12} (0.01)/0.14 (0.01) & 0.22 (0.00)/0.25 (0.03)\\ 
$10, 1000, 1$ & \textbf{0.08} (0.00)/0.08 (0.00) & 0.10 (0.00)/0.24 (0.03) & \textbf{0.10} (0.01)/0.10 (0.01) & 0.15 (0.00)/0.25 (0.03)\\ 
$10, 1500, 1$ & \textbf{0.07} (0.00)/0.07 (0.00) & 0.08 (0.00)/0.24 (0.03) & \textbf{0.09} (0.01)/0.09 (0.01) & 0.12 (0.00)/0.25 (0.03)\\ 
$10, 500, 3$ & \textbf{0.13} (0.01)/0.14 (0.01) & 0.14 (0.00)/0.27 (0.02) & \textbf{0.15} (0.01)/0.16 (0.01) & 0.22 (0.00)/0.27 (0.03)\\ 
$10, 1000, 3$ & 0.11 (0.00)/0.11 (0.00) & \textbf{0.10} (0.00)/0.26 (0.02) & \textbf{0.12} (0.01)/0.13 (0.01) & 0.15 (0.00)/0.26 (0.03)\\ 
$10, 1500, 3$ & 0.09 (0.00)/0.09 (0.00) & \textbf{0.08} (0.00)/0.26 (0.02) & \textbf{0.11} (0.01)/0.12 (0.00) & 0.12 (0.00)/0.26 (0.03)\\ 
$20, 500, 0.1$ & \textbf{0.07} (0.00)/0.08 (0.01) & 0.14 (0.00)/0.22 (0.02) & \textbf{0.09} (0.01)/0.11 (0.01) & 0.22 (0.00)/0.23 (0.02)\\ 
$20, 1000, 0.1$ & \textbf{0.06} (0.00)/0.06 (0.00) & 0.10 (0.00)/0.22 (0.02) & \textbf{0.07} (0.01)/0.08 (0.01) & 0.15 (0.00)/0.22 (0.02)\\ 
$20, 1500, 0.1$ & \textbf{0.05} (0.00)/0.05 (0.00) & 0.08 (0.00)/0.21 (0.02) & \textbf{0.07} (0.00)/0.07 (0.00) & 0.13 (0.00)/0.22 (0.02)\\ 
$20, 500, 1$ & \textbf{0.08} (0.01)/0.09 (0.01) & 0.14 (0.00)/0.23 (0.02) & \textbf{0.10} (0.01)/0.11 (0.01) & 0.22 (0.00)/0.23 (0.02)\\ 
$20, 1000, 1$ & \textbf{0.07} (0.00)/0.07 (0.00) & 0.10 (0.00)/0.23 (0.02) & \textbf{0.08} (0.01)/0.08 (0.01) & 0.15 (0.00)/0.23 (0.02)\\ 
$20, 1500, 1$ & \textbf{0.06} (0.00)/0.06 (0.00) & 0.08 (0.00)/0.23 (0.02) & \textbf{0.08} (0.00)/0.08 (0.00) & 0.12 (0.00)/0.23 (0.02)\\ 
$20, 500, 3$ & \textbf{0.12} (0.00)/0.13 (0.00) & 0.14 (0.00)/0.25 (0.02) & \textbf{0.13} (0.01)/0.14 (0.01) & 0.22 (0.00)/0.25 (0.02)\\ 
$20, 1000, 3$ & \textbf{0.10} (0.00)/0.11 (0.00) & 0.10 (0.00)/0.25 (0.02) & \textbf{0.11} (0.01)/0.11 (0.01) & 0.15 (0.00)/0.24 (0.02)\\ 
$20, 1500, 3$ & \textbf{0.08} (0.00)/0.09 (0.00) & 0.08 (0.00)/0.25 (0.02) & \textbf{0.09} (0.00)/0.10 (0.00) & 0.12 (0.00)/0.24 (0.02)\\ 
       \midrule
        \bottomrule
     \end{tabular}
  \end{threeparttable}}
\end{table}

\section{Proofs} \label{sec:proofs}
This section proves Theorem \ref{thm:alphaconvergence}, Corollaries \ref{crl:rate_improvement} and \ref{crl:alphaconvergencel2}, Theorem \ref{thm:asympnormal}, and Corollary \ref{crl:asympnormal2}. The transfer-step proof follows Chen et al.\ (\citeyear{chen2019spectral}) and Fan et al.\ (\citeyear{fan2024uncertainty}) and uses a regularised MLE, gradient descent and a leave-one-out argument. The main difference is that observations in the pooled likelihood may come from different secondary-attribute models. The primary correction is handled separately using the ridge form of the second step.

\noindent We first prove two results for the comparison graph and its Laplacian. Section \ref{sec:transfer-convergence} then shows that
\[
\|\hat{\bm u}-\bm u^*\|_\infty
\lesssim
\kappa_1^2
\sqrt{\frac{\log M}{Mp(n_0+n_{\mathcal S})}}.
\]
Section \ref{sec:primary-correction-proof} then establishes the deterministic ridge comparison
\[
\|\hat{\bm\alpha}_{\lambda_\delta}-\bm\alpha^*\|_\infty
\le
\|\hat{\bm u}-\bm\alpha^*\|_\infty
+
\frac{\|\nabla L_0(\bm\alpha^*)\|_\infty}{\lambda_\delta},
\]
which is subsequently combined with the transfer bound and a concentration result for the primary score.
\begin{lemma}
\label{lemma:lemma1_joint}
Suppose that Assumption \ref{ass:1} holds. Let $d_i$ be the degree of node $i$, $d_{\min} = \min_{1\leq i \leq M} d_i$ and $d_{\max} = \max_{1\leq i \leq M} d_i$. If $Mp > c_p \log M$ for some sufficiently large constant $c_p > 0$ and $M$ is sufficiently large, then the following event
\begin{equation*}
E_0 = \left\{ \frac{Mp}{2} \leq d_{\min} \leq d_{\max} \leq \frac{3Mp}{2}\right\}
\end{equation*}
occurs with probability at least $1 - O(M^{-10})$.
\begin{proof}
For each vertex $i$, $d_i\sim\mathrm{Binomial}(M-1,p)$ with mean $(M-1)p$. For sufficiently large $M$, the thresholds $Mp/2$ and $3Mp/2$ are fixed constant-factor deviations below and above this mean. Chernoff's inequality therefore gives
\[
\mathbb P\!\left(d_i<\frac{Mp}{2}\right)
+
\mathbb P\!\left(d_i>\frac{3Mp}{2}\right)
\le 2e^{-cMp}
\]
for an absolute constant $c>0$. A union bound over $i=1,\ldots,M$ and the condition $Mp\ge c_p\log M$, with $c_p$ sufficiently large, yield the stated probability.
\end{proof}
\end{lemma}

\noindent To analyse the Hessians of the negative log-likelihood functions, we use the Laplacian matrix of the common theoretical comparison graph (cf.\ Chen et al., \citeyear{chen2019spectral}; Fan et al., \citeyear{fan2024uncertainty}),
\[
\bm L_{\mathcal G}
=
\sum_{(j,l)\in\mathcal E,\,j>l}
(\bm e_j-\bm e_l)(\bm e_j-\bm e_l)^T,
\]
where $\bm e_1,\ldots,\bm e_M$ are the canonical basis vectors in $\mathbb R^M$. The following lemma provides bounds on the eigenvalues of $\bm L_{\mathcal G}$.
\begin{lemma}
\label{lemma:lemma2_joint}
Suppose Assumption \ref{ass:1} holds and $Mp > c_p \log M$ for some sufficiently large constant $c_p > 0$. Then the following event
\begin{equation*}
E_1 = \left\{\frac{Mp}{2} \leq \lambda_{\min, \perp}(\bm L_{\mathcal G}) \leq \|\bm L_{\mathcal G}\| \leq 2 Mp\right\}
\end{equation*}
occurs with probability at least $1 - O(M^{-10})$ for large $M$.
\begin{proof}
Let $A_{jl}=\mathbbm1\{(j,l)\in\mathcal E\}$ and $\bm x_{jl}=\bm e_j-\bm e_l$. Since
\[
\sum_{j>l}\bm x_{jl}\bm x_{jl}^T
=M\bm I_M-\bm1_M\bm1_M^T,
\]
we have
\[
\mathbb E\bm L_{\mathcal G}
=p(M\bm I_M-\bm1_M\bm1_M^T),
\]
whose nonzero eigenvalues are all $Mp$. Write
\[
\bm L_{\mathcal G}-\mathbb E\bm L_{\mathcal G}
=
\sum_{j>l}(A_{jl}-p)\bm x_{jl}\bm x_{jl}^T.
\]
Each summand has operator norm at most $2$, and, because $(\bm x_{jl}\bm x_{jl}^T)^2=2\bm x_{jl}\bm x_{jl}^T$, the matrix-variance parameter is at most $2Mp$. Matrix Bernstein therefore gives
\[
\|\bm L_{\mathcal G}-\mathbb E\bm L_{\mathcal G}\|
\lesssim
\sqrt{Mp\log M}+\log M
\]
with probability at least $1-O(M^{-10})$, after increasing the numerical constant in the tail bound. If $Mp\ge c_p\log M$ with $c_p$ sufficiently large, the right-hand side is at most $Mp/2$. Weyl's inequality then gives
\[
\lambda_{\min,\perp}(\bm L_{\mathcal G})\ge\frac{Mp}{2},
\qquad
\|\bm L_{\mathcal G}\|\le\frac{3Mp}{2}\le2Mp,
\]
which proves the result.
\end{proof}
\end{lemma}
\noindent Under Assumption \ref{ass:1}, the primary and pooled theoretical samples share the same Erd\H{o}s--R\'enyi comparison graph. Hence the degree and Laplacian events in Lemmas \ref{lemma:lemma1_joint} and \ref{lemma:lemma2_joint} apply to both the primary and pooled negative log-likelihoods.

\subsection{Transfer convergence} \label{sec:transfer-convergence}
Throughout this subsection, $L$ denotes the pooled negative log-likelihood $L_{\mathrm{tr}}=L_{\mathcal D_{\mathrm{tr}}}$ under Assumptions \ref{ass:1}, \ref{ass:2} and \ref{ass:pooledtarget}. Directly obtaining an entrywise rate for $\hat{\bm u}$ is difficult. We therefore first study a regularised version of the pooled estimator and subsequently show that the same rate is attained by the unregularised MLE. The regularised objective is smooth and strongly convex, which allows us to use the gradient-descent argument of Chen et al.\ (\citeyear{chen2019spectral}) and Fan et al.\ (\citeyear{fan2024uncertainty}); see also Bubeck (\citeyear{bubeck2015convex}).

\noindent Define $B_u:=1+\log\kappa_1$, where the addition of one avoids division by zero when $\kappa_1=1$. Since $\bm u^*$ is centred and has range $\log\kappa_1$, $\|\bm u^*\|_\infty\le\log\kappa_1$ and $\|\bm u^*\|_2\le\sqrt M\log\kappa_1$.

\begin{theorem}[Auxiliary regularised pooled estimator]\label{thm:convergence_step1_alpha}
Suppose Assumptions \ref{ass:1}, \ref{ass:2} and \ref{ass:pooledtarget} hold, $Mp\ge C_\kappa\kappa_1^4\log M$ for a sufficiently large $C_\kappa$, and $n_0 + n_{\mathcal{S}} \leq c_2 M^{c_3}$ for some $c_2, c_3 >0$. Set $\lambda_u = c_{\lambda_{u}}B_u^{-1}\sqrt{\frac{p\log M}{n_0 + n_{\mathcal{S}}}}$ for some $c_{\lambda_{u}} > 0$. Then, with probability at least $1-O(M^{-5})$,
\begin{equation*}
\|\hat{\bm{u}}_{\lambda_u} - \bm{u}^*\|_\infty \lesssim \kappa_1^2\sqrt{\frac{\log M}{M p(n_0 + n_\mathcal{S})}},
\end{equation*}
\noindent where $\hat{\bm{u}}_{\lambda_u} = \argmin_{\bm{u} \in \mathbb{R}^M:\,\bm1_M^T\bm u=0}L_{\lambda_u}(\bm{u})$, $L_{\lambda_u}(\bm{u}) = L(\bm{u}) + \frac{\lambda_u}{2}\|\bm{u}\|_2^2$.
\end{theorem}

\noindent To establish Theorem \ref{thm:convergence_step1_alpha}, we first record the gradient and Hessian of $L_{\lambda_u}(\bm u)$, together with the concentration and spectral bounds used in the proof. The gradient of $L_{\lambda_u}(\bm u)$ is given by
\begin{gather}
\nabla L_{\lambda_u}(\bm{u}) =  \sum_{(j,l)\in \mathcal E, j>l}\left\{-y_{l,j} + \frac{\exp(u_j)}{\exp(u_j) + \exp(u_l)}\right\}(\bm e_j - \bm e_l) + \lambda_u \bm{u}\notag \\
= \frac{1}{n_0 + n_{\mathcal{S}}}\sum_{(j,l)\in \mathcal E, j>l}\sum_{i = 1}^{n_0 + n_{\mathcal{S}}}\left\{-y_{l,j}^{(i)} + \frac{\exp(u_j)}{\exp(u_j) + \exp(u_l)}\right\}(\bm e_j - \bm e_l) + \lambda_u \bm{u},
\end{gather}
and the Hessian of $L_{\lambda_u}(\bm{u})$ is
\begin{equation}
\nabla^2 L_{\lambda_u}(\bm{u})
=
\sum_{(j,l)\in \mathcal E,\,j>l}
\frac{\exp(u_j+u_l)}{[\exp(u_j)+\exp(u_l)]^2}
(\bm e_j-\bm e_l)(\bm e_j-\bm e_l)^T
+\lambda_u\bm I_M.
\label{eq:hessian_transfer_correct}
\end{equation}
The Hessian contains no response term because the derivative of the linear term $-y_{l,j}(u_j-u_l)$ is constant. This is the same weighted-graph-Laplacian Hessian used in the standard Bradley--Terry analysis.
\\
\\
\noindent The pooled observations need not be identically distributed across attributes. Let
\[
\pi_{j,l}^{(i)}
:=
\mathbb P(y_{l,j}^{(i)}=1\mid\mathcal G),
\qquad
z_{j,l}^{(i)}
:=
\{\pi_{j,l}^{(i)}-y_{l,j}^{(i)}\}(\bm e_j-\bm e_l).
\]
Then $\mathbb E[z_{j,l}^{(i)}\mid\mathcal G]=0$ for every observed edge. Assumption \ref{ass:pooledtarget} gives, edge by edge,
\[
\frac1{N_{\mathrm{tr}}}\sum_{i=1}^{N_{\mathrm{tr}}}\pi_{j,l}^{(i)}
=
\frac{\exp(u_j^*)}{\exp(u_j^*)+\exp(u_l^*)}.
\]
Consequently,
\begin{equation}
\nabla L(\bm u^*)
=
\frac1{N_{\mathrm{tr}}}
\sum_{(j,l)\in\mathcal E,\,j>l}
\sum_{i=1}^{N_{\mathrm{tr}}}
z_{j,l}^{(i)},
\label{eq:pseudotrue_empirical_score}
\end{equation}
so the likelihood score at $\bm u^*$ is exactly centred. This is the point at which the pooled Bradley--Terry compatibility assumption is used.

\noindent The next three lemmas provide concentration bounds for the gradient and spectral bounds for the Hessian; they are used repeatedly throughout the leave-one-out proof.
\begin{lemma}
\label{lemma:lemma3_step1}
Suppose Assumptions \ref{ass:1}, \ref{ass:2} and \ref{ass:pooledtarget} hold and $Mp\ge c\log M$ for a sufficiently large constant $c>0$. For $\lambda_u = c_{\lambda_{u}}B_u^{-1}\sqrt{\frac{p\log M}{n_0 + n_{\mathcal{S}}}}$, the following event
\begin{equation*}
E_2 = \left\{\|\nabla L_{\lambda_u}(\bm{u}^*)\|_2 \leq c_0 \sqrt{\frac{M^2 p \log M}{n_0 + n_\mathcal{S}}}\right\}
\end{equation*}
occurs with probability at least $1 - O(M^{-10})$ for a sufficiently large constant $c_0>0$, which may depend on $c_{\lambda_{u}}$.
\begin{proof}
Conditional on the graph, the unregularised score in Equation \eqref{eq:pseudotrue_empirical_score} is a sum of independent centred bounded variables. For coordinate $k$, each of the $d_k$ incident edges contributes $N_{\mathrm{tr}}$ observations, so on the degree event $E_0$ Bernstein's inequality gives
\[
\mathbb P\!\left(
|\{\nabla L(\bm u^*)\}_k|>t\mid\mathcal G
\right)
\le
2\exp\!\left[-c\min\left\{
\frac{N_{\mathrm{tr}}t^2}{Mp},
N_{\mathrm{tr}}t
\right\}\right].
\]
Taking $t=C\sqrt{Mp\log M/N_{\mathrm{tr}}}$ and using $MpN_{\mathrm{tr}}\gtrsim\log M$ yields, after a union bound over $k$,
\[
\|\nabla L(\bm u^*)\|_\infty
\lesssim
\sqrt{\frac{Mp\log M}{N_{\mathrm{tr}}}}
\]
with probability at least $1-O(M^{-10})$. Hence
\[
\|\nabla L(\bm u^*)\|_2
\le\sqrt M\,\|\nabla L(\bm u^*)\|_\infty
\lesssim
M\sqrt{\frac{p\log M}{N_{\mathrm{tr}}}}.
\]
The regularised score satisfies
\[
\nabla L_{\lambda_u}(\bm u^*)
=
\nabla L(\bm u^*)+\lambda_u\bm u^*.
\]
Using $\|\bm u^*\|_2\le\sqrt M\log\kappa_1$ and
\[
\lambda_u
=c_{\lambda_u}B_u^{-1}\sqrt{\frac{p\log M}{N_{\mathrm{tr}}}},
\qquad B_u=1+\log\kappa_1,
\]
we obtain
\[
\lambda_u\|\bm u^*\|_2
\lesssim \sqrt{\frac{Mp\log M}{N_{\mathrm{tr}}}}
\le M\sqrt{\frac{p\log M}{N_{\mathrm{tr}}}}.
\]
Combining the conditional concentration bound with $\mathbb P(E_0^c)=O(M^{-10})$ proves the stated unconditional probability for a sufficiently large constant $c_0$.
\end{proof}
\end{lemma}

\noindent From this point on, we work on the high-probability event $E_2$ from Lemma \ref{lemma:lemma3_step1}, together with the graph events stated above. 

\begin{lemma}
\label{lemma:lemma4_step1}
Suppose $E_1$ in Lemma \ref{lemma:lemma2_joint} occurs. Then
\begin{equation*}
\lambda_{\max}(\nabla^2 L_{\lambda_u}(\bm{u})) \leq \lambda_u + Mp, \quad \forall \bm{u} \in \mathbb{R}^M \text{ s.t. } \bm1_M^T \bm{u} = 0.
\end{equation*}
\begin{proof}
Since $\frac{\exp(u_j + u_l)}{[\exp(u_j) + \exp(u_l)]^2} \leq 0.25$,
\begin{equation*}
\lambda_{\max}(\nabla^2 L_{\lambda_u}(\bm{u})) \leq \lambda_u + \frac{1}{4}\|\bm{L}_{\mathcal G}\| \leq  \lambda_u + Mp, \quad \forall \bm{u} \in \mathbb{R}^M  \text{ s.t. } \bm1_M^T \bm{u} = 0.
\end{equation*}
\end{proof}
\end{lemma}

\begin{lemma}
\label{lemma:lemma5_step1}
Suppose $E_1$ (see Lemma \ref{lemma:lemma2_joint}) occurs. For any $r\ge0$ and every $\bm u$ satisfying $\|\bm u-\bm u^*\|_\infty\le r$,
\begin{equation*}
\lambda_{\min,\perp}(\nabla^2L(\bm u))
\ge \frac{Mp}{8\kappa_1e^{2r}},
\qquad
\lambda_{\min,\perp}(\nabla^2L_{\lambda_u}(\bm u))
\ge \lambda_u+\frac{Mp}{8\kappa_1e^{2r}}.
\end{equation*}
\begin{proof}
For every $j\ne l$,
\[
\frac{\exp(u_j+u_l)}{[\exp(u_j)+\exp(u_l)]^2}
=
\frac{e^{-|u_j-u_l|}}{(1+e^{-|u_j-u_l|})^2}
\ge \frac14e^{-|u_j-u_l|}.
\]
Moreover,
\[
|u_j-u_l|
\le |u_j^*-u_l^*|+|u_j-u_j^*|+|u_l-u_l^*|
\le \log\kappa_1+2r.
\]
Hence every edge weight is at least $(4\kappa_1e^{2r})^{-1}$. Using $\lambda_{\min,\perp}(\bm L_{\mathcal G})\ge Mp/2$ on $E_1$ gives
\[
\lambda_{\min,\perp}(\nabla^2L(\bm u))
\ge \frac{Mp}{8\kappa_1e^{2r}}.
\]
Adding $\lambda_u\bm I_M$ yields the regularised bound.
\end{proof}
\end{lemma}

\subsubsection{Gradient Descent}
We now consider gradient descent on the centred parameter space. The Bradley--Terry likelihood gradient is orthogonal to $\bm1_M$, while the ridge gradient $\lambda_u\bm u$ is centred whenever $\bm u$ is centred. Since the initial value is centred, all subsequent iterates therefore remain centred and no explicit projection is required. Let $\{\bm u^{[t]}\}_{t=0,1,\ldots}$ denote the resulting sequence, with step size $\eta=(\lambda_u+Mp)^{-1}$ and $T=\lceil M^{C_T}\rceil$ iterations for a sufficiently large fixed constant $C_T$. The polynomial number of iterations is used only in the proof, to make the optimisation error negligible relative to the statistical error.

\begin{algorithm}
\caption{Gradient descent for regularised MLE}\label{alg:gdrmle_step1}
  \begin{algorithmic}[1]
  	\STATE Initialise $\bm{u}^{[0]} = \bm{u}^*$
      \FOR{$t = 0$ to $T-1$}
		\STATE $\bm{u}^{[t+1]} = \bm{u}^{[t]} - \eta \nabla L_{\lambda_u}(\bm{u}^{[t]})$
              \ENDFOR
  \end{algorithmic}
\end{algorithm}

\begin{lemma}
\label{lemma:lemma6_step1}
Suppose $E_0$ in Lemma \ref{lemma:lemma1_joint} occurs. Then
\begin{equation*}
\|\bm{u}^{[t]} - \hat{\bm{u}}_{\lambda_u}\|_2 \leq \rho^{t}\|\bm{u}^{[0]} - \hat{\bm{u}}_{\lambda_u}\|_2, 
\end{equation*}
where $\rho = 1 - \frac{\lambda_u}{\lambda_u + Mp}$.
\begin{proof}
On $E_0$, $\|\bm L_{\mathcal G}\|\le2d_{\max}\le3Mp$. Since every logistic weight is at most $1/4$,
\[
\lambda_{\max}(\nabla^2L_{\lambda_u}(\bm u))
\le
\lambda_u+\frac14\|\bm L_{\mathcal G}\|
\le
\lambda_u+Mp
\]
for every $\bm u$. Moreover, $L_{\lambda_u}$ is globally $\lambda_u$-strongly convex because its unregularised Hessian is positive semidefinite. Its unique unconstrained minimiser is centred: taking the inner product of $\nabla L_{\lambda_u}(\bm u)=\bm0$ with $\bm1_M$ gives $\bm1_M^T\bm u=0$. Hence this minimiser is exactly $\hat{\bm u}_{\lambda_u}$ as defined above. Standard gradient-descent contraction for a $\lambda_u$-strongly convex, $(\lambda_u+Mp)$-smooth function with step size $\eta=(\lambda_u+Mp)^{-1}$ therefore gives
\[
\|\bm u^{[t]}-\hat{\bm u}_{\lambda_u}\|_2
\le
\left(1-\frac{\lambda_u}{\lambda_u+Mp}\right)^t
\|\bm u^{[0]}-\hat{\bm u}_{\lambda_u}\|_2.
\]
\end{proof}
\end{lemma}

\begin{lemma}
\label{lemma:lemma7_step1}
Suppose $E_2$ in Lemma \ref{lemma:lemma3_step1} occurs. Then
\begin{equation*}
\|\bm{u}^{[0]} - \hat{\bm{u}}_{\lambda_u}\|_2 = \|\hat{\bm{u}}_{\lambda_u} - \bm{u}^*\|_2 \leq C_7B_u M
\end{equation*}
\begin{proof}
Since $\hat{\bm{u}}_{\lambda_u}$ is the minimiser, $L_{\lambda_u}(\bm{u}^*) \geq L_{\lambda_u}(\hat{\bm{u}}_{\lambda_u})$. By the mean value theorem, for some $\bm{u}'$ between $\bm{u}^*$ and $\hat{\bm{u}}_{\lambda_u}$, we have
\begin{equation*}
L_{\lambda_u}(\hat{\bm{u}}_{\lambda_u}) = L_{\lambda_u}(\bm{u}^*) + \nabla L_{\lambda_u}(\bm{u}^*)^T (\hat{\bm{u}}_{\lambda_u} - \bm{u}^*) + \frac{1}{2}(\hat{\bm{u}}_{\lambda_u} - \bm{u}^*)^T\nabla^2 L_{\lambda_u}(\bm{u}')(\hat{\bm{u}}_{\lambda_u} - \bm{u}^*).
\end{equation*}
Consequently,
\begin{equation*}
\begin{gathered}
L_{\lambda_u}(\bm{u}^*) \geq L_{\lambda_u}(\bm{u}^*) + \nabla L_{\lambda_u}(\bm{u}^*)^T (\hat{\bm{u}}_{\lambda_u} - \bm{u}^*) + \frac{1}{2}(\hat{\bm{u}}_{\lambda_u} - \bm{u}^*)^T\nabla^2 L_{\lambda_u}(\bm{u}')(\hat{\bm{u}}_{\lambda_u} - \bm{u}^*)\\
\geq L_{\lambda_u}(\bm{u}^*) + \nabla L_{\lambda_u}(\bm{u}^*)^T (\hat{\bm{u}}_{\lambda_u} - \bm{u}^*) + \frac{\lambda_u}{2}\|\hat{\bm{u}}_{\lambda_u} - \bm{u}^*\|_2^2.
\end{gathered}
\end{equation*}
Therefore,
\begin{equation*}
\begin{gathered}
\frac{\lambda_u}{2}\|\hat{\bm{u}}_{\lambda_u} - \bm{u}^*\|_2^2 \leq - \nabla L_{\lambda_u}(\bm{u}^*)^T (\hat{\bm{u}}_{\lambda_u} - \bm{u}^*)\\
\leq \|\nabla L_{\lambda_u}(\bm{u}^*)\|_2\|\hat{\bm{u}}_{\lambda_u} - \bm{u}^*\|_2
\end{gathered}
\end{equation*}
Finally, suppose $E_2$ occurs. Then
\begin{equation*}
\|\hat{\bm{u}}_{\lambda_u} - \bm{u}^*\|_2 \leq \frac{2\|\nabla L_{\lambda_u}(\bm{u}^*)\|_2}{\lambda_u} \leq \frac{2c_0\sqrt{\frac{M^2 p \log M}{n_0 + n_{\mathcal{S}}}}}{c_{\lambda_{u}}B_u^{-1}\sqrt{\frac{p\log M}{n_0 + n_{\mathcal{S}}}}} \leq C_7B_u M,
\end{equation*}
for $C_7 \geq \frac{2c_0}{c_{\lambda_{u}}}$.
\end{proof}
\end{lemma}

\begin{lemma}
\label{lemma:lemma8_step1}
Suppose $E_0$ and $E_2$ occur, $Mp\ge C_\kappa\kappa_1^4\log M$, and $n_0+n_{\mathcal S}\le c_2M^{c_3}$. If $T=\lceil M^{C_T}\rceil$ with $C_T>1+c_3/2$, then there exists a constant $C_0>0$, independent of $M$, such that, for all sufficiently large $M$,
\begin{equation*}
\|\bm u^{[T]}-\hat{\bm u}_{\lambda_u}\|_\infty
\le
C_0\kappa_1^2\sqrt{\frac{\log M}{Mp(n_0+n_{\mathcal S})}}.
\end{equation*}
\begin{proof}
Lemma \ref{lemma:lemma6_step1} and Lemma \ref{lemma:lemma7_step1} give
\[
\|\bm u^{[T]}-\hat{\bm u}_{\lambda_u}\|_2
\le
\left(1-\frac{\lambda_u}{\lambda_u+Mp}\right)^T C_7B_uM
\le
C_7B_uM\exp\left\{-\frac{T\lambda_u}{\lambda_u+Mp}\right\}.
\]
Here
\[
\lambda_u
=c_{\lambda_u}B_u^{-1}\sqrt{\frac{p\log M}{n_0+n_{\mathcal S}}},
\qquad Mp\gtrsim\log M.
\]
The standing condition also implies $\kappa_1^4\lesssim Mp/\log M\le M/\log M$, and hence $B_u\lesssim\log M$. For large $M$, $\lambda_u\le Mp$, so
\[
\frac{\lambda_u}{\lambda_u+Mp}
\ge
\frac{\lambda_u}{2Mp}
\gtrsim
\frac{1}{M^{1+c_3/2}\sqrt{\log M}},
\]
where we used $n_0+n_{\mathcal S}\le c_2M^{c_3}$ and $p\le1$. Thus, if $T=\lceil M^{C_T}\rceil$ with any fixed $C_T>1+c_3/2$, the exponent $T\lambda_u/(\lambda_u+Mp)$ diverges polynomially. Moreover, since $\kappa_1\ge1$, $p\le1$ and $n_0+n_{\mathcal S}\le c_2M^{c_3}$, the target rate is bounded below by a fixed inverse-polynomial order in $M$. Hence the exponentially small optimisation error is eventually bounded by
$C_0\kappa_1^2\sqrt{\log M/[Mp(n_0+n_{\mathcal S})]}$ for a constant $C_0$ independent of $M$. Finally $\|v\|_\infty\le\|v\|_2$ gives the claimed result.
\end{proof}
\end{lemma}

\subsection{Leave-one-out technique}
For notational convenience, for $j\ne k$ define the pooled population comparison probability
\[
y_{k,j}^*
:=
\frac{\exp(u_j^*)}{\exp(u_j^*)+\exp(u_k^*)}.
\]
By Assumption \ref{ass:pooledtarget}, this equals the average probability that object $j$ is preferred to object $k$ among observations entering the pooled transfer sample. We write $y_{k,j}:=N_{\mathrm{tr}}^{-1}\sum_{i=1}^{N_{\mathrm{tr}}}y_{k,j}^{(i)}$ for the corresponding pooled empirical proportion. The leave-one-out construction below is therefore the same population replacement used in Chen et al.\ (\citeyear{chen2019spectral}), except that $y_{k,j}^*$ represents the pooled comparison probability.

\noindent We next study the statistical error of $\hat{\bm{u}}_{\lambda_u}$. We use the leave-one-out technique of Chen et al.\ (\citeyear{chen2019spectral}), combined with a proof by induction to show that $\bm{u}^{[T]}$ does not deviate too strongly from $\bm{u}^*$, even after many iterations of Algorithm \ref{alg:gdrmle_step1}. For $1 \leq k \leq M$, define the following leave-one-out objective.
\begin{equation*}
\begin{gathered}
L^{(k)}(\bm{u}) = \sum_{(j,l)\in \mathcal E, j > l, j \neq k, l \neq k}\left\{-y_{l,j}\left(u_j - u_l\right) + \log\left[1 + \exp(u_j - u_l)\right]\right\}\\
+ p \sum_{j \neq k}\left\{- \frac{\exp(u_j^*)}{\exp(u_j^*) + \exp(u_k^*)}(u_j - u_k) + \log\left[1 + \exp(u_j - u_k)\right]\right\}\\
L^{(k)}_{\lambda_u}(\bm{u}) = L^{(k)}(\bm{u}) + \frac{\lambda_u}{2}\|\bm{u}\|_2^2.
\end{gathered}
\end{equation*}

\noindent Algorithm \ref{alg:loos_step1} then describes how the leave-one-out sequences $\{\bm{u}^{[t], (k)}\}_{t = 0,1,\ldots}$ are constructed.

\begin{algorithm}[H]
\caption{Leave-one-out sequence constructor.}\label{alg:loos_step1}
  \begin{algorithmic}[1]
  	\STATE Initialise $\bm{u}^{[0], (k)} = \bm{u}^*$
      \FOR{$t = 0$ to $T-1$}
		\STATE $\bm{u}^{[t+1], (k)} = \bm{u}^{[t], (k)} - \eta \nabla L_{\lambda_u}^{(k)}(\bm{u}^{[t], (k)})$
              \ENDFOR
  \end{algorithmic}
\end{algorithm}

\noindent The unregularised part of $L^{(k)}_{\lambda_u}$ depends only on log-worth differences, so its gradient is orthogonal to $\bm1_M$, while the ridge gradient is $\lambda_u\bm u$. Since $\bm u^*$ is centred, every leave-one-out iterate remains in $\bm1_M^\perp$, just as for the full gradient-descent sequence.

\noindent Here $\bm{u}^{[t],(k)}$ denotes the $t$-th iterate of the $k$-th leave-one-out sequence, while $u_k^{[t],(k)}$ denotes its $k$-th coordinate. The previous section showed that the gradient-descent iterates converge to the regularised MLE. We now show that $\bm{u}^{[T]}$ remains close to $\bm{u}^*$.

Fix $A>C_T+12$, and let $\mathcal H_t$ be the event that all four induction bounds in Equations \eqref{eq:bound1_step1}--\eqref{eq:bound4_step1} hold at iteration $t$. The only new concentration argument at each iteration is needed when comparing the full and leave-one-out sequences in Lemma \ref{lemma:lemma12_step1}. Conditional on the relevant leave-one-out sequence, the coefficients in the incident-edge fluctuation term are fixed, while the incident edge indicators remain independent Bernoulli variables. Bernstein's inequality can therefore be applied conditionally. Hoeffding's inequality is used similarly for the outcome terms.

The constants in these concentration bounds can be chosen so that, after a union bound over $k=1,\ldots,M$, the probability that $\mathcal H_t$ holds but the third induction bound fails at iteration $t+1$ is at most $O(M^{-A})$ on the standing graph events. A further union bound over $T=\lceil M^{C_T}\rceil=O(M^{C_T})$ iterations gives total failure probability

$$
O(M^{-A+C_T})=O(M^{-12}).
$$

\noindent The remaining induction steps are deterministic once the standing graph, score and previous induction events hold.
\\
\\
\noindent Throughout Lemmas \ref{lemma:lemma9_step1}--\ref{lemma:lemma13_step1}, we retain the standing assumptions of Theorem \ref{thm:convergence_step1_alpha}, including $Mp\ge C_\kappa\kappa_1^4\log M$, and use the step size $0<\eta\le(\lambda_u+Mp)^{-1}$. The following bounds are proved by induction in Lemmas \ref{lemma:lemma10_step1}, \ref{lemma:lemma11_step1}, \ref{lemma:lemma12_step1} and \ref{lemma:lemma13_step1}, respectively:
\begin{gather}
\|\bm{u}^{[t]} - \bm{u}^*\|_2 \leq C_1\kappa_1 \sqrt{\frac{\log M}{p (n_0 + n_\mathcal{S})}} \label{eq:bound1_step1},\\
\max_{1 \leq k \leq M}|u^{[t], (k)}_k - u_k^*|  \leq C_2\kappa_1^2\sqrt{\frac{\log M}{M p (n_0 + n_\mathcal{S})}} \label{eq:bound2_step1},\\
\max_{1 \leq k \leq M}\|\bm{u}^{[t]} - \bm{u}^{[t], (k)}\|_2  \leq C_3\kappa_1\sqrt{\frac{\log M}{M p (n_0 + n_\mathcal{S})}} \label{eq:bound3_step1},\\
\|\bm{u}^{[t]} - \bm{u}^*\|_\infty \leq C_4\kappa_1^2 \sqrt{\frac{\log M}{M p (n_0 + n_\mathcal{S})}} \label{eq:bound4_step1}.
\end{gather}
At $t=0$, all four bounds hold because $\bm u^{[0]}=\bm u^{[0],(k)}=\bm u^*$ for every $k$. The constants can be fixed without circularity: choose $C_1$ and $C_3$ first, then $C_6$, $C_2$, $C_4$ and $C_5$ in that order, and finally choose $C_\kappa$ sufficiently large relative to these fixed constants. None of these constants depends on $M$. The following lemmas show that the four bounds are preserved from iteration $t$ to iteration $t+1$.

\begin{lemma}
\label{lemma:lemma9_step1}
Suppose that the bounds provided in Equations \eqref{eq:bound1_step1}--\eqref{eq:bound4_step1} hold for iteration $t$. Then there exist constants $C_5,C_6>0$ such that 
\begin{gather}
\max_{1 \leq k \leq M}\|\bm{u}^{[t], (k)} - \bm{u}^*\|_\infty  \leq C_5\kappa_1^2\sqrt{\frac{\log M}{M p (n_0 + n_\mathcal{S})}}, \label{eq:lemma9_step1_1}\\
\max_{1 \leq k \leq M}\|\bm{u}^{[t], (k)} - \bm{u}^*\|_2  \leq C_6\kappa_1\sqrt{\frac{\log M}{p (n_0 + n_\mathcal{S})}} \label{eq:lemma9_step1_2}.
\end{gather}
\begin{proof}
For the first bound,
\begin{equation*}
\begin{gathered}
\max_{1 \leq k \leq M}\|\bm{u}^{[t], (k)} - \bm{u}^*\|_\infty \leq \max_{1 \leq k \leq M}\|\bm{u}^{[t]} - \bm{u}^{[t], (k)}\|_2 + \|\bm{u}^{[t]} - \bm{u}^*\|_\infty \text{ (Triangle inequality)}\\
\leq C_3\kappa_1\sqrt{\frac{\log M}{M p (n_0 + n_\mathcal{S})}} + C_4\kappa_1^2 \sqrt{\frac{\log M}{M p (n_0 + n_\mathcal{S})}} \text{ (Equations \eqref{eq:bound3_step1} and \eqref{eq:bound4_step1})}\\
\leq C_5\kappa_1^2 \sqrt{\frac{\log M}{M p (n_0 + n_\mathcal{S})}},
\end{gathered}
\end{equation*}
where the last inequality holds when $C_5 \geq C_3 + C_4$. The second bound follows from
\begin{equation*}
\begin{gathered}
\max_{1 \leq k \leq M}\|\bm{u}^{[t], (k)} - \bm{u}^*\|_2 \leq \max_{1 \leq k \leq M}\|\bm{u}^{[t]} - \bm{u}^{[t], (k)}\|_2 + \|\bm{u}^{[t]} - \bm{u}^*\|_2 \text{ (Triangle inequality)}\\
\leq C_3\kappa_1\sqrt{\frac{\log M}{M p (n_0 + n_\mathcal{S})}} + C_1\kappa_1 \sqrt{\frac{\log M}{p (n_0 + n_\mathcal{S})}} \text{ (Equations \eqref{eq:bound3_step1} and \eqref{eq:bound1_step1})}\\
\leq C_6\kappa_1 \sqrt{\frac{\log M}{ p (n_0 + n_\mathcal{S})}},
\end{gathered}
\end{equation*}
where the last inequality holds when $C_6 \geq C_1 + C_3$.
\end{proof}
\end{lemma}

\begin{lemma}
\label{lemma:lemma10_step1}
Suppose that the bounds provided in Equations \eqref{eq:bound1_step1}--\eqref{eq:bound4_step1} hold for iteration $t$ and that $E_1$ and $E_2$ occur. Then
\begin{equation*}
\|\bm{u}^{[t+1]} - \bm{u}^*\|_2 \leq C_1\kappa_1 \sqrt{\frac{\log M}{p (n_0 + n_\mathcal{S})}}.
\end{equation*}
This bound holds provided that $0 < \eta \leq \frac{1}{\lambda_{u} + M p}$ and $C_1$ is sufficiently large.
\begin{proof}
By definition,
\begin{equation*}
\begin{gathered}
\bm{u}^{[t+1]} - \bm{u}^* = \bm{u}^{[t]} - \eta \nabla L_{\lambda_u}(\bm{u}^{[t]}) - \bm{u}^*\\
= (\bm{u}^{[t]} - \bm{u}^*) - [\eta \nabla L_{\lambda_u}(\bm{u}^{[t]}) - \eta \nabla L_{\lambda_u}(\bm{u}^*)] - \eta \nabla L_{\lambda_u}(\bm{u}^*)
\end{gathered}
\end{equation*}
Consider $\bm{u}(\tau) = \bm{u}^* + \tau\left(\bm{u}^{[t]} - \bm{u}^*\right)$. By the fundamental theorem of calculus, $\nabla L_{\lambda_u}(\bm{u}^{[t]}) - \nabla L_{\lambda_u}(\bm{u}^*) = \int_{0}^1 \nabla^2 L_{\lambda_u}(\bm{u}(\tau))(\bm{u}^{[t]} - \bm{u}^*)d\tau$. Therefore,
\begin{equation*}
\begin{gathered}
(\bm{u}^{[t]} - \bm{u}^*) - \eta[\nabla L_{\lambda_u}(\bm{u}^{[t]}) - \nabla L_{\lambda_u}(\bm{u}^*)] - \eta \nabla L_{\lambda_u}(\bm{u}^*)\\
= \left(\bm I_M - \eta\int_{0}^1 \nabla^2 L_{\lambda_u}(\bm{u}(\tau))d\tau\right)\left(\bm{u}^{[t]} - \bm{u}^*\right) - \eta \nabla L_{\lambda_u}(\bm{u}^*).
\end{gathered}
\end{equation*}
Choose the fixed value $\epsilon_0=1/5<\log(5/4)$. By Equation \eqref{eq:bound4_step1},
\[
\sup_{\tau\in[0,1]}\|\bm u(\tau)-\bm u^*\|_\infty
\le C_4\kappa_1^2\sqrt{\frac{\log M}{Mp(n_0+n_{\mathcal S})}}
\le \frac{C_4}{\sqrt{C_\kappa}}.
\]
Choosing $C_\kappa$ sufficiently large makes this at most $\epsilon_0/2$. Lemma \ref{lemma:lemma5_step1}, with $r=\epsilon_0/2$, and Lemma \ref{lemma:lemma4_step1} therefore give
\begin{equation*}
\frac{Mp}{10\kappa_1}+\lambda_u
\le \frac{Mp}{8\kappa_1e^{\epsilon_0}}+\lambda_u
\le \lambda_{\min,\perp}(\nabla^2L_{\lambda_u}(\bm u(\tau)))
\le \lambda_{\max}(\nabla^2L_{\lambda_u}(\bm u(\tau)))
\le \lambda_u+Mp.
\end{equation*}
Let $\bm{A} = \int_{0}^1 \nabla^2 L_{\lambda_u}(\bm{u}(\tau))d\tau$. Then
\begin{equation*}
 \lambda_u + \frac{Mp}{10 \kappa_1} \leq \lambda_{\min, \perp}\left(\bm{A}\right) \leq \lambda_{\max}\left(\bm{A}\right) \leq \lambda_u + Mp.
\end{equation*}
This gives the upper bound
\begin{equation*}
\begin{gathered}
\|\left(\bm I_M - \eta\bm A\right)\left(\bm u^{[t]} - \bm u^*\right)\|_2
\leq
\max_{\lambda\in[\lambda_{\min,\perp}(\bm A),\lambda_{\max}(\bm A)]}|1-\eta\lambda|\,
\|\bm u^{[t]}-\bm u^*\|_2\\
\leq \left(1 - \frac{\eta Mp}{10\kappa_1}\right)\|\bm{u}^{[t]} - \bm{u}^*\|_2.
\end{gathered}
\end{equation*}
Suppose $E_2$ occurs. Then, by the triangle inequality,
\begin{equation*}
\begin{gathered}
\|\bm{u}^{[t+1]} - \bm{u}^*\|_2 \leq \|\left(\bm I_M - \eta\bm{A}\right)\left(\bm{u}^{[t]} - \bm{u}^*\right)\|_2 + \eta\| \nabla L_{\lambda_u}(\bm{u}^*)\|_2\\
\leq \left(1 - \frac{\eta Mp}{10\kappa_1}\right)C_1\kappa_1\sqrt{\frac{\log M}{p (n_0 + n_\mathcal{S})}} + \eta c_0 \sqrt{\frac{M^2 p \log M}{(n_0 + n_\mathcal{S})}} \text{ (Lemma \ref{lemma:lemma3_step1} and Equation \eqref{eq:bound1_step1})}\\
\leq C_1\kappa_1 \sqrt{\frac{\log M}{p (n_0 + n_\mathcal{S})}},
\end{gathered}
\end{equation*}
provided, for example, that $C_1\ge 10c_0$ (up to the fixed constants hidden in the preceding bounds).
\end{proof}
\end{lemma}

\begin{lemma}
\label{lemma:lemma11_step1}
Suppose that the bounds provided in Equations \eqref{eq:bound1_step1}--\eqref{eq:bound4_step1} hold for iteration $t$. Then
\begin{equation*}
\max_{1 \leq k \leq M}|u_k^{[t+1], (k)} -  u_k^*| \leq C_2\kappa_1^2\sqrt{\frac{\log M}{M p (n_0 + n_\mathcal{S})}} 
\end{equation*}
This bound holds provided that $0 < \eta \leq \frac{1}{\lambda_{u} +  M p}$ and $C_2 \gtrsim C_6 + c_{\lambda_u}$.
\begin{proof}
Algorithm \ref{alg:loos_step1} gives the update
\begin{equation*}
\bm{u}^{[t+1], (k)} = \bm{u}^{[t], (k)} - \eta \nabla L_{\lambda_u}^{(k)}(\bm{u}^{[t], (k)}).
\end{equation*}
Hence,
\begin{gather}
u_k^{[t+1], (k)} - u_k^* = u_k^{[t], (k)} - \eta \left(\nabla L_{\lambda_u}^{(k)}(\bm{u}^{[t], (k)})\right)_k - u_k^* \notag\\
= u_k^{[t], (k)} - u_k^* - \eta \left\{p \sum_{j:j \neq k}\left[\frac{\exp(u_j^*)}{\exp(u_j^*) + \exp(u_k^*)} - \frac{\exp(u_j^{[t], (k)})}{\exp(u_j^{[t], (k)}) + \exp(u_k^{[t], (k)})}\right] \right\} - \eta\lambda_u u_k^{[t], (k)}\label{eq:lemma11_step1}
\end{gather}
By the mean value theorem, the inner term in the last equation can be rewritten as
\begin{equation*}
\begin{gathered}
\frac{\exp(u_j^*)}{\exp(u_j^*) + \exp(u_k^*)} - \frac{\exp(u_j^{[t], (k)})}{\exp(u_j^{[t], (k)}) + \exp(u_k^{[t], (k)})}\\
= \frac{1}{1 + \exp(u_k^* - u_j^*)} - \frac{1}{1 + \exp(u_k^{[t],(k)} - u_j^{[t],(k)})}\\
= -\frac{\exp(c_j)}{[1 + \exp(c_j)]^2}\left[u^*_k - u_j^* - \left(u_k^{[t],(k)} - u_j^{[t],(k)}\right)\right],
\end{gathered}
\end{equation*}
with $c_j$ between $u^*_k - u_j^*$ and $u_k^{[t],(k)} - u_j^{[t],(k)}$. Substituting this result into Equation \eqref{eq:lemma11_step1} gives
\begin{equation*}
\begin{gathered}
u_k^{[t+1], (k)} - u_k^* = \left(1 - \eta\lambda_u - \eta p \sum_{j:j\neq k}\frac{\exp(c_j)}{[1+\exp(c_j)]^2}\right)\left(u_k^{[t], (k)} - u_k^*\right)\\
 + \eta p \sum_{j:j\neq k}\frac{\exp(c_j)}{[1+\exp(c_j)]^2}\left(u_j^{[t], (k)} - u_j^*\right) - \eta\lambda_u u_k^*
\end{gathered}
\end{equation*}
Taking absolute values gives 
\begin{equation*}
\begin{gathered}
\left|u_k^{[t+1], (k)} - u_k^*\right| \leq \left|1 - \eta\lambda_u - \eta p \sum_{j:j\neq k}\frac{\exp(c_j)}{[1+\exp(c_j)]^2}\right|\left|u_k^{[t], (k)} - u_k^*\right|\\
 + \eta p \sum_{j:j\neq k}\frac{\exp(c_j)}{[1+\exp(c_j)]^2}\left|u_j^{[t], (k)} - u_j^*\right| + \eta\lambda_u |u_k^*|\\
 \leq \left|1 - \eta\lambda_u - \eta p \sum_{j:j\neq k}\frac{\exp(c_j)}{[1+\exp(c_j)]^2}\right|\left|u_k^{[t], (k)} - u_k^*\right|\\
 + \frac{\eta p}{4} \sum_{j:j\neq k}\left|u_j^{[t], (k)} - u_j^*\right| + \eta\lambda_u \|\bm{u}^*\|_\infty \left(\text{As } \frac{\exp(c_j)}{[1+\exp(c_j)]^2} \leq \frac{1}{4}\right)\\
  \leq \left|1 - \eta\lambda_u - \eta p \sum_{j:j\neq k}\frac{\exp(c_j)}{[1+\exp(c_j)]^2}\right|\left|u_k^{[t], (k)} - u_k^*\right|\\
 + \frac{\eta p}{4}\sqrt{M}\|\bm{u}^{[t], (k)} - \bm{u}^*\|_2 + \eta\lambda_u \|\bm{u}^*\|_\infty (\text{Cauchy-Schwarz}).
\end{gathered}
\end{equation*}
Also, since $\frac{\exp(c_j)}{[1+\exp(c_j)]^2} \leq \frac{1}{4}$,
\begin{equation*}
1 - \eta\lambda_u - \eta p\sum_{j:j\neq k}\frac{\exp(c_j)}{[1+\exp(c_j)]^2} \geq 1 - \eta\lambda_u - \frac{\eta Mp}{4} \geq 1 - \eta(M p + \lambda_u) \geq 0
\end{equation*}
and
\begin{equation*}
\begin{gathered}
\left|1 - \eta\lambda_u - \eta p\sum_{j:j\neq k}\frac{\exp(c_j)}{[1+\exp(c_j)]^2}\right| = 1 - \eta\lambda_u - \eta p\sum_{j:j\neq k}\frac{\exp(c_j)}{[1+\exp(c_j)]^2}\\
\leq 1 - \eta p (M - 1)\min_{j:j \neq k}\frac{\exp(c_j)}{[1+\exp(c_j)]^2}
\end{gathered}
\end{equation*}
To sharpen the upper bound on $\left|u_k^{[t+1], (k)} - u_k^*\right|$, we lower-bound the coefficient above. From Equation \eqref{eq:lemma9_step1_1},
\begin{equation*}
\begin{gathered}
\max_{j:j\neq k}\left|c_j\right| \leq \max_{j:j\neq k}\left|u_k^* - u_j^*\right| + \max_{j:j\neq k}\left|u_k^* - u_j^* - \left(u_k^{[t],(k)} - u_j^{[t],(k)}\right)\right|\\
\leq \log \kappa_1 + 2\|\bm{u}^{[t],(k)} - \bm{u}^*\|_\infty\\
\leq \log \kappa_1 + \epsilon,
\end{gathered}
\end{equation*}
For the same fixed $\epsilon_0=1/5$, the standing condition $Mp\ge C_\kappa\kappa_1^4\log M$ gives
\[
2C_5\kappa_1^2\sqrt{\frac{\log M}{Mp(n_0+n_{\mathcal S})}}
\le
\frac{2C_5}{\sqrt{C_\kappa(n_0+n_{\mathcal S})}}
\le
\frac{2C_5}{\sqrt{C_\kappa}}.
\]
Choosing $C_\kappa$ sufficiently large relative to $C_5$ and $\epsilon_0$ therefore allows us to take $\epsilon=\epsilon_0$, and
\begin{equation*}
\frac{\exp(c_j)}{[1+\exp(c_j)]^2} = \frac{\exp(-|c_j|)}{[1+\exp(-|c_j|)]^2} \geq \frac{\exp(-|c_j|)}{4} \geq \frac{1}{4\exp(\epsilon_0)\kappa_1} \geq \frac{1}{5\kappa_1}.
\end{equation*}
Since $M\ge2$, $(M-1)/5\ge M/10$. Combining this with the preceding coefficient bound yields the contraction factor $1-\eta Mp/(10\kappa_1)$. Combining these bounds gives
\begin{equation*}
\begin{gathered}
\left|u_k^{[t+1], (k)} - u_k^*\right| \leq \left(1 - \frac{\eta M p}{10\kappa_1}\right)\left|u_k^{[t], (k)} - u_k^*\right| + \frac{\eta p \sqrt{M}}{4}\|\bm{u}^{[t],(k)} - \bm{u}^*\|_2 + \eta\lambda_u \|\bm{u}^*\|_\infty\\
\leq \left(1 - \frac{\eta M p}{10\kappa_1}\right)C_2\kappa_1^2\sqrt{\frac{\log M}{M p (n_0 + n_\mathcal{S})}} + \frac{\eta p \sqrt{M}}{4}C_6\kappa_1\sqrt{\frac{\log M}{p (n_0 + n_\mathcal{S})}} + c_{\lambda_u}\eta\sqrt{\frac{M p \log M}{n_0 + n_\mathcal{S}}}\\
\leq C_2\kappa_1^2\sqrt{\frac{\log M}{M p (n_0 + n_\mathcal{S})}},
\end{gathered}
\end{equation*}
Indeed, the contraction reduces the preceding target bound by
\[
\frac{C_2}{10}\eta\kappa_1\sqrt{\frac{Mp\log M}{n_0+n_{\mathcal S}}},
\]
whereas the two added terms are bounded, up to fixed numerical constants, by
\[
C_6\eta\kappa_1\sqrt{\frac{Mp\log M}{n_0+n_{\mathcal S}}}
\quad\text{and}\quad
c_{\lambda_u}\eta\kappa_1\sqrt{\frac{Mp\log M}{n_0+n_{\mathcal S}}}.
\]
Thus the last inequality holds when $C_2$ is chosen sufficiently large relative to $C_6$ and $c_{\lambda_u}$.
\end{proof}
\end{lemma}

\begin{lemma}
\label{lemma:lemma12_step1}
Let $\mathcal H_t$ be the induction event defined above, and let $\mathcal B_{t+1}^{(3)}$ denote the event
\begin{equation*}
\max_{1 \leq k \leq M}\|\bm{u}^{[t+1]} -  \bm{u}^{[t+1], (k)}\|_2 \leq C_3\kappa_1\sqrt{\frac{\log M}{M p (n_0 + n_\mathcal{S})}}.
\end{equation*}
For each fixed iteration $t$,
\[
\mathbb P\!\left(E_0\cap E_1\cap\mathcal H_t\cap(\mathcal B_{t+1}^{(3)})^c\right)=O(M^{-A}),
\]
provided the standing assumptions of Theorem \ref{thm:convergence_step1_alpha} hold, $0<\eta\le(\lambda_u+Mp)^{-1}$, and $C_3$ is sufficiently large.
\begin{proof}
By definition,
\begin{equation*}
\bm{u}^{[t+1]} - \bm{u}^{[t+1], (k)} = \bm{u}^{[t]} - \bm{u}^{[t], (k)} - \eta \nabla L_{\lambda_u}\left(\bm{u}^{[t]}\right) + \eta \nabla L_{\lambda_u}^{(k)}\left(\bm{u}^{[t], (k)}\right)\end{equation*}
Consider $\bm{u}(\tau) = \bm{u}^{[t], (k)} + \tau\left(\bm{u}^{[t]} - \bm{u}^{[t], (k)}\right)$. By the fundamental theorem of calculus, $\nabla L_{\lambda_u}(\bm{u}^{[t]}) - \nabla L_{\lambda_u}(\bm{u}^{[t], (k)}) = \int_{0}^1 \nabla^2 L_{\lambda_u}(\bm{u}(\tau))\left(\bm{u}^{[t]} - \bm{u}^{[t], (k)}\right)d\tau$. Therefore,
\begin{equation*}
\begin{gathered}
\bm{u}^{[t]} - \bm{u}^{[t], (k)} - \eta \nabla L_{\lambda_u}\left(\bm{u}^{[t]}\right) + \eta \nabla L_{\lambda_u}^{(k)}\left(\bm{u}^{[t], (k)}\right)\\
= \bm{u}^{[t]} - \bm{u}^{[t], (k)} - \eta \nabla L_{\lambda_u}\left(\bm{u}^{[t]}\right) + \eta \nabla L_{\lambda_u}^{(k)}\left(\bm{u}^{[t], (k)}\right) + \eta \nabla L_{\lambda_u}\left(\bm{u}^{[t], (k)}\right) - \eta \nabla L_{\lambda_u}\left(\bm{u}^{[t], (k)}\right)\\
= \left(\bm I_M - \eta\int_{0}^1 \nabla^2 L_{\lambda_u}(\bm{u}(\tau))d\tau\right)(\bm{u}^{[t]} - \bm{u}^{[t], (k)}) - \eta\left(\nabla L_{\lambda_u}\left(\bm{u}^{[t], (k)}\right) - \nabla L_{\lambda_u}^{(k)}\left(\bm{u}^{[t], (k)}\right)\right),
\end{gathered}
\end{equation*}
Define $\bm{v}_1 = \left(\bm I_M - \eta\int_{0}^1 \nabla^2 L_{\lambda_u}(\bm{u}(\tau))d\tau\right)(\bm{u}^{[t]} - \bm{u}^{[t], (k)})$ and $\bm{v}_2 = \eta\left(\nabla L_{\lambda_u}\left(\bm{u}^{[t], (k)}\right) - \nabla L_{\lambda_u}^{(k)}\left(\bm{u}^{[t], (k)}\right)\right)$. It remains to bound $\|\bm{v}_1\|_2$ and $\|\bm{v}_2\|_2$. On the induction event, Equation \eqref{eq:bound4_step1} and Lemma \ref{lemma:lemma9_step1} imply uniformly in $k$ and $\tau\in[0,1]$ that
\[
\|\bm u(\tau)-\bm u^*\|_\infty
\le \max\{C_4,C_5\}\kappa_1^2\sqrt{\frac{\log M}{Mp(n_0+n_{\mathcal S})}}.
\]
The standing condition allows $C_\kappa$ to be chosen so that this radius is at most $\epsilon_0/2$. Since both full and leave-one-out iterates are centred, Lemmas \ref{lemma:lemma4_step1}--\ref{lemma:lemma5_step1} give the same contraction as in Lemma \ref{lemma:lemma10_step1}: 
\begin{equation*}
\|\bm{v}_1\|_2 \leq \left(1 - \frac{\eta Mp}{10\kappa_1}\right)\|\bm{u}^{[t]} - \bm{u}^{[t], (k)}\|_2,
\end{equation*}
This holds whenever $\eta \leq \frac{1}{\lambda_u + Mp}$.
We next bound $\|\bm{v}_2\|_2$.
\begin{equation*}
\begin{gathered}
\frac{1}{\eta}\bm{v}_2\\
= \sum_{j:j \neq k}\left\{\left(-y_{k,j} + \frac{\exp(u_j^{[t], (k)})}{\exp(u_j^{[t], (k)}) + \exp(u_k^{[t], (k)})}\right)\mathbbm{1}\{(j,k)\in \mathcal E\}
-p\left(-y_{k,j}^* + \frac{\exp(u_j^{[t], (k)})}{\exp(u_j^{[t], (k)}) + \exp(u_k^{[t], (k)})}\right)\right\}\\
\cdot(\bm e_j - \bm e_k)\\
= \sum_{j:j \neq k}\left\{\left(-\frac{\exp(u_j^*)}{\exp(u_j^*) + \exp(u_k^*)} + \frac{\exp(u_j^{[t], (k)})}{\exp(u_j^{[t], (k)}) + \exp(u_k^{[t], (k)})}\right)\{\mathbbm{1}[(j,k)\in \mathcal E] - p\}\right\}(\bm e_j - \bm e_k)\\
+ \frac{1}{n_0 + n_{\mathcal{S}}}\sum_{(j,k)\in \mathcal E}\sum_{i = 1}^{n_0 + n_{\mathcal{S}}}\left(-y_{k,j}^{(i)} + \frac{\exp(u_j^*)}{\exp(u_j^*) + \exp(u_k^*)}\right)(\bm e_j - \bm e_k),\\
\end{gathered}
\end{equation*}
where the last two terms are denoted by $\bm v^k$ and $\bm z^k$, respectively. By definition,
\begin{equation*}
z^k_j = \begin{cases}
 \frac{1}{n_0 + n_{\mathcal{S}}}\sum_{i = 1}^{n_0 + n_{\mathcal{S}}}\left(-y_{k,j}^{(i)} + \frac{\exp(u_j^*)}{\exp(u_j^*) + \exp(u_k^*)}\right) & \text{if } (j,k) \in \mathcal E;\\
  \frac{1}{n_0 + n_{\mathcal{S}}}\sum_{r:(r,k)\in \mathcal E}\sum_{i = 1}^{n_0 + n_{\mathcal{S}}}\left(y_{k,r}^{(i)} - \frac{\exp(u_r^*)}{\exp(u_r^*) + \exp(u_k^*)}\right) & \text{if } j = k;\\
  0, & \text{otherwise}.
\end{cases}
\end{equation*}
For a fixed incident edge $(j,k)$, the individual pooled observations need not have the same mean. Nevertheless, Assumption \ref{ass:pooledtarget} gives
\[
\sum_{i=1}^{N_{\mathrm{tr}}}
\mathbb E\!\left[-y_{k,j}^{(i)}+\frac{\exp(u_j^*)}{\exp(u_j^*)+\exp(u_k^*)}\right]=0.
\]
Thus each displayed edge sum is centred as a whole. Since the summands are independent and bounded by one, and $|\{j:(j,k)\in\mathcal E\}|\asymp Mp$ on $E_0$, Hoeffding's inequality and a union bound give, for all $1\le k\le M$,
\begin{equation*}
|z_k^k| \lesssim \sqrt{\frac{M p \log M}{n_0 + n_{\mathcal{S}}}} \text{ and } |z_j^k| \lesssim \sqrt{\frac{\log M}{n_0 + n_{\mathcal{S}}}} \quad \forall j \text{ s.t. } (j,k) \in \mathcal E,
\end{equation*}
resulting in
\begin{equation*}
\|\bm{z}^k\|_2 \leq |z_k^k| + \sqrt{\sum_{j:(j,k) \in \mathcal E}(z_j^k)^2} \lesssim \sqrt{\frac{Mp \log M}{n_0 + n_{\mathcal{S}}}}, \quad \forall 1 \leq k \leq M.
\end{equation*}
We now turn to $\bm v^k$, for which
\begin{equation*}
v_j^k = \begin{cases}
 \xi_j(1-p) & \text{if } (j,k) \in \mathcal E;\\
-\sum_{r:r\neq k}\xi_r(\mathbbm{1}[(r,k)\in \mathcal E] - p)& \text{if } j = k;\\
  -\xi_j p, & \text{otherwise},
\end{cases}
\end{equation*}
where 
\begin{equation*}
\begin{gathered}
\xi_j = -\frac{\exp(u_j^*)}{\exp(u_j^*) + \exp(u_k^*)} + \frac{\exp(u_j^{[t], (k)})}{\exp(u_j^{[t], (k)}) + \exp(u_k^{[t], (k)})}\\
= -\frac{1}{1+\exp(u_k^* - u_j^*)} + \frac{1}{1+\exp(u_k^{[t], (k)} - u_j^{[t], (k)})}
\end{gathered}
\end{equation*}
To bound $\xi_j$, define $f(x) = (1+\exp(x))^{-1}$. Hence,
\begin{equation*}
\begin{gathered}
|\xi_j| = \left|f\left(u_k^{[t], (k)} - u_j^{[t], (k)}\right) - f\left(u_k^* - u_j^*\right)\right|\\
\leq \left|\left(u_k^{[t], (k)} - u_j^{[t], (k)}\right) - \left(u_k^* - u_j^*\right)\right|\\
\leq \left|u_j^* - u_j^{[t], (k)}\right| + \left|u_k^* - u_k^{[t], (k)}\right|,
\end{gathered}
\end{equation*}
and hence
\begin{equation*}
|\xi_j| \leq 2\|\bm{u}^{[t], (k)} - \bm{u}^*\|_\infty \text{ for }j\ne k,
\qquad
\sum_{j\ne k} \xi_j^2 \leq 4M\|\bm{u}^{[t], (k)} - \bm{u}^*\|_\infty^2.
\end{equation*}
By construction, $\bm u^{[t],(k)}$ is independent of the incident edge indicators $\{\mathbbm1[(j,k)\in\mathcal E]:j\ne k\}$. Conditional on the leave-one-out sequence, the $\xi_j$ are therefore fixed and the centred Bernoulli variables $\mathbbm1[(j,k)\in\mathcal E]-p$ are independent. Bernstein's inequality gives
\begin{equation*}
\begin{gathered}
|v^k_k| \lesssim \sqrt{\left(p \sum_{j\ne k} \xi_j^2 \right)\log M} + \max_{j\ne k}|\xi_j| \log M\\
\lesssim \left(\sqrt{M p \log M} + \log M\right)\|\bm{u}^{[t], (k)} - \bm{u}^*\|_\infty,
\end{gathered}
\end{equation*}
with high probability. Therefore,
\begin{equation*}
\begin{gathered}
\|\bm{v}^k\|_2 \leq |v^k_k| + \sqrt{\sum_{j:(j,k) \in \mathcal E} (v_j^k)^2} + \sqrt{\sum_{j:(j,k) \not\in \mathcal E \text{ and } j \neq k} (v_j^k)^2}\\
\lesssim \left(\sqrt{M p \log M} + \log M\right)\|\bm{u}^{[t], (k)} - \bm{u}^*\|_\infty + \sqrt{Mp}\|\bm{u}^{[t], (k)} - \bm{u}^*\|_\infty + p\sqrt{M}\|\bm{u}^{[t], (k)} - \bm{u}^*\|_\infty\\
\lesssim \left(\sqrt{M p \log M} + \log M\right)\|\bm{u}^{[t], (k)} - \bm{u}^*\|_\infty\\
\lesssim \left(\sqrt{M p \log M}\right)\|\bm{u}^{[t], (k)} - \bm{u}^*\|_\infty.
\end{gathered}
\end{equation*}
The final inequality uses the standing condition $Mp\ge C_\kappa\kappa_1^4\log M$: since $\kappa_1\ge1$, this implies $Mp\gtrsim\log M$ and hence $\log M\lesssim\sqrt{Mp\log M}$. Combining these bounds yields
\begin{equation*}
\|\bm{v}_2\|_2 \lesssim \eta\left(\sqrt{\frac{M p \log M}{n_0 + n_{\mathcal{S}}}} + \left(\sqrt{M p \log M}\right)\|\bm{u}^{[t], (k)} - \bm{u}^*\|_\infty\right).
\end{equation*} 
\begin{equation*}
\begin{gathered}
\max_{1 \leq k \leq M}\|\bm{u}^{[t+1]} -  \bm{u}^{[t+1], (k)}\|_2 \leq \left(1 - \frac{\eta M p}{10\kappa_1}\right)\max_{1 \leq k \leq M}\|\bm{u}^{[t]} - \bm{u}^{[t], (k)}\|_2\\
+ C\eta\left(\sqrt{\frac{M p \log M}{n_0 + n_{\mathcal{S}}}} + \left(\sqrt{M p \log M}\right)\|\bm{u}^{[t], (k)} - \bm{u}^*\|_\infty\right)\\
\leq \left(1 - \frac{\eta M p}{10\kappa_1}\right)C_3\kappa_1\sqrt{\frac{\log M}{M p (n_0 + n_{\mathcal{S}})}}\\
+ C\eta\left(\sqrt{\frac{M p \log M}{n_0 + n_{\mathcal{S}}}} + \left(\sqrt{M p \log M}\right)C_5\kappa_1^2\sqrt{\frac{\log M}{M p (n_0 + n_{\mathcal{S}})}}\right)\\
\leq C_3\kappa_1\sqrt{\frac{\log M}{M p (n_0 + n_{\mathcal{S}})}},
\end{gathered}
\end{equation*}
for some $C>0$. The contraction decreases the first term by
$C_3\eta\sqrt{Mp\log M/(n_0+n_{\mathcal S})}/10$. Moreover,
\[
\kappa_1^2\frac{\log M}{\sqrt{n_0+n_{\mathcal S}}}
=
\left(\kappa_1^2\sqrt{\frac{\log M}{Mp}}\right)
\sqrt{\frac{Mp\log M}{n_0+n_{\mathcal S}}}
\le
\frac{1}{\sqrt{C_\kappa}}
\sqrt{\frac{Mp\log M}{n_0+n_{\mathcal S}}},
\]
The last inequality follows from $Mp\ge C_\kappa\kappa_1^4\log M$. Thus the two perturbation terms are absorbed by the contraction after choosing $C_3$ and then $C_\kappa$ sufficiently large. The constants in the Hoeffding and Bernstein thresholds can be enlarged, depending only on the fixed $A$, so that these simultaneous concentration bounds fail with probability at most $O(M^{-A})$, as required.
\end{proof}
\end{lemma}

\begin{lemma}
\label{lemma:lemma13_step1}
Suppose that the bounds provided in Equations \eqref{eq:bound1_step1}--\eqref{eq:bound4_step1} hold for iteration $t$, and suppose that the conclusions of Lemmas \ref{lemma:lemma11_step1} and \ref{lemma:lemma12_step1} hold at iteration $t+1$. Then
\begin{equation*}
\|\bm{u}^{[t+1]} -  \bm{u}^*\|_\infty \leq C_4\kappa_1^2\sqrt{\frac{\log M}{M p (n_0 + n_\mathcal{S})}} 
\end{equation*}
This bound holds provided that $C_4 \geq C_2 + C_3$.
\begin{proof}
The triangle inequality gives
\begin{equation*}
\begin{gathered}
\left|u_k^{[t+1]} -  u_k^*\right| \leq \left|u_k^{[t+1]} -  u_k^{[t+1], (k)}\right| + \left|u_k^{[t+1], (k)} -  u_k^*\right|\\
\leq \|\bm{u}^{[t+1]} -  \bm{u}^{[t+1], (k)}\|_2 + \left|u_k^{[t+1], (k)} -  u_k^*\right|\\
\leq C_3\kappa_1\sqrt{\frac{\log M}{M p (n_0 + n_{\mathcal{S}})}} + C_2\kappa_1^2\sqrt{\frac{\log M}{M p (n_0 + n_{\mathcal{S}})}}\\
\leq C_4\kappa_1^2\sqrt{\frac{\log M}{M p (n_0 + n_{\mathcal{S}})}}.
\end{gathered}
\end{equation*}
\end{proof}
\end{lemma}

\noindent\textbf{Proof of Theorem \ref{thm:convergence_step1_alpha}.}
\begin{proof}
At $t=0$, the event $\mathcal H_0$ holds deterministically. On $E_0\cap E_1\cap E_2$, Lemmas \ref{lemma:lemma10_step1} and \ref{lemma:lemma11_step1} are deterministic induction implications. Conditional on the comparison bound in Lemma \ref{lemma:lemma12_step1} holding at iteration $t+1$, Lemma \ref{lemma:lemma13_step1} is also deterministic. Lemma \ref{lemma:lemma12_step1} can fail at a given iteration with probability at most $O(M^{-A})$. Since $T=\lceil M^{C_T}\rceil=O(M^{C_T})$ and $A>C_T+12$, a union bound over the iterations gives failure probability $O(M^{-12})$. Together with $\mathbb P(E_0^c\cup E_1^c\cup E_2^c)=O(M^{-10})$, this yields
\begin{equation*}
\|\bm{u}^{[T]} -  \bm{u}^*\|_\infty \leq C_4\kappa_1^2\sqrt{\frac{\log M}{M p (n_0 + n_\mathcal{S})}}
\end{equation*}
with probability at least $1-O(M^{-10})$, and hence certainly at least $1-O(M^{-5})$. Combining this result with Lemma \ref{lemma:lemma8_step1} gives
\begin{equation*}
\|\hat{\bm{u}}_{\lambda_u} - \bm{u}^*\|_\infty \leq \|\bm{u}^{[T]} - \bm{u}^*\|_\infty + \|\bm{u}^{[T]} - \hat{\bm{u}}_{\lambda_u}\|_\infty \leq (C_0 + C_4)\kappa_1^2 \sqrt{\frac{\log M}{M p (n_0 + n_\mathcal{S})}}.
\end{equation*}
\noindent This proves Theorem \ref{thm:convergence_step1_alpha}, namely $\|\hat{\bm{u}}_{\lambda_u} - \bm{u}^*\|_\infty \lesssim \kappa_1^2 \sqrt{\frac{\log M}{M p (n_0 + n_\mathcal{S})}}$.
\end{proof}

\noindent We now show that the unregularised MLE has the same rate as the regularised estimator used in the proof. Following Fan et al.\ (\citeyear{fan2024uncertainty}), we use a localisation argument. We first minimise the unregularised negative log-likelihood over a small $\ell_\infty$ neighbourhood of $\bm u^*$. On this neighbourhood, the Bradley--Terry Hessian is uniformly strongly convex on the identifiable subspace $\bm1_M^\perp$. We then show that the constrained minimiser lies strictly inside the neighbourhood. It is therefore also a stationary point of the unconstrained problem. Convexity and connectivity of the comparison graph imply that this point is the unique centred unregularised MLE. The resulting statement is as follows.

\begin{theorem}[Unregularised pooled estimator]\label{thm:convergence_normal_step1}
Suppose Assumptions \ref{ass:1}, \ref{ass:2} and \ref{ass:pooledtarget} hold, $Mp\ge C_\kappa\kappa_1^4\log M$ for a sufficiently large $C_\kappa$, and $n_0 + n_{\mathcal{S}} \leq c_2 M^{c_3}$ for some $c_2, c_3 >0$. Then, with probability at least $1-O(M^{-5})$,
\begin{equation*}
\|\hat{\bm{u}} - \bm{u}^*\|_\infty \lesssim \kappa_1^2\sqrt{\frac{\log M}{M p(n_0 + n_\mathcal{S})}}.
\end{equation*}
\end{theorem}
\begin{proof}[Proof of Theorem \ref{thm:convergence_normal_step1}]
Define the local constrained MLE
\begin{equation}
\tilde{\bm u}
:=
\argmin_{\substack{\bm1_M^T\bm u=0\\
\|\bm u-\bm u^*\|_\infty\le \epsilon_1}}
L(\bm u),
\qquad \epsilon_1:=\frac15.
\label{eq:constrainedmle_alpha}
\end{equation}
The feasible set in Equation \eqref{eq:constrainedmle_alpha} is nonempty and compact, and $L$ is continuous, so $\tilde{\bm u}$ exists. The inequality constraint is only a proof device. Once we show that the minimiser lies strictly inside the neighbourhood, the constraint can be removed, as in Fan et al.\ (\citeyear{fan2024uncertainty}).
We work on the intersection of the graph event $E_1$ and the high-probability event from Theorem \ref{thm:convergence_step1_alpha}. By the probability bounds established above, this intersection has probability at least $1-O(M^{-5})$.
Fix any constant $c_{\lambda_u}>0$ and use the auxiliary ridge parameter $\lambda_u=c_{\lambda_u}B_u^{-1}\sqrt{p\log M/(n_0+n_{\mathcal S})}$ from Theorem \ref{thm:convergence_step1_alpha}. This parameter is used only in the proof. By Theorem \ref{thm:convergence_step1_alpha} and the condition $Mp\ge C_\kappa\kappa_1^4\log M$,
\[
\|\hat{\bm u}_{\lambda_u}-\bm u^*\|_\infty
\lesssim
\kappa_1^2\sqrt{\frac{\log M}{Mp(n_0+n_{\mathcal S})}}
\lesssim
\frac{1}{\sqrt{C_\kappa(n_0+n_{\mathcal S})}}.
\]
The sufficiently large constant $C_\kappa$ can therefore be chosen so that $\|\hat{\bm u}_{\lambda_u}-\bm u^*\|_\infty\le\epsilon_1=1/5$. Hence $\hat{\bm u}_{\lambda_u}$ is feasible for Equation \eqref{eq:constrainedmle_alpha}, and optimality of $\tilde{\bm u}$ gives $L(\tilde{\bm u})\le L(\hat{\bm u}_{\lambda_u})$.
\\
\\
\noindent Taylor's theorem now gives
\begin{equation*}
L(\tilde{\bm{u}}) = L(\hat{\bm{u}}_{\lambda_u}) + \nabla L(\hat{\bm{u}}_{\lambda_u})^T(\tilde{\bm{u}} - \hat{\bm{u}}_{\lambda_u}) + \frac{1}{2}(\tilde{\bm{u}} - \hat{\bm{u}}_{\lambda_u})^T\nabla^2 L(\bm{c})(\tilde{\bm{u}} - \hat{\bm{u}}_{\lambda_u})
\end{equation*}
for a convex combination $\bm c=\zeta\tilde{\bm u}+(1-\zeta)\hat{\bm u}_{\lambda_u}$ with $\zeta\in[0,1]$. Thus
\begin{equation}
\label{eq:taylor_alpha}
\nabla L(\hat{\bm{u}}_{\lambda_u})^T(\tilde{\bm{u}} - \hat{\bm{u}}_{\lambda_u}) + \frac{1}{2}(\tilde{\bm{u}} - \hat{\bm{u}}_{\lambda_u})^T\nabla^2 L(\bm{c})(\tilde{\bm{u}} - \hat{\bm{u}}_{\lambda_u}) \leq 0.
\end{equation}
Since both $\tilde{\bm u}$ and $\hat{\bm u}_{\lambda_u}$ lie in the radius-$1/5$ neighbourhood of $\bm u^*$, the same is true of $\bm c$:
\[
\|\bm c-\bm u^*\|_\infty\le\frac15.
\]
Lemma \ref{lemma:lemma5_step1}, with $r=1/5$, therefore gives
\begin{equation}
\label{eq:w_conev_alpha}
\lambda_{\min,\perp}(\nabla^2L(\bm c))
\ge \frac{Mp}{8\kappa_1e^{2/5}}
\ge \frac{Mp}{12\kappa_1}.
\end{equation}
If we now combine Equation \eqref{eq:taylor_alpha} and Equation \eqref{eq:w_conev_alpha}, we obtain
\begin{equation*}
\begin{gathered}
\frac{Mp}{24 \kappa_1}\|\tilde{\bm{u}} - \hat{\bm{u}}_{\lambda_u}\|_2^2 \leq \frac{1}{2}(\tilde{\bm{u}} - \hat{\bm{u}}_{\lambda_u})^T\nabla^2 L(\bm{c})(\tilde{\bm{u}} - \hat{\bm{u}}_{\lambda_u})\\
\leq - \nabla L(\hat{\bm{u}}_{\lambda_u})^T(\tilde{\bm{u}} - \hat{\bm{u}}_{\lambda_u})\\
\leq \|\nabla L(\hat{\bm{u}}_{\lambda_u})\|_2 \|\tilde{\bm{u}} - \hat{\bm{u}}_{\lambda_u}\|_2.
\end{gathered}
\end{equation*}
Since the regularised optimum is centred and the Bradley--Terry likelihood gradient is orthogonal to $\bm1_M$, its KKT equation is $\nabla L(\hat{\bm u}_{\lambda_u})=-\lambda_u\hat{\bm u}_{\lambda_u}$. Moreover, Theorem \ref{thm:convergence_step1_alpha} and the preceding localisation give
\[
\|\hat{\bm u}_{\lambda_u}\|_2
\le
\sqrt M\left(\|\bm u^*\|_\infty+\|\hat{\bm u}_{\lambda_u}-\bm u^*\|_\infty\right)
\lesssim B_u\sqrt M.
\]
It follows that
\begin{gather}
\|\tilde{\bm{u}} - \hat{\bm{u}}_{\lambda_u}\|_2 \leq  \frac{24 \kappa_1}{Mp}\|\nabla L(\hat{\bm{u}}_{\lambda_u})\|_2 \notag\\
\leq \frac{24 \kappa_1 \lambda_u}{Mp}\|\hat{\bm{u}}_{\lambda_u}\|_2 \notag\\
\lesssim \frac{24 \kappa_1 \lambda_u}{Mp}B_u\sqrt{M} \notag\\
\lesssim  \kappa_1 \sqrt{\frac{\log M}{M p (n_0 + n_{\mathcal{S}})}} \label{eq:w_tilde_w_lam_alpha}.
\end{gather}
Combining the preceding bounds gives
\begin{equation*}
\begin{gathered}
\|\tilde{\bm{u}} - \bm{u}^*\|_\infty  = \|(\tilde{\bm{u}} - \hat{\bm{u}}_{\lambda_u}) + (\hat{\bm{u}}_{\lambda_u} - \bm{u}^*)\|_\infty\\
\leq \|\tilde{\bm{u}} - \hat{\bm{u}}_{\lambda_u}\|_\infty + \|\hat{\bm{u}}_{\lambda_u} - \bm{u}^*\|_\infty \text{ (triangle inequality)}\\
\leq \|\tilde{\bm{u}} - \hat{\bm{u}}_{\lambda_u}\|_2 + \|\hat{\bm{u}}_{\lambda_u} - \bm{u}^*\|_\infty\\
\lesssim \kappa_1 \sqrt{\frac{\log M}{M p (n_0 + n_{\mathcal{S}})}} + \kappa_1^2\sqrt{\frac{\log M}{Mp (n_0 + n_{\mathcal{S}})}} \text{ (Equation \eqref{eq:w_tilde_w_lam_alpha} and Theorem \ref{thm:convergence_step1_alpha})}.
\end{gathered}
\end{equation*}
Under the standing condition $Mp\ge C_\kappa\kappa_1^4\log M$ and $n_0+n_{\mathcal S}\ge1$, the last display is $O(C_\kappa^{-1/2})$. Increasing $C_\kappa$ if necessary therefore makes the display strictly smaller than $\epsilon_1=1/5$. Hence $\tilde{\bm u}$ lies in the interior of the artificial $\ell_\infty$ constraint in Equation \eqref{eq:constrainedmle_alpha}. The inactive inequality constraint leaves only the centring constraint, so the KKT equation is $\nabla L(\tilde{\bm u})+\nu\bm1_M=\bm0$ for some scalar $\nu$. Since every Bradley--Terry likelihood gradient is orthogonal to $\bm1_M$, multiplication by $\bm1_M^T$ gives $\nu=0$, and hence $\nabla L(\tilde{\bm u})=\bm0$. Convexity of $L$ therefore makes $\tilde{\bm u}$ a global minimiser on the centred parameter space. Moreover, on the graph event $E_1$, $\lambda_{\min,\perp}(\bm L_{\mathcal G})>0$, so $\mathcal G$ is connected. For every finite $\bm u$ the Bradley--Terry Hessian is a weighted Laplacian with strictly positive edge weights, and is therefore positive definite on $\bm1_M^\perp$. Thus $L$ is strictly convex on the centred parameter space, the global minimiser is unique, and $\tilde{\bm u}=\hat{\bm u}$. In particular, this argument establishes existence and finiteness of the centred unregularised pooled MLE on the event considered. We conclude the proof by noting that
\begin{equation*}
\begin{gathered}
\|\hat{\bm{u}} - \bm{u}^*\|_\infty = \|\tilde{\bm{u}} - \bm{u}^*\|_\infty \lesssim \kappa_1^2 \sqrt{\frac{\log M}{M p (n_0 + n_\mathcal{S})}}.
\end{gathered}
\end{equation*}
\end{proof}
\noindent This concludes the proof of Theorem \ref{thm:convergence_normal_step1}, as well as the proof of the first part of Theorem \ref{thm:alphaconvergence}.

\subsection{Primary-attribute correction and transfer improvement}\label{sec:primary-correction-proof}

We next consider the primary-attribute correction step. In contrast to the transfer step, the relevant likelihood is the primary Bradley--Terry likelihood and its curvature is therefore governed by the primary parameter $\bm\alpha^*$, rather than by the discrepancy parameter. The improvement obtained from transfer comes from the ridge term, which keeps the correction close to the pooled estimate. Recall the equivalent parameterisation in Equation \eqref{eq:alpha_ridge_form}:
\begin{equation}
\hat{\bm\alpha}_{\lambda_\delta}
=
\argmin_{\bm1_M^T\bm\alpha=0}
\left\{
L_0(\bm\alpha)
+
\frac{\lambda_\delta}{2}
\|\bm\alpha-\hat{\bm u}\|_2^2
\right\}.
\label{eq:alpha_ridge_proof}
\end{equation}
This is exactly equivalent to estimating $\bm\delta$ in line 2 of Algorithm \ref{alg:oraclebtl} and setting $\hat{\bm\alpha}_{\lambda_\delta}=\hat{\bm u}+\hat{\bm\delta}_{\lambda_\delta}$.

\begin{lemma}[Primary score concentration]\label{lemma:primary_score}
Suppose the degree event $E_0$ in Lemma \ref{lemma:lemma1_joint} holds and $Mp\ge c\log M$ for a sufficiently large constant $c>0$. Then, conditional on $\mathcal G$, with probability at least $1-O(M^{-10})$,
\begin{equation}
\|\nabla L_0(\bm\alpha^*)\|_\infty
\lesssim
\sqrt{\frac{Mp\log M}{n_0}},
\label{eq:primary_score_inf}
\end{equation}
and
\begin{equation}
\|\nabla L_0(\bm\alpha^*)\|_2
\lesssim
M\sqrt{\frac{p\log M}{n_0}}.
\label{eq:primary_score_l2}
\end{equation}
\begin{proof}
For coordinate $k$,
\[
\{\nabla L_0(\bm\alpha^*)\}_k
=
\frac1{n_0}
\sum_{j:(j,k)\in\mathcal E}
\sum_{i=1}^{n_0}
\varepsilon_{jk}^{(i)},
\]
up to the deterministic orientation sign, where the $\varepsilon_{jk}^{(i)}$ are conditionally independent, mean-zero and bounded by one. On $E_0$, $d_k\lesssim Mp$. Bernstein's inequality therefore yields
\[
\mathbb P\left(
|\{\nabla L_0(\bm\alpha^*)\}_k|>t
\mid\mathcal G
\right)
\le
2\exp\left[
-c\min\left\{
\frac{n_0t^2}{Mp},
n_0t
\right\}
\right].
\]
Taking $t=C\sqrt{Mp\log M/n_0}$, the assumption $Mpn_0\gtrsim\log M$ ensures that both terms in the Bernstein exponent are of order at least $\log M$. A union bound over $k\in[M]$ therefore proves \eqref{eq:primary_score_inf}. The second display follows from
\[
\|\nabla L_0(\bm\alpha^*)\|_2
\le
\sqrt M\,
\|\nabla L_0(\bm\alpha^*)\|_\infty.
\]
\end{proof}
\end{lemma}

\noindent The following deterministic lemma provides the main bound needed for the correction step.

\begin{lemma}[Ridge interpolation for Bradley--Terry]\label{lemma:ridge_interpolation}
Let $\bm b\in\mathbb R^M$ satisfy $\bm1_M^T\bm b=0$ and let
\[
\hat{\bm\alpha}_{\lambda}
=
\argmin_{\bm1_M^T\bm\alpha=0}
\left\{
L_0(\bm\alpha)
+
\frac{\lambda}{2}\|\bm\alpha-\bm b\|_2^2
\right\},
\qquad \lambda>0.
\]
Then
\begin{equation}
\|\hat{\bm\alpha}_{\lambda}-\bm\alpha^*\|_\infty
\le
\|\bm b-\bm\alpha^*\|_\infty
+
\frac{\|\nabla L_0(\bm\alpha^*)\|_\infty}{\lambda},
\label{eq:ridge_interp_inf_det}
\end{equation}
and
\begin{equation}
\|\hat{\bm\alpha}_{\lambda}-\bm\alpha^*\|_2
\le
\|\bm b-\bm\alpha^*\|_2
+
\frac{\|\nabla L_0(\bm\alpha^*)\|_2}{\lambda}.
\label{eq:ridge_interp_l2_det}
\end{equation}

\begin{proof}
The objective is coercive and $\lambda$-strongly convex, so the centred minimiser exists uniquely. Write
\[
\bm e
:=
\hat{\bm\alpha}_{\lambda}-\bm\alpha^*.
\]
The KKT equation for the equality-constrained problem is
\[
\nabla L_0(\hat{\bm\alpha}_{\lambda})
+
\lambda(\hat{\bm\alpha}_{\lambda}-\bm b)
+
\nu\bm1_M
=
\bm0.
\]
Every Bradley--Terry likelihood gradient is orthogonal to $\bm1_M$, and both $\hat{\bm\alpha}_{\lambda}$ and $\bm b$ are centred. Multiplying by $\bm1_M^T$ therefore gives $\nu=0$. By the fundamental theorem of calculus,
\[
\nabla L_0(\hat{\bm\alpha}_{\lambda})
-
\nabla L_0(\bm\alpha^*)
=
\bar{\bm H}\bm e,
\]
where
\begin{equation}
\bar{\bm H}
:=
\int_0^1
\nabla^2L_0(\bm\alpha^*+t\bm e)\,dt.
\label{eq:Hbar_ridge}
\end{equation}
Hence
\begin{equation}
(\bar{\bm H}+\lambda\bm I_M)\bm e
=
-\nabla L_0(\bm\alpha^*)
+
\lambda(\bm b-\bm\alpha^*).
\label{eq:ridge_exact_identity}
\end{equation}

\noindent For every argument, the Bradley--Terry Hessian is a weighted graph Laplacian. Consequently $\bar{\bm H}$ has non-positive off-diagonal entries, non-negative diagonal entries and $\bar{\bm H}\bm1_M=\bm0$. Let
\[
\bm A:=\bar{\bm H}+\lambda\bm I_M.
\]
The matrix $-\bar{\bm H}$ is the generator of a continuous-time Markov chain, so $\exp(-t\bar{\bm H})$ has non-negative entries and row sums one. The resolvent identity gives
\[
\bm A^{-1}
=
\int_0^\infty
e^{-\lambda t}
e^{-t\bar{\bm H}}\,dt.
\]
Therefore $\bm A^{-1}$ has non-negative entries and each of its rows sums to $1/\lambda$. It follows that
\begin{equation}
\|\bm A^{-1}\|_\infty
=
\frac1\lambda.
\label{eq:Mmatrix_inverse_inf}
\end{equation}
Applying \eqref{eq:Mmatrix_inverse_inf} to \eqref{eq:ridge_exact_identity} gives
\[
\|\bm e\|_\infty
\le
\frac{\|\nabla L_0(\bm\alpha^*)\|_\infty}{\lambda}
+
\|\bm b-\bm\alpha^*\|_\infty,
\]
which is \eqref{eq:ridge_interp_inf_det}.

\noindent For the $\ell_2$ result, $\bar{\bm H}\succeq0$ implies
$\|\bm A^{-1}\|_2\le1/\lambda$. Applying this to
\eqref{eq:ridge_exact_identity} proves \eqref{eq:ridge_interp_l2_det}.
\end{proof}
\end{lemma}

\begin{theorem}[Primary correction bound]\label{thm:convergence_step2_alpha}
Suppose Assumptions \ref{ass:1}, \ref{ass:2} and \ref{ass:pooledtarget} hold, $Mp\ge C_\kappa\kappa_1^4\log M$ for a sufficiently large $C_\kappa$, and $N_{\mathrm{tr}}\le c_2M^{c_3}$ for some constants $c_2,c_3>0$. On the intersection of the transfer event
\[
\|\hat{\bm u}-\bm u^*\|_\infty
\lesssim r_{u,\infty}
\],
the degree event $E_0$, and the primary-score event in Lemma \ref{lemma:primary_score}, the estimator in Equation \eqref{eq:alpha_ridge_proof} satisfies, simultaneously for all $\lambda_\delta>0$,
\begin{equation}
\|\hat{\bm\alpha}_{\lambda_\delta}-\bm\alpha^*\|_\infty
\lesssim
d_{\mathcal S,\infty}
+
r_{u,\infty}
+
\frac1{\lambda_\delta}
\sqrt{\frac{Mp\log M}{n_0}}.
\label{eq:step2_transfer_bound}
\end{equation}
This intersection has probability at least $1-O(M^{-5})$; no independence among these events is required.
\end{theorem}
\begin{proof}[Proof of Theorem \ref{thm:convergence_step2_alpha}]
Apply Lemma \ref{lemma:ridge_interpolation} with $\bm b=\hat{\bm u}$. By the triangle inequality,
\[
\|\hat{\bm u}-\bm\alpha^*\|_\infty
\le
\|\hat{\bm u}-\bm u^*\|_\infty
+
\|\bm u^*-\bm\alpha^*\|_\infty
\lesssim
r_{u,\infty}+d_{\mathcal S,\infty}.
\]
Combining this with Lemma \ref{lemma:primary_score} yields \eqref{eq:step2_transfer_bound}; Lemma \ref{lemma:ridge_interpolation} is deterministic and holds for every $\lambda_\delta>0$.
\end{proof}

\noindent\textbf{Proof of Theorem \ref{thm:alphaconvergence}.}
\begin{proof}
The transfer bound \eqref{eq:u_main_rate} is Theorem \ref{thm:convergence_normal_step1}. The second part follows from Theorem \ref{thm:convergence_step2_alpha}. If
\[
\lambda_\delta
\ge
c_{\lambda_\delta}
\frac{\sqrt{Mp\log M/n_0}}
{r_{\mathrm{tr},\infty}},
\]
then the last term in \eqref{eq:step2_transfer_bound} is at most a constant multiple of $r_{\mathrm{tr},\infty}$. Therefore
\[
\|\hat{\bm\alpha}_{\lambda_\delta}-\bm\alpha^*\|_\infty
\lesssim
r_{\mathrm{tr},\infty}
=
d_{\mathcal S,\infty}
+
\kappa_1^2
\sqrt{\frac{\log M}{Mp(n_0+n_{\mathcal S})}},
\]
as claimed.
\end{proof}

\noindent\textbf{Proof of Corollary \ref{crl:rate_improvement}.}
\begin{proof}
The primary-only rate \eqref{eq:target_only_rate} follows from Proposition \ref{prop:target_only_rate}, whose proof is the primary-only specialization of the same Chen-style leave-one-out argument. Comparing \eqref{eq:main_linf_transfer} and \eqref{eq:target_only_rate} gives \eqref{eq:improvement_condition_general}. If every informative secondary attribute contributes $n_s=n_0$ comparisons per edge, then
\[
n_{\mathcal S}=|\mathcal S|n_0,
\qquad
n_0+n_{\mathcal S}=(1+|\mathcal S|)n_0,
\]
which gives \eqref{eq:equal_source_rate} and the sufficient conditions in \eqref{eq:equal_source_improvement}.
\end{proof}

\noindent\textbf{Proof of Corollary \ref{crl:alphaconvergencel2}.}
\begin{proof}
The $\ell_\infty$ bound in Theorem \ref{thm:alphaconvergence} immediately gives the conservative $\ell_2$ bound
\[
\|\hat{\bm u}-\bm u^*\|_2
\le \sqrt M\|\hat{\bm u}-\bm u^*\|_\infty
\lesssim
\kappa_1^2
\sqrt{\frac{\log M}{p(n_0+n_{\mathcal S})}}.
\]
Therefore
\[
\|\hat{\bm u}-\bm\alpha^*\|_2
\lesssim
\sqrt{h_{\mathrm{agg}}}
+
\kappa_1^2
\sqrt{\frac{\log M}{p(n_0+n_{\mathcal S})}}
=
r_{\mathrm{tr},2}.
\]
Lemma \ref{lemma:ridge_interpolation} and Equation \eqref{eq:primary_score_l2} yield
\[
\|\hat{\bm\alpha}_{\lambda_\delta}-\bm\alpha^*\|_2
\lesssim
r_{\mathrm{tr},2}
+
\frac{M\sqrt{p\log M/n_0}}{\lambda_\delta}.
\]
Under \eqref{eq:lambda_l2}, the second term is of order $r_{\mathrm{tr},2}$, proving \eqref{eq:main_l2_transfer}. These bounds hold on the same joint transfer and primary-score event used in Theorem \ref{thm:alphaconvergence}, whose probability is at least $1-O(M^{-5})$. If the conditions of Proposition \ref{prop:target_only_rate} also hold, then
\[
\|\hat{\bm\alpha}^{(0)}-\bm\alpha^*\|_2
\le
\sqrt M\|\hat{\bm\alpha}^{(0)}-\bm\alpha^*\|_\infty
\lesssim
\kappa_3^2\sqrt{\frac{\log M}{pn_0}}.
\]
Comparing this conservative primary-only upper bound with \eqref{eq:main_l2_transfer} gives \eqref{eq:l2_improvement_condition}.
\end{proof}

\section*{Asymptotic normality}

Throughout this section, $L$ denotes the negative log-likelihood $L_{\mathrm{inf}}=L_{\mathcal D_{\mathrm{inf}}}$ constructed from the separate inference data set. Conditional on the common comparison graph $\mathcal G$, the training estimator $\hat{\bm\alpha}_{\lambda_\delta}$ is independent of the inference outcomes used to construct $L$. We use this conditional independence for the score concentration and Berry--Esseen steps, but the final asymptotic-normality theorem accounts for all graph and outcome randomness. The proof follows ideas from Liu et al.\ (\citeyear{liu2023lagrangian}).

\begin{lemma}
\label{lemma:lemma25}
Suppose that the conditions of Theorem \ref{thm:alphaconvergence} and the inference-sample construction in Section \ref{sec:statistical-properties} hold. On the degree event $E_0$, conditional on $\mathcal G$, with probability at least $1-O(M^{-10})$,
\begin{equation*}
\left\|\nabla L(\bm\alpha^*)\right\|_\infty \lesssim \sqrt{\frac{Mp\log M}{n_{\mathrm{inf}}}}.
\end{equation*}
\begin{proof}
For coordinate $k$,
\[
\{\nabla L(\bm\alpha^*)\}_k
=
\frac1{n_{\mathrm{inf}}}
\sum_{j:(j,k)\in\mathcal E}
\sum_{i=1}^{n_{\mathrm{inf}}}
\varepsilon_{jk}^{(i)},
\]
up to the deterministic orientation sign, where the $\varepsilon_{jk}^{(i)}$ are conditionally independent, mean-zero and bounded by one. On the degree event of Lemma \ref{lemma:lemma1_joint}, $d_k\lesssim Mp$. Bernstein's inequality therefore gives
\[
\mathbb P\!\left(
|\{\nabla L(\bm\alpha^*)\}_k|>t
\mid\mathcal G
\right)
\le
2\exp\!\left[
-c\min\left\{
\frac{n_{\mathrm{inf}}t^2}{Mp},
n_{\mathrm{inf}}t
\right\}
\right].
\]
Taking $t=C\sqrt{Mp\log M/n_{\mathrm{inf}}}$ with $C$ sufficiently large, the standing condition $Mp\gtrsim\log M$ and $n_{\mathrm{inf}}\ge1$ make both terms in the Bernstein exponent at least a constant multiple of $\log M$. A union bound over $k\in[M]$ then proves the result. This is the same coordinatewise Bernstein argument used for the primary score in Lemma \ref{lemma:primary_score}.
\end{proof}
\end{lemma}

\begin{lemma}
\label{lemma:lemma26}
Suppose Assumption \ref{ass:rmax} holds. Then, with probability at least $1-O(M^{-5})$,
\begin{equation*}
\left\|
\nabla L(\hat{\bm\alpha}_{\lambda_\delta})-\nabla L(\bm\alpha^*)
-\nabla^2L(\bm\alpha^*)(\hat{\bm\alpha}_{\lambda_\delta}-\bm\alpha^*)
\right\|_\infty
\lesssim Mp\,R_{\alpha,\lambda}^2.
\end{equation*}
\begin{proof}
Using a mean-value expansion, we write $\nabla L(\hat{\bm{\alpha}}_{\lambda_\delta}) - \nabla L(\bm{\alpha}^*)$ as
\begin{equation*}
\begin{gathered}
\nabla L(\hat{\bm{\alpha}}_{\lambda_\delta}) - \nabla L(\bm{\alpha}^*) = \sum_{(j,l)\in \mathcal E, j>l}\left(\frac{\exp(\hat{\alpha}_{\lambda_\delta,j})}{\exp(\hat{\alpha}_{\lambda_\delta,j}) + \exp(\hat{\alpha}_{\lambda_\delta,l})} - \frac{\exp(\alpha_j^*)}{\exp(\alpha_j^*) + \exp(\alpha_l^*)}\right)(\bm e_j - \bm e_l)\\
=  \sum_{(j,l)\in \mathcal E, j>l}\frac{\exp(c_{jl})}{[1+\exp(c_{jl})]^2}\left[\hat{\alpha}_{\lambda_\delta,j} - \hat{\alpha}_{\lambda_\delta,l} - \left(\alpha_j^* - \alpha_l^*\right)\right](\bm e_j - \bm e_l)\\
=  \sum_{(j,l)\in \mathcal E, j>l}\frac{\exp(c_{jl})}{[1+\exp(c_{jl})]^2}(\bm e_j - \bm e_l)(\bm e_j - \bm e_l)^T\left(\hat{\bm{\alpha}}_{\lambda_\delta} - \bm{\alpha}^*\right).
\end{gathered}
\end{equation*}
Here $c_{jl}\in\mathbb R$ lies between $\hat\alpha_{\lambda_\delta,j}-\hat\alpha_{\lambda_\delta,l}$ and $\alpha_j^*-\alpha_l^*$. Define
\[
w(x):=\frac{\exp(x)}{[1+\exp(x)]^2}.
\]
Since $w$ is globally Lipschitz and
\[
|c_{jl}-(\alpha_j^*-\alpha_l^*)|
\le
|\hat\alpha_{\lambda_\delta,l}-\alpha_l^*|+|\hat\alpha_{\lambda_\delta,j}-\alpha_j^*|
\le 2\|\hat{\bm\alpha}_{\lambda_\delta}-\bm\alpha^*\|_\infty,
\]
we obtain
\[
|w(c_{jl})-w(\alpha_j^*-\alpha_l^*)|
\lesssim
\|\hat{\bm\alpha}_{\lambda_\delta}-\bm\alpha^*\|_\infty.
\]
Therefore
\begin{equation*}
\nabla L(\hat{\bm\alpha}_{\lambda_\delta})-\nabla L(\bm\alpha^*)
-\nabla^2L(\bm\alpha^*)(\hat{\bm\alpha}_{\lambda_\delta}-\bm\alpha^*)
=\bm D(\hat{\bm\alpha}_{\lambda_\delta}-\bm\alpha^*),
\end{equation*}
where
\[
\bm D
:=
\sum_{(j,l)\in\mathcal E,\,j>l}
\{w(c_{jl})-w(\alpha_j^*-\alpha_l^*)\}
(\bm e_j-\bm e_l)(\bm e_j-\bm e_l)^T.
\]
For row $k$, each incident edge contributes once to the diagonal and once to an off-diagonal entry. Hence, on the degree event in Lemma \ref{lemma:lemma1_joint},
\begin{align*}
\|\bm D\|_\infty
&\le
2d_{\max}
\max_{(j,l)\in\mathcal E}
|w(c_{jl})-w(\alpha_j^*-\alpha_l^*)|\\
&\lesssim
Mp\|\hat{\bm\alpha}_{\lambda_\delta}-\bm\alpha^*\|_\infty.
\end{align*}
Thus, on the intersection of the degree event $E_0$ and the preliminary-estimator event from Theorem \ref{thm:alphaconvergence},
\begin{equation*}
\left\|
\nabla L(\hat{\bm\alpha}_{\lambda_\delta})-\nabla L(\bm\alpha^*)
-\nabla^2L(\bm\alpha^*)(\hat{\bm\alpha}_{\lambda_\delta}-\bm\alpha^*)
\right\|_\infty
\lesssim Mp\|\hat{\bm\alpha}_{\lambda_\delta}-\bm\alpha^*\|_\infty^2
\lesssim MpR_{\alpha,\lambda}^2.
\end{equation*}
The complement of this intersection has probability $O(M^{-5})$, which proves the lemma.
\end{proof}
\end{lemma}

\begin{lemma}
\label{lemma:lemma27}
Suppose Assumption \ref{ass:rmax} holds. Then, with probability at least $1-O(M^{-5})$,
\begin{equation*}
\|\nabla^2L(\hat{\bm\alpha}_{\lambda_\delta})-\nabla^2L(\bm\alpha^*)\|_\infty
\lesssim Mp\,R_{\alpha,\lambda}.
\end{equation*}
\begin{proof}
On the intersection of the preliminary-estimator event from Theorem \ref{thm:alphaconvergence} and the degree event $E_0$,
\begin{equation*}
\begin{gathered}
\|\nabla^2L(\hat{\bm\alpha}_{\lambda_\delta})-\nabla^2L(\bm\alpha^*)\|_\infty
\lesssim
Mp\|\hat{\bm\alpha}_{\lambda_\delta}-\bm\alpha^*\|_\infty
\lesssim MpR_{\alpha,\lambda}.
\end{gathered}
\end{equation*}
The complement of this intersection has probability $O(M^{-5})$.
\end{proof}
\end{lemma}

\begin{lemma}
\label{lemma:lemma28}
Suppose Assumption \ref{ass:rmax} holds and let
\[
\bm H(\bm\alpha):=\nabla^2L(\bm\alpha),
\qquad
\bm K(\bm\alpha):=
\begin{pmatrix}
\bm H(\bm\alpha)&\bm1_M\\
\bm1_M^T&0
\end{pmatrix},
\]
where $\bm\alpha$ is either $\hat{\bm\alpha}_{\lambda_\delta}$ or $\bm\alpha^*$. Under Assumption \ref{ass:rmax}, $\|\hat{\bm\alpha}_{\lambda_\delta}-\bm\alpha^*\|_\infty\lesssim R_{\alpha,\lambda}=o(1)$ with probability at least $1-O(M^{-5})$; no localisation is needed when $\bm\alpha=\bm\alpha^*$. On the intersection of this localisation event with $E_1$ (an event of probability at least $1-O(M^{-5})$),
\begin{equation}
\|\bm K(\bm\alpha)^{-1}\|_2
=
\max\left\{
\frac{1}{\lambda_{\min,\perp}(\bm H(\bm\alpha))},
\frac{1}{\sqrt M}
\right\},
\label{eq:bordered_inverse_norm}
\end{equation}
and the upper-left block of $\bm K(\bm\alpha)^{-1}$ equals $\bm H(\bm\alpha)^+$. In particular,
\begin{equation}
\|\bm H(\bm\alpha)^+\|_2
=
\frac{1}{\lambda_{\min,\perp}(\bm H(\bm\alpha))}
\lesssim \frac{\kappa_3}{Mp}.
\label{eq:pseudoinverse_norm}
\end{equation}
\begin{proof}
The Bradley--Terry Hessian satisfies $\bm H(\bm\alpha)\bm1_M=\bm0$. Let
$\lambda_1\ge\cdots\ge\lambda_{M-1}>0=\lambda_M$
be its eigenvalues with orthonormal eigenvectors
$\bm v_1,\ldots,\bm v_{M-1},\bm1_M/\sqrt M$. Since all nonzero eigenvectors are orthogonal to $\bm1_M$,
\[
\bm K(\bm\alpha)
\binom{\bm v_l}{0}
=
\lambda_l\binom{\bm v_l}{0},
\qquad l=1,\ldots,M-1.
\]
Furthermore,
\[
\bm K(\bm\alpha)
\binom{\bm1_M}{\sqrt M}
=
\sqrt M\binom{\bm1_M}{\sqrt M},
\qquad
\bm K(\bm\alpha)
\binom{\bm1_M}{-\sqrt M}
=
-\sqrt M\binom{\bm1_M}{-\sqrt M}.
\]
Hence the eigenvalues of the symmetric bordered matrix are
\[
\lambda_1,\ldots,\lambda_{M-1},\sqrt M,-\sqrt M.
\]
Taking reciprocals gives \eqref{eq:bordered_inverse_norm}; the two bordered directions contribute singular values $1/\sqrt M$.
\\
\\
\noindent To identify the upper-left block, write the spectral decomposition
\[
\bm H(\bm\alpha)^+
=
\sum_{l=1}^{M-1}\lambda_l^{-1}\bm v_l\bm v_l^T.
\]
Direct multiplication shows
\[
\bm K(\bm\alpha)^{-1}
=
\begin{pmatrix}
\bm H(\bm\alpha)^+&M^{-1}\bm1_M\\
M^{-1}\bm1_M^T&0
\end{pmatrix}.
\]
Finally, on $E_1$, if $\|\bm\alpha-\bm\alpha^*\|_\infty\le r$ then every edge difference satisfies
\[
|\alpha_j-\alpha_l|
\le |\alpha_j^*-\alpha_l^*|+2r
\le \log\kappa_3+2r.
\]
Hence its logistic Hessian weight is at least
\[
\frac14e^{-|\alpha_j-\alpha_l|}
\ge \frac{e^{-2r}}{4\kappa_3}.
\]
Under Assumption \ref{ass:rmax}, $r=O(R_{\alpha,\lambda})=o(1)$ for $\bm\alpha=\hat{\bm\alpha}_{\lambda_\delta}$, while $r=0$ for $\bm\alpha=\bm\alpha^*$. Therefore, for all sufficiently large $M$, the edge weights are uniformly bounded below by $c/\kappa_3$ for a fixed $c>0$. Combining this with $\lambda_{\min,\perp}(\bm L_{\mathcal G})\ge Mp/2$ gives
\[
\lambda_{\min,\perp}(\bm H(\bm\alpha))
\gtrsim \frac{Mp}{\kappa_3},
\]
which yields \eqref{eq:pseudoinverse_norm}.
\end{proof}
\end{lemma}

\begin{lemma}
\label{lemma:lemma29}
Suppose Assumption \ref{ass:rmax} holds. Let
\[
\hat{\bm\Theta}_{11}:=[\nabla^2L(\hat{\bm\alpha}_{\lambda_\delta})]^+,
\qquad
\bm\Theta_{11}^*:=[\nabla^2L(\bm\alpha^*)]^+.
\]
Then, with probability at least $1-O(M^{-5})$,
\begin{equation}
\|\hat{\bm\Theta}_{11}-\bm\Theta_{11}^*\|_2
\lesssim
\frac{\kappa_3^2}{Mp}R_{\alpha,\lambda}
=
D_{\Theta,\lambda}.
\label{eq:theta_perturbation}
\end{equation}
\begin{proof}
Both Hessians have the same nullspace $\operatorname{span}(\bm1_M)$. Restricted to $\bm1_M^\perp$ they are invertible, so the usual inverse resolvent identity applies on that subspace:
\[
\hat{\bm\Theta}_{11}-\bm\Theta_{11}^*
=
\hat{\bm\Theta}_{11}
\{\nabla^2L(\bm\alpha^*)-\nabla^2L(\hat{\bm\alpha}_{\lambda_\delta})\}
\bm\Theta_{11}^*.
\]
Lemma \ref{lemma:lemma28} gives
\[
\|\hat{\bm\Theta}_{11}\|_2\vee\|\bm\Theta_{11}^*\|_2
\lesssim\frac{\kappa_3}{Mp}.
\]
It remains to bound the Hessian perturbation in operator norm. Let
\[
w_{jl}(\bm\alpha)
=
\frac{\exp(\alpha_j+\alpha_l)}
{[\exp(\alpha_j)+\exp(\alpha_l)]^2}.
\]
Since $w_{jl}(\bm\alpha)=w(\alpha_j-\alpha_l)$ with $w(t)=e^t/(1+e^t)^2$ and $w'$ is bounded,
\[
|w_{jl}(\hat{\bm\alpha}_{\lambda_\delta})-w_{jl}(\bm\alpha^*)|
\lesssim
|\hat\alpha_{\lambda_\delta,j}-\alpha_j^*|+|\hat\alpha_{\lambda_\delta,l}-\alpha_l^*|
\lesssim R_{\alpha,\lambda}.
\]
Consequently, on $E_1$,
\[
\begin{aligned}
\|\nabla^2L(\hat{\bm\alpha}_{\lambda_\delta})-\nabla^2L(\bm\alpha^*)\|_2
&\le
\max_{(j,l)\in\mathcal E}
|w_{jl}(\hat{\bm\alpha}_{\lambda_\delta})-w_{jl}(\bm\alpha^*)|
\,\|\bm L_{\mathcal G}\|_2\\
&\lesssim MpR_{\alpha,\lambda}.
\end{aligned}
\]
Combining the last three displays yields
\[
\|\hat{\bm\Theta}_{11}-\bm\Theta_{11}^*\|_2
\lesssim
\left(\frac{\kappa_3}{Mp}\right)^2
(MpR_{\alpha,\lambda})
=
\frac{\kappa_3^2}{Mp}R_{\alpha,\lambda}.
\]
The preliminary localisation and $E_1$ fail with probability at most $O(M^{-5})$ in total.
\end{proof}
\end{lemma}

\begin{lemma}
\label{lemma:lemma30}
Suppose Assumption \ref{ass:rmax} holds, and let $\bm\alpha$ be either $\hat{\bm\alpha}_{\lambda_\delta}$ or $\bm\alpha^*$. Define
\begin{equation}
\label{eq:lemma30_1}
\begin{pmatrix}
\bm{\Theta}_{11} & \frac{1}{M}\bm1_M\\
\frac{1}{M}\bm1_M^T & 0
\end{pmatrix}
:=
\begin{pmatrix}
\nabla^2 L(\bm{\alpha}) & \bm1_M\\
\bm1_M^T & 0
\end{pmatrix}^{-1}.
\end{equation}
Then $\bm\Theta_{11}=[\nabla^2L(\bm\alpha)]^+$. On the localisation and graph event used in Lemma \ref{lemma:lemma28}, which has probability at least $1-O(M^{-5})$,
\begin{equation}
\frac{c}{Mp}
\le
\|\bm\Theta_{11}\|_2
\le
\frac{C\kappa_3}{Mp},
\qquad
\frac{c}{Mp}
\le
[\bm\Theta_{11}]_{kk}
\le
\frac{C\kappa_3}{Mp},
\quad k=1,\ldots,M,
\label{eq:theta_diag_bounds}
\end{equation}
for constants $c,C>0$.
\begin{proof}
The block representation in \eqref{eq:lemma30_1} follows from Lemma \ref{lemma:lemma28}. Equivalently, if
\[
\begin{pmatrix}
\bm\Theta_{11}&\bm\beta\\
\bm\beta^T&\gamma
\end{pmatrix}
=
\begin{pmatrix}
\nabla^2L(\bm\alpha)&\bm1_M\\
\bm1_M^T&0
\end{pmatrix}^{-1},
\]
then multiplication by the bordered matrix gives
\[
\bm\Theta_{11}\bm1_M=0,\qquad
\bm\Theta_{11}\nabla^2L(\bm\alpha)+\bm\beta\bm1_M^T=\bm I_M.
\]
Multiplying the latter identity by $\bm1_M$ gives $\bm\beta=M^{-1}\bm1_M$, and the remaining block equations give $\gamma=0$. Thus $\bm\Theta_{11}$ is the Moore--Penrose pseudoinverse.
\\
\\
\noindent Let $\lambda_1\ge\cdots\ge\lambda_{M-1}>0$ denote the positive eigenvalues of the Hessian. Then the nonzero eigenvalues of $\bm\Theta_{11}$ are
$1/\lambda_1,\ldots,1/\lambda_{M-1}$. The Hessian bounds imply
\[
c_HMp/\kappa_3
\le
\lambda_{M-1}
\le
\lambda_1
\le
C_HMp,
\]
so
\[
\frac{1}{C_HMp}
\le
\|\bm\Theta_{11}\|_2
=
\frac{1}{\lambda_{M-1}}
\le
\frac{\kappa_3}{c_HMp}.
\]

\noindent For a diagonal element, expand
\[
\bm e_k=\sum_{l=1}^{M-1}\beta_l\bm v_l+\frac1M\bm1_M.
\]
Orthogonality gives $\sum_{l=1}^{M-1}\beta_l^2=1-1/M$. Hence
\[
[\bm\Theta_{11}]_{kk}
=
\sum_{l=1}^{M-1}\frac{\beta_l^2}{\lambda_l}.
\]
Using the upper and lower Hessian eigenvalue bounds yields
\[
\frac{1-1/M}{C_HMp}
\le
[\bm\Theta_{11}]_{kk}
\le
\frac{\kappa_3(1-1/M)}{c_HMp},
\]
which proves \eqref{eq:theta_diag_bounds}.
\end{proof}
\end{lemma}

\noindent Set
\[
\bm\Delta_\alpha:=\hat{\bm\alpha}_{\lambda_\delta}-\bm\alpha^*.
\]
On the graph and localisation event in Lemma \ref{lemma:lemma28}, $\nabla^2L(\hat{\bm\alpha}_{\lambda_\delta})$ has nullspace $\operatorname{span}(\bm1_M)$. Both $\hat{\bm\alpha}_{\lambda_\delta}$ and $\bm\alpha^*$ are centred, so $\bm\Delta_\alpha\in\bm1_M^\perp$, and the Moore--Penrose identity gives
$\hat{\bm\Theta}_{11}\nabla^2L(\hat{\bm\alpha}_{\lambda_\delta})\bm\Delta_\alpha=\bm\Delta_\alpha$.
Consequently, on this event Equation \eqref{eq:debiased_corrected} gives the exact decomposition
\begin{equation}
\begin{aligned}
\hat{\bm\alpha}^{\mathrm{db}}-\bm\alpha^*
={}&
\underbrace{(\hat{\bm\Theta}_{11}-\bm\Theta_{11}^*)
\left\{-\nabla L(\hat{\bm\alpha}_{\lambda_\delta})
+\nabla^2L(\hat{\bm\alpha}_{\lambda_\delta})\bm\Delta_\alpha\right\}}_{I_1}
\\
&+
\underbrace{\bm\Theta_{11}^*
\left\{\nabla L(\bm\alpha^*)-\nabla L(\hat{\bm\alpha}_{\lambda_\delta})
+\nabla^2L(\hat{\bm\alpha}_{\lambda_\delta})\bm\Delta_\alpha\right\}}_{I_2}
-
\bm\Theta_{11}^*\nabla L(\bm\alpha^*).
\end{aligned}
\label{eq:difference_dbtrue}
\end{equation}
\noindent We next bound the terms $I_1$ and $I_2$.

\begin{lemma}
\label{lemma:lemma31}
Suppose Assumption \ref{ass:rmax} holds and that the events in Lemmas \ref{lemma:lemma25}--\ref{lemma:lemma29} occur. For $I_1,I_2\in\mathbb R^M$ defined in Equation \eqref{eq:difference_dbtrue},
\begin{align}
\|I_1\|_\infty
&\lesssim
\sqrt M\,D_{\Theta,\lambda}
\left\{
\sqrt{\frac{Mp\log M}{n_{\mathrm{inf}}}}
+MpR_{\alpha,\lambda}^2
\right\},
\label{eq:I1_bound_corrected}\\
\|I_2\|_\infty
&\lesssim
\sqrt M\,\kappa_3 R_{\alpha,\lambda}^2.
\label{eq:I2_bound_corrected}
\end{align}
Here $R_{\alpha,\lambda}$ and $D_{\Theta,\lambda}$ are defined in Section \ref{sec:statistical-properties}. The term $I_1$ accounts for replacing the population inverse information matrix by its estimated version, whereas $I_2$ is the second-order Taylor remainder after multiplication by the population inverse information matrix.
\begin{proof}
We first bound $I_1$. By Equation \eqref{eq:difference_dbtrue},
\begin{equation}
I_1
=
(\hat{\bm\Theta}_{11}-\bm\Theta_{11}^*)
\left\{-\nabla L(\hat{\bm\alpha}_{\lambda_\delta})
+\nabla^2L(\hat{\bm\alpha}_{\lambda_\delta})(\hat{\bm\alpha}_{\lambda_\delta}-\bm\alpha^*)\right\}.
\label{eq:I1_alpha_exact}
\end{equation}
Recall $\bm\Delta_\alpha=\hat{\bm\alpha}_{\lambda_\delta}-\bm\alpha^*$. We decompose the vector in braces as
\begin{align}
&-\nabla L(\hat{\bm\alpha}_{\lambda_\delta})+\nabla^2L(\hat{\bm\alpha}_{\lambda_\delta})\bm\Delta_\alpha \notag\\
&\qquad=
-\nabla L(\bm\alpha^*)
-\left\{\nabla L(\hat{\bm\alpha}_{\lambda_\delta})-\nabla L(\bm\alpha^*)-\nabla^2L(\bm\alpha^*)\bm\Delta_\alpha\right\}
+\left\{\nabla^2L(\hat{\bm\alpha}_{\lambda_\delta})-\nabla^2L(\bm\alpha^*)\right\}\bm\Delta_\alpha.
\label{eq:I1_vector_decomposition}
\end{align}
Lemma \ref{lemma:lemma25} gives
\begin{equation}
\|\nabla L(\bm\alpha^*)\|_\infty
\lesssim \sqrt{\frac{Mp\log M}{n_{\mathrm{inf}}}}.
\label{eq:I1_score_part}
\end{equation}
By Lemma \ref{lemma:lemma26},
\begin{equation}
\left\|\nabla L(\hat{\bm\alpha}_{\lambda_\delta})-\nabla L(\bm\alpha^*)
-\nabla^2L(\bm\alpha^*)\bm\Delta_\alpha\right\|_\infty
\lesssim MpR_{\alpha,\lambda}^2.
\label{eq:I1_taylor_part}
\end{equation}
Finally, because each row of a weighted graph Laplacian has at most order $Mp$ nonzero mass on the high-probability degree event and the logistic weight is Lipschitz, the proof of Lemma \ref{lemma:lemma27} also gives the induced $\ell_\infty$ operator bound
\begin{equation}
\|\nabla^2L(\hat{\bm\alpha}_{\lambda_\delta})-\nabla^2L(\bm\alpha^*)\|_\infty
\lesssim MpR_{\alpha,\lambda}.
\label{eq:I1_hessian_part}
\end{equation}
Since $\|\bm\Delta_\alpha\|_\infty\lesssim R_{\alpha,\lambda}$, the last term in \eqref{eq:I1_vector_decomposition} is therefore $O(MpR_{\alpha,\lambda}^2)$ in $\ell_\infty$. Combining \eqref{eq:I1_score_part}--\eqref{eq:I1_hessian_part},
\begin{equation}
\left\|-\nabla L(\hat{\bm\alpha}_{\lambda_\delta})
+\nabla^2L(\hat{\bm\alpha}_{\lambda_\delta})\bm\Delta_\alpha\right\|_\infty
\lesssim
\sqrt{\frac{Mp\log M}{n_{\mathrm{inf}}}}+MpR_{\alpha,\lambda}^2.
\label{eq:I1_vector_bound}
\end{equation}
Using $\|A v\|_\infty\le \|A v\|_2\le \|A\|_2\|v\|_2\le \sqrt M\|A\|_2\|v\|_\infty$ and Lemma \ref{lemma:lemma29}, Equation \eqref{eq:I1_alpha_exact} yields
\begin{align*}
\|I_1\|_\infty
&\le
\sqrt M\,\|\hat{\bm\Theta}_{11}-\bm\Theta_{11}^*\|_2
\left\|-\nabla L(\hat{\bm\alpha}_{\lambda_\delta})
+\nabla^2L(\hat{\bm\alpha}_{\lambda_\delta})\bm\Delta_\alpha\right\|_\infty\\
&\lesssim
\sqrt M\,D_{\Theta,\lambda}
\left\{\sqrt{\frac{Mp\log M}{n_{\mathrm{inf}}}}+MpR_{\alpha,\lambda}^2\right\},
\end{align*}
which proves \eqref{eq:I1_bound_corrected}.
\\
\\
\noindent We next bound $I_2$. By Equation \eqref{eq:difference_dbtrue},
\begin{equation}
I_2
=
\bm\Theta_{11}^*
\left\{\nabla L(\bm\alpha^*)-\nabla L(\hat{\bm\alpha}_{\lambda_\delta})
+\nabla^2L(\hat{\bm\alpha}_{\lambda_\delta})\bm\Delta_\alpha\right\}.
\label{eq:I2_alpha_exact}
\end{equation}
Add and subtract $\nabla^2L(\bm\alpha^*)\bm\Delta_\alpha$ to obtain
\begin{align}
&\nabla L(\bm\alpha^*)-\nabla L(\hat{\bm\alpha}_{\lambda_\delta})
+\nabla^2L(\hat{\bm\alpha}_{\lambda_\delta})\bm\Delta_\alpha \notag\\
&\qquad=
-\left\{\nabla L(\hat{\bm\alpha}_{\lambda_\delta})-\nabla L(\bm\alpha^*)
-\nabla^2L(\bm\alpha^*)\bm\Delta_\alpha\right\}
+\left\{\nabla^2L(\hat{\bm\alpha}_{\lambda_\delta})-\nabla^2L(\bm\alpha^*)\right\}\bm\Delta_\alpha.
\label{eq:I2_vector_decomposition}
\end{align}
The first term on the right is $O(MpR_{\alpha,\lambda}^2)$ by Lemma \ref{lemma:lemma26}, and the second is of the same order by \eqref{eq:I1_hessian_part} and $\|\bm\Delta_\alpha\|_\infty\lesssim R_{\alpha,\lambda}$. Hence the vector in braces in \eqref{eq:I2_alpha_exact} has $\ell_\infty$ norm $O(MpR_{\alpha,\lambda}^2)$. Lemma \ref{lemma:lemma28} gives
\begin{equation*}
\|\bm\Theta_{11}^*\|_2\lesssim \frac{\kappa_3}{Mp}.
\end{equation*}
Consequently,
\begin{align*}
\|I_2\|_\infty
&\le
\sqrt M\,\|\bm\Theta_{11}^*\|_2
\left\|\nabla L(\bm\alpha^*)-\nabla L(\hat{\bm\alpha}_{\lambda_\delta})
+\nabla^2L(\hat{\bm\alpha}_{\lambda_\delta})\bm\Delta_\alpha\right\|_\infty\\
&\lesssim
\sqrt M\frac{\kappa_3}{Mp}(MpR_{\alpha,\lambda}^2)
=
\sqrt M\,\kappa_3R_{\alpha,\lambda}^2,
\end{align*}
which proves \eqref{eq:I2_bound_corrected}.
\end{proof}
\end{lemma}

\noindent Equation \eqref{eq:difference_dbtrue} can therefore be written as
\begin{equation}
\label{eq:difdefcomment}
\hat{\bm\alpha}^{\mathrm{db}}-\bm\alpha^*
=
-\bm\Theta_{11}^*\nabla L(\bm\alpha^*)+\bm r,
\end{equation}
where $\bm r=I_1+I_2$. Lemma \ref{lemma:lemma31} therefore yields
\begin{equation}
\|\bm r\|_\infty
\lesssim
\sqrt M\,D_{\Theta,\lambda}
\left\{\sqrt{\frac{Mp\log M}{n_{\mathrm{inf}}}}+MpR_{\alpha,\lambda}^2\right\}
+
\sqrt M\,\kappa_3R_{\alpha,\lambda}^2.
\label{eq:asymp_remainder_bound}
\end{equation}
Multiplying Equation \eqref{eq:asymp_remainder_bound} by $\sqrt{Mpn_{\mathrm{inf}}}$ and substituting $D_{\Theta,\lambda}=\kappa_3^2R_{\alpha,\lambda}/(Mp)$ gives
\begin{align*}
\sqrt{Mpn_{\mathrm{inf}}}\,\|\bm r\|_\infty
\lesssim{}&
\kappa_3^2R_{\alpha,\lambda}\sqrt{M\log M}
+M\kappa_3^2R_{\alpha,\lambda}^3\sqrt{pn_{\mathrm{inf}}}\\
&+M\kappa_3R_{\alpha,\lambda}^2\sqrt{pn_{\mathrm{inf}}}.
\end{align*}
The first and third terms are $o(1)$ by Assumption \ref{ass:rmax}. Moreover, the second condition in Equation \eqref{eq:inference_conditions_simple} implies $\kappa_3R_{\alpha,\lambda}=o(1)$, so the middle term is $o(1)$ relative to the third. Therefore, on an event whose probability tends to one,
\begin{equation}
\sqrt{Mpn_{\mathrm{inf}}}\,\|\bm r\|_\infty=o(1).
\label{eq:asymp_remainder_small}
\end{equation}
Consequently the left-hand side of Equation \eqref{eq:asymp_remainder_small} converges to zero in probability, and the leading term in \eqref{eq:difdefcomment} is the projected Fisher-score term $-\bm\Theta_{11}^*\nabla L(\bm\alpha^*)$.

\begin{lemma}[Conditional Berry--Esseen bound]
\label{lemma:lemma32}
Suppose the assumptions of Theorem \ref{thm:asympnormal} hold. Conditional on any realisation of the common comparison graph $\mathcal G$ satisfying the degree and Laplacian events $E_0\cap E_1$, for every fixed coordinate $j$,
\begin{equation}
\sup_{x\in\mathbb R}
\left|
\mathbb P\!\left(
\frac{\sqrt{n_{\mathrm{inf}}}\,[\bm\Theta_{11}^*]_j\nabla L(\bm\alpha^*)}
{\sqrt{[\bm\Theta_{11}^*]_{jj}}}
\le x\,\middle|\,\mathcal G
\right)-\Phi(x)
\right|
\lesssim
\frac{\kappa_3}{\sqrt{Mpn_{\mathrm{inf}}}}.
\label{eq:lemma32_1}
\end{equation}
For every fixed pair $j\ne l$,
\begin{equation}
\sup_{x\in\mathbb R}
\left|
\mathbb P\!\left(
\frac{\sqrt{n_{\mathrm{inf}}}\,([\bm\Theta_{11}^*]_j-[\bm\Theta_{11}^*]_l)\nabla L(\bm\alpha^*)}
{\sqrt{(\bm e_j-\bm e_l)^T\bm\Theta_{11}^*(\bm e_j-\bm e_l)}}
\le x\,\middle|\,\mathcal G
\right)-\Phi(x)
\right|
\lesssim
\frac{\kappa_3}{\sqrt{Mpn_{\mathrm{inf}}}}.
\label{eq:lemma32_pairwise}
\end{equation}
Here $\Phi$ is the standard normal distribution function. In particular, both conditional normal approximations are asymptotically valid under Assumption \ref{ass:rmax}.
\begin{proof}
We prove \eqref{eq:lemma32_1}; the proof of \eqref{eq:lemma32_pairwise} is identical after replacing the $j$-th row of $\bm\Theta_{11}^*$ by the row contrast $([\bm\Theta_{11}^*]_j-[\bm\Theta_{11}^*]_l)$.
\\
\\
\noindent For $(k,q)\in\mathcal E$, $k>q$, let
\[
\pi_{q,k}^*
:=
\frac{\exp(\alpha_k^*)}{\exp(\alpha_k^*)+\exp(\alpha_q^*)},
\qquad
b_{q,k,j}:=[\bm\Theta_{11}^*]_j(\bm e_k-\bm e_q),
\]
and define
\[
x_{q,k}^{(i)}:=b_{q,k,j}\{\pi_{q,k}^*-y_{q,k}^{(i)}\}.
\]
Conditional on $\mathcal G$, the variables $x_{q,k}^{(i)}$ are independent and centred. Since the inference negative log-likelihood uses the average of the $n_{\mathrm{inf}}$ repeated comparisons,
\begin{equation}
[\bm\Theta_{11}^*]_j\nabla L(\bm\alpha^*)
=
\frac1{n_{\mathrm{inf}}}
\sum_{i=1}^{n_{\mathrm{inf}}}
\sum_{(k,q)\in\mathcal E,\,k>q}
x_{q,k}^{(i)}.
\label{eq:BE_score_sum}
\end{equation}
For one repetition,
\begin{align}
\sum_{(k,q)\in\mathcal E,\,k>q}
\operatorname{Var}(x_{q,k}^{(i)}\mid\mathcal G)
&=
[\bm\Theta_{11}^*]_j
\nabla^2L(\bm\alpha^*)
[\bm\Theta_{11}^*]_j^T\\
&=[\bm\Theta_{11}^*]_{jj}.
\label{eq:BE_variance_identity}
\end{align}
The second equality follows from the Moore--Penrose identities. Indeed, Lemma \ref{lemma:lemma28} gives $\bm\Theta_{11}^*=[\nabla^2L(\bm\alpha^*)]^+$, and hence
\[
\bm\Theta_{11}^*\nabla^2L(\bm\alpha^*)\bm\Theta_{11}^*
=\bm\Theta_{11}^*.
\]
This is the projected Fisher-information identity; it replaces the ordinary inverse identity, which cannot hold because the Bradley--Terry Hessian is singular in the direction $\bm1_M$.
\\
\\
\noindent For the third absolute moment, a centred Bernoulli random variable satisfies
\begin{align}
\mathbb E\left|x_{q,k}^{(i)}\right|^3
&=
|b_{q,k,j}|^3\,
\pi_{q,k}^*(1-\pi_{q,k}^*)
\left\{(1-\pi_{q,k}^*)^2+(\pi_{q,k}^*)^2\right\}\\
&\le
|b_{q,k,j}|\,b_{q,k,j}^2
\pi_{q,k}^*(1-\pi_{q,k}^*).
\label{eq:BE_third_moment_edge}
\end{align}
Summing over the graph and using \eqref{eq:BE_variance_identity},
\begin{equation}
\sum_{(k,q)\in\mathcal E,\,k>q}
\mathbb E|x_{q,k}^{(i)}|^3
\le
[\bm\Theta_{11}^*]_{jj}
\max_{(k,q)\in\mathcal E}|b_{q,k,j}|.
\label{eq:BE_third_sum}
\end{equation}
By Lemma \ref{lemma:lemma28},
\[
|b_{q,k,j}|
\le
\|[\bm\Theta_{11}^*]_j\|_2\|\bm e_k-\bm e_q\|_2
\le
\sqrt2\,\|\bm\Theta_{11}^*\|_2
\lesssim \frac{\kappa_3}{Mp}.
\]
Consequently,
\begin{equation}
\sum_{(k,q)\in\mathcal E,\,k>q}
\mathbb E|x_{q,k}^{(i)}|^3
\lesssim
[\bm\Theta_{11}^*]_{jj}\frac{\kappa_3}{Mp}.
\label{eq:BE_third_final}
\end{equation}
Applying the Berry--Esseen theorem to the $n_{\mathrm{inf}}|\mathcal E|$ independent summands in \eqref{eq:BE_score_sum} and using \eqref{eq:BE_variance_identity} gives
\begin{align*}
&\sup_x\left|
\mathbb P\left(
\frac{\sqrt{n_{\mathrm{inf}}}[\bm\Theta_{11}^*]_j\nabla L(\bm\alpha^*)}
{\sqrt{[\bm\Theta_{11}^*]_{jj}}}
\le x\,\middle|\,\mathcal G\right)-\Phi(x)
\right|\\
&\qquad\lesssim
\frac{n_{\mathrm{inf}}[\bm\Theta_{11}^*]_{jj}\kappa_3/(Mp)}
{\{n_{\mathrm{inf}}[\bm\Theta_{11}^*]_{jj}\}^{3/2}}.
\end{align*}
Lemma \ref{lemma:lemma30} gives $[\bm\Theta_{11}^*]_{jj}\gtrsim1/(Mp)$, and hence the last display is
\[
O\left(\frac{\kappa_3}{\sqrt{Mpn_{\mathrm{inf}}}}\right)=o(1)
\]
under Assumption \ref{ass:rmax}. This proves the bound in \eqref{eq:lemma32_1}.
\\
\\
\noindent For the pairwise contrast, put $\bm v=\bm e_j-\bm e_l\in\bm1_M^\perp$. Its variance for one repetition is $\bm v^T\bm\Theta_{11}^*\bm v$. Since the positive eigenvalues of $\bm\Theta_{11}^*$ are the reciprocals of those of the Hessian and $\lambda_{\max}(\nabla^2L)\lesssim Mp$, we have $\bm v^T\bm\Theta_{11}^*\bm v\gtrsim1/(Mp)$. Moreover, for each edge coefficient, $|\bm v^T\bm\Theta_{11}^*(\bm e_k-\bm e_q)|\lesssim\kappa_3/(Mp)$ by Lemma \ref{lemma:lemma28}. Repeating the preceding third-moment argument therefore gives the bound in \eqref{eq:lemma32_pairwise}.
\end{proof}
\end{lemma}

\noindent We are now ready to prove Theorem \ref{thm:asympnormal} and Corollary \ref{crl:asympnormal2}.\\
\noindent \textbf{Proof of Theorem \ref{thm:asympnormal}}
\begin{proof}
Equation \eqref{eq:difdefcomment} gives, coordinatewise,
\[
\hat\alpha_j^{\mathrm{db}}-\alpha_j^*
=-[\bm\Theta_{11}^*]_j\nabla L(\bm\alpha^*)+r_j.
\]
By Equation \eqref{eq:asymp_remainder_small}, the normalised remainder satisfies
\[
\frac{\sqrt{n_{\mathrm{inf}}}\,|r_j|}{\sqrt{[\bm\Theta_{11}^*]_{jj}}}=o(1)
\]
on an event whose probability tends to one, because Lemma \ref{lemma:lemma30} gives $[\bm\Theta_{11}^*]_{jj}\gtrsim1/(Mp)$.

Let
\[
Z_j
:=
\frac{\sqrt{n_{\mathrm{inf}}}[\bm\Theta_{11}^*]_j\nabla L(\bm\alpha^*)}
{\sqrt{[\bm\Theta_{11}^*]_{jj}}}.
\]
Let $E_{\mathcal G}:=E_0\cap E_1$, and extend $Z_j$ arbitrarily on $E_{\mathcal G}^c$. Lemma \ref{lemma:lemma32} gives, on $E_{\mathcal G}$,
\[
\sup_x\left|\mathbb P(Z_j\le x\mid\mathcal G)-\Phi(x)\right|
\lesssim
\frac{\kappa_3}{\sqrt{Mpn_{\mathrm{inf}}}}.
\]
Integrating this conditional bound over $\mathcal G$ and bounding the contribution of $E_{\mathcal G}^c$ by $\mathbb P(E_{\mathcal G}^c)=O(M^{-10})$ shows that
\[
\sup_x|\mathbb P(Z_j\le x)-\Phi(x)|
\lesssim
\frac{\kappa_3}{\sqrt{Mpn_{\mathrm{inf}}}}+O(M^{-10})
=o(1).
\]
Thus $Z_j\xrightarrow{d}N(0,1)$ after all sources of randomness are taken into account. Since the standard normal distribution is symmetric, $-Z_j\xrightarrow{d}N(0,1)$ as well. Slutsky's theorem together with the negligible remainder proves Equation \eqref{eq:asympnormal_corrected}.
\end{proof}

\noindent \textbf{Proof of Corollary \ref{crl:asympnormal2}}
\begin{proof}
Lemma \ref{lemma:lemma29} gives
\[
\|\hat{\bm\Theta}_{11}-\bm\Theta_{11}^*\|_2
\lesssim
D_{\Theta,\lambda}
=
\frac{\kappa_3^2}{Mp}R_{\alpha,\lambda}.
\]
For each fixed $j$, Lemma \ref{lemma:lemma30} gives $[\bm\Theta_{11}^*]_{jj}\gtrsim1/(Mp)$. Hence
\[
\left|
\frac{[\hat{\bm\Theta}_{11}]_{jj}}
{[\bm\Theta_{11}^*]_{jj}}-1
\right|
\lesssim
\kappa_3^2R_{\alpha,\lambda}
=o(1)
\]
with probability tending to one under Assumption \ref{ass:rmax}. Therefore
$[\hat{\bm\Theta}_{11}]_{jj}^{1/2}/[\bm\Theta_{11}^*]_{jj}^{1/2}\to1$ in probability, and the studentised result follows from Theorem \ref{thm:asympnormal} and Slutsky's theorem.
\end{proof}

\end{document}